\documentclass[journal,article,submit,pdftex,moreauthors]{Definitions/mdpi} 
\usepackage{amsmath, amssymb}
\usepackage{amsfonts}
\usepackage{amssymb}
\usepackage{color}

\newcommand{\ms}{\overline{\text{\tiny MS}}}
\newcommand{\os}{\overline{\text{\tiny os}}}
\usepackage{graphics}
\usepackage{dsfont}

\newcommand{\nn}{\nonumber}
\firstpage{1} 
\pubvolume{1}
\issuenum{1}
\articlenumber{0}
\pubyear{2025}
\copyrightyear{2025}
\datereceived{ } 
\daterevised{ } 
\dateaccepted{ } 
\datepublished{ } 
\hreflink{https://doi.org/} 

\Title{Quark-meson diquark model and color superconductivity in dense quark matter}

\TitleCitation{Title}

\Author{Jens O. Andersen $^{1,\dagger,\ddagger}$\orcidA{}and Mathias P. Nødtvedt $^{1,\ddagger}$}

\AuthorNames{Jens O. Andersen and Mathias Nødtvedt}

\isAPAStyle{%
       \AuthorCitation{Lastname, F., Lastname, F., \& Lastname, F.}
         }{%
        \isChicagoStyle{%
        \AuthorCitation{Lastname, Firstname, Firstname Lastname, and Firstname Lastname.}
        }{
        \AuthorCitation{Lastname, F.; Lastname, F.; Lastname, F.}
        }
}

\address{%
$^{1}$ \quad Department of Physics, Faculty of Natural Sciences,NTNU,
Norwegian University of Science and Technology, Høgskoleringen 5, N-7491 Trondheim, Norway; jensoa@ntnu.no; mathias.p.nodtvedt@ntnu.no}

\corres{Correspondence: jensoa@ntnu.no}

\secondnote{The authors contributed equally to this work.}

\abstract{
We consider the two- and three-flavor QMD models as renormalizable low-energy models for QCD at finite quark chemical potentials with quarks, mesons, and diquarks as effective degrees of freedom.
Using the on-shell scheme the parameters in the scalar sector can be fixed and expressed in terms of observed meson masses and decay constants. The remaining parameters can be varied.
In the QMD models, all the symmetries are global, including the $SU(N_c)$ symmetry. 
The breaking of the global symmetries gives rise to a number of Goldstone bosons depending on the symmetry-breaking pattern, i.e. whether the system is in the 2SC phase or the color-flavor-locked (CFL) phase.
This is in contrast to perturbative QCD, where some of the gauge bosons become massive via the Higgs mechanism. We classify the Goldstone bosons and show that their type and number are in accordance with general counting rules. 
The thermodynamic potential $\Omega$ is calculated in the mean-field approximation, where we include quark loops, while mesons and diquarks are treated at tree level.
As important applications, we study the properties of the 
pion-condensed phase at finite isospin chemical potential, and the 2SC and CFL phases at finite baryon chemical potential. We present a few numerical results focusing on the speed of sound, gaps, and condensates. It is shown that the BCS gaps approaches a constant for large isospin and baryon chemical potentials and that the speed of sound approaches the conformal value from above in the same limit.
}


\begin{document}

\setcounter{section}{0} 


\section{Introduction}
QCD is the modern theory that describes the interactions between quarks and gluons, which are the constituents of hadrons. The behavior of bulk hadronic matter has been studied for decades, largely motivated by attempts to describe neutron stars.
It is expected that hadronic matter undergoes a transition from nuclear degrees of freedom to deconfined quarks as the density increases. The behavior of ultradense matter is relevant for the description of the most massive neutron stars in the universe, where the fundamental question is: Are the cores of these stars so dense that they contain one or more phases of quark matter?

The thermodynamic properties of strongly interacting matter at asymptotically large (baryon) densities can, due to asymptotic freedom, be calculated in perturbation theory. 
Perturbative calculations in dense and cold QCD are almost as old as the theory itself, starting with the ${\cal O}(\alpha_s^2)$-calculations of the thermodynamic potential in massless QCD in the late 1970s~\cite{freed0,freed1,freed2,baluni}.~\footnote{To obtain results to a given order
in $\alpha_s$ requires resummation of diagrams from all order of perturbation theory.} In recent years there has been significant progress in the calculation of
thermodynamic quantities in cold and dense QCD. 
Finite quark-mass effects have been included in Refs.~\cite{quarkmass1,quarkmass2}.
An ${\cal O}(\alpha_s^3\log\alpha_s)$ result was obtained a few years ago~\cite{gorda}.
Finally, an outline of how to obtain the complete $\alpha_s^3$ result 
has been given in Ref.~\cite{alex} and requires the development of highly sophisticated calculational techniques.

The perturbative calculations mentioned above are expansions about the trivial ground state, which means that there are no condensates.
However, this is not
the correct ground state at high density. It was realized a long time ago~\cite{color1,color2,bailin} that quark matter at high density involves
fermion pairing.~\footnote{There are some recent two-loop calculations of the pressure that take 
the quark gap into account~\cite{brownie0,minato,brownie}.}
Due to an attractive interaction channel provided by one-gluon
exchange, the Fermi surface is rendered unstable, which
leads to the formation of Cooper pairs. A Cooper pair of two quarks is a colored object and therefore not a gauge-invariant quantity. One may therefore think that the gauge symmetry of QCD is broken by Cooper pairs, however, according to
Elitzur's theorem local symmetries cannot be broken spontaneously~\cite{elitzur}. In perturbative calculations, one first fixes the gauge and then introduces a
condensate, which "breaks" the symmetry~\cite{frankie}. However, the gap in the fermionic quasiparticle spectrum depends on the magnitude of the condensate, which is a gauge-invariant quantity.

At asymptotically high densities, one can ignore the mass of the $s$-quark, and quarks of all flavors and colors participate in the pairing on an equal footing.
There are a left-handed and a right-handed condensate, which are invariant under $SU(3)_{L+c}$ and $SU(3)_{R+c}$, respectively. This phase is called the color-flavor-locked (CFL) phase, as the ground state
is symmetric under the group $SU(3)_{L+R+c}$.~\footnote{In the QCD vacuum $SU(3)_L$ and $SU(3)_R$ rotations are similarly locked, but now by a nonzero quark-antiquark condensate.} When the baryon density is lowered, one can no longer ignore the 
the effects of the $s$-quark mass. The nonzero $m_s$ puts stress on the BCS pairing caused by a splitting of the Fermi momenta of the different quark flavors.
If this stress is moderate, the system responds by a flavor rotation of the ground state, which can be interpreted as kaon condensation~\cite{bedak0}.
Decreasing the density further, it is not energetically favorable that all colors and flavors participate in the pairing. One possibility is then the
so-called 2SC phase, where the red and green $u$ and the $d$ quarks form Cooper pairs.
The $s$ quarks of any color and the blue $u$ and $d$
quarks are unpaired.
However, another possibility is a transition directly from the CFL phase to  unpaired or normal quark matter without going via the 2SC phase. In NJL model calculations, there is no 2SC window if the diquark coupling is weak~\cite{markraja}. However, for larger values of the coupling, there is a 2SC
region in the phase diagram~\cite{abuki,rusten}.

A related problem is the behavior of strongly interacting matter at finite isospin  $\mu_I$. One of the reasons for the interest in QCD at finite isospin is that it is amenable to Monte Carlo methods.
The determinant of the Dirac operator is
manifestly real, and there is no sign problem. The same remark applies to two-color QCD at finite baryon chemical potential and QCD in a magnetic field at $\mu_B=0$.
It is then possible to confront the results of low-energy models and theories with the predictions of lattice simulations.
In recent years, a number of high-precision lattice simulations at finite $\mu_I$
have been performed~\cite{isobrandt,abbott}, and it is now established that at $T=0$, there is a second-order transition from the vacuum to a pion-condensed phase, exactly at $\mu_I=m_{\pi}$, where $m_{\pi}$ is the physical pion mass. The critical exponents are of the $O(2)$ model, as expected. The speed of sound $c_s$, which is obtained from the equation of state,
is perhaps the most interesting quantity. It turns out that $c_s$ has a peak structure as a function of $\mu_I$ and that it approaches the Stefan-Boltzmann
result or the conformal limit $c_s^2={1\over3}$ from above. 
Chiral perturbation theory ($\chi$pt) is a low-energy effective theory for QCD~\cite{gasser1,gasser2}, which gives model-independent predictions. In the context of finite isospin density, it is
expected to give reliable results within its domain of validity, which is small values
of $\mu_I$. It has been shown to agree very well with lattice simulations 
for chemical potentials up to 1.2-1.3 $m_{\pi}$ depending on the quantity in question.
On the other hand, $\chi$pt completely fails at large $\mu_I$.
In the limit $\mu_I\rightarrow\infty$, the speed of sound approaches one. The problem is that $\chi$pt is based on mesonic degrees of freedom, which are not suitable at large chemical potentials. In this regime, perturbation theory is expected to be valid and the zeroth-order result is simply a free Fermi gas
with a Fermi sphere. In analogy with color superconductivity, there is an attractive interaction due to one-gluon exchange, which renders the Fermi surface unstable. This gives rise to loosely bound Cooper pairs of $u$ and $\bar{d}$ quarks 
(or $d$ and $\bar{u}$ quarks depending on the sign of $\mu_I$).
Thus, these degrees of freedom are suitable, rather than tightly bound pions. The quark-meson (QM) model, or the linear sigma model coupled to quarks~\cite{qmrischke}, 
is a low-energy effective model that has quark and meson degrees of freedom. We shall see later that the QM model is reproducing the behavior of $c_s^2$ as found in lattice simulations surprisingly well over a very wide range of values of 
$\mu_I$~\cite{kojo,brandtqm,us2,isofarias}.

There is another aspect of strongly interacting matter that we have not yet addressed, namely that of strong magnetic fields~\cite{reviewb}.
It is known that neutron stars have very strong magnetic
fields on their surface. For neutron stars, these fields can be as strong as $10^{12}$ Gauss, while for magnetars, they are even higher, up to $10^{16}$ Gauss~\cite{duncan}. In the interior of neutron stars, it is expected that the magnetic fields can be several magnitudes larger. Introducing a magnetic field gives rise to a new energy scale in the system.
It is important to know the response of nuclear matter and color-superconducting matter to a strong magnetic field, since the equation of state may be altered significantly and possibly 
affect the mass-radius relation for hybrid stars.
Although we do not present any calculations involving magnetic fields, we will briefly discuss possible phases of QCD and what to expect at the end of the paper.

In this paper, we discuss the quark-meson diquark model as a low-energy effective model for two- and three-flavor QCD at finite quark chemical potentials, extending our previous work~\cite{us,us2,us3},
see also~\cite{bjs}. 
This model was introduced in the
context of two-color QCD at finite temperature and baryon density in~\cite{smekal}.
The model was studied in the mean-field approximation, as well as using the
functional renormalization group (FRG), in which bosonic fluctuations are included.
The QMD model was also discussed in some detail in~\cite{pawlow,braun2}
and further investigated in~\cite{stoll}.
We write down a low-energy effective Lagrangian with renormalizable
interactions, i. e. we include operators that are at most dimension four.
The fields are canonically normalized and all the terms in the Lagrangian are multiplied by a parameter, either a bare mass or a bare coupling.
We organize the parameters into two groups: if the corresponding operator only
contains mesonic fields, it belongs to the scalar-pseudoscalar sector. If it contains
at least one diquark field, it belongs to the diquark sector.
The masses and couplings in the scalar-pseudoscalar sector can be related
to physical meson masses and meson decay constants.
At tree level, this is straightforward. In~\cite{us}, it was shown how one can determine masses and coupling beyond tree level by combining the $\overline{\rm MS}$
and on-shell schemes in a consistent way. For three flavors, these calculations
are significantly more involved than for two flavors, although exploiting some
useful relations makes the problem more tractable.
The remaining masses and couplings are in the diquark sector and are undetermined by 
the parameter fixing in the vacuum. They can be varied in the same way that
diquark and vector couplings in the NJL model are varied.
Instead, we simply choose numerical values for the running parameters at a 
fixed renormalization scale $\Lambda_0$. In Ref.~\cite{us}, it was shown that
physically sound results were obtained by making judicious choices of these parameters. It turns out that the results are mainly sensitive to the values of the diquark mass and quark-diquark coupling.

The paper is organized as follows. In Sections 2 and 3, we discuss the properties
of the two-and three-flavor QMD model. Section 4 is devoted to
a discussion of chemical potentials and charge neutrality.
In Section 5, we consider pion condensation
in the two-flavor QMD model at finite isospin density, while Section 6 is devoted to the 2SC and CFl phases in dense QCD. In Section 7, we take a closer look at the Goldstone bosons 
and their properties in these phases. In Section 8, we summarize
and discuss future directions.
In Appendices A-E, we present some calculations details of 
dimensional regularization, parameter fixing, and renormalization of 
the thermodynamic potential.

\section{Two-flavor quark-meson diquark model}
In this section, we discuss the degrees of freedom and the Lagrangian of the two-flavor quark-meson diquark model. The purely mesonic sector for different $N_f$ was discussed in detail in~\cite{roder}.
The Lagrangian is written as 
\begin{eqnarray}
\nonumber
{\cal L}&=&{\cal L}_{\rm scalar} +{\cal L}_{\rm quark}
+ {\cal L}_{\rm scalar-quark} +
{\cal L}_{\rm diquark}+{\cal L}_{\rm scalar-diquark}
+{\cal L}_{\rm quark-diquark}
\\ &&
+\delta{\cal L}_{\rm scalar}\;.
\end{eqnarray}
We discuss each term separately below.

Conventionally, the scalar sector in the two-flavor quark-meson model
is restricted to the four fields
$\sigma$ and $\pi_a$. The fields can be organized in the $2\times2$ matrix 
\begin{eqnarray}
\Sigma&=&{1\over2}(\sigma+i\tau_a\pi_a)
={1\over2}\left(
\begin{array}{cc}
\sigma+i\pi^0&i\sqrt{2}\pi^+\\
i\sqrt{2}\pi^-&\sigma-i\pi^0\\
\end{array}\right)
\;,    
\label{sigmi}
\end{eqnarray}
where $\tau_a$ are the Pauli matrices, $\pi^0=\pi_3$ and $\pi^{\pm}={1\over\sqrt{2}}(\pi_1\mp i\pi_2)$. Under $SU(2)_L\times SU(2)_R$ transformations the 
matrix Eq.~(\ref{sigmi}) transforms as
\begin{eqnarray}
\label{trafo}
\Sigma&\rightarrow&L\Sigma R^{\dagger}\;,    
\end{eqnarray}
where $L$ is an $SU(2)_L$ rotation and $R$ is an $SU(2)_R$ rotation.
In the following, we denote the trace of a matrix $A$ by $\langle A\rangle$.
Using the cyclicity of the trace, it then follows from the transformation property Eq.~(\ref{trafo}) that
$\langle\Sigma\rangle$ is invariant only if $L=R$ or 
under vector rotations $SU(2)_{L+R}$. Knowing the transformation properties of
$\Sigma$, we can construct terms that are invariant under $SU(2)_L\times SU(2)_R$.
Again using the cyclic property of the trace, we find that the mass term 
$\langle\Sigma^{\dagger}\Sigma\rangle$ is invariant. 
Similarly, the kinetic term $\langle D_{\mu}\Sigma^{\dagger}D^{\mu}\Sigma\rangle$,
where $D_{\mu}$ is the covariant derivative, is invariant. 
We shall have more to say about $D_{\mu}$ below.
It also follows that any power $n$ of $\langle\Sigma^{\dagger}\Sigma\rangle$ is invariant. Restricting ourselves to renormalizable terms, $n=2$ is the highest
power. One might think that the term 
$(\langle\Sigma^{\dagger}\Sigma)^2\rangle$ should also be included; however, for two flavors, this is not an independent operator. In fact, explicit
calculations show that it is equal to $\langle\Sigma^{\dagger}\Sigma\rangle^2$. 
In the same vein, one can easily show that 
$\det\Sigma+\det\Sigma^{\dagger}=\langle\Sigma^{\dagger}\Sigma\rangle$ and is not an independent
term. The terms in the scalar sector including the explicit symmetry breaking term are

\begin{eqnarray}
{\cal L}_{\rm scalar}&=&
\langle D_{\mu}\Sigma^{\dagger}D^{\mu}\Sigma\rangle-m^2\langle\Sigma^{\dagger}\Sigma\rangle-
{\lambda_1}\langle\Sigma^{\dagger}\Sigma\rangle^2
+\langle H(\Sigma+\Sigma^{\dagger})\rangle\;,
\label{lscalar}
\end{eqnarray}
where $m^2$ is the mass parameter, $\lambda_1$ is the scalar self-coupling, and $H$
is the matrix
\begin{eqnarray}
\label{bryt1}
H&=&{1\over2}h_0+{1\over2}h_3\tau_3\;.    
\end{eqnarray}
A nonzero $h_0$ in Eq.~(\ref{bryt1}), explicitly breaks $SU(2)_L\times SU(2)_R$
down to $SU(2)_V$, whereas a nonzero $h_3$ explicitly breaks the isospin
symmetry $SU(2)_{I_3}$ down to $U(1)_V$.
The covariant derivative of $\Sigma$ in Eq.~(\ref{lscalar})
contains the appropriate chemical potentials as the zeroth component of a
gauge field in the usual way. For $\sigma$ and $\pi^0$, these vanish, while
for $\pi^{\pm}$, the covariant derivative is 
$D^{\mu}\pi^{\pm}=(\partial^{\mu}\pm i\mu_I\delta^{0\mu})\pi^{\pm}$, where
$\mu_I=\mu_u-\mu_d$ is the isospin chemical potential.

We next consider the quark terms and introduce the notation
\begin{eqnarray}
\psi&=&( \psi_i^a)^T = (\psi_u^r,\psi_d^r,\psi_u^g,\psi_d^g,\psi_u^b,\psi_d^b)^T\;,    
\end{eqnarray}
where  $i=u,d$ is a flavor index and $a=r,g,b$ is a color index.
$\psi$ is a flavor doublet and a color triplet.
The quark term is 
\begin{eqnarray}
{\cal L}_{\rm quark}&=&
\bar{\psi}(i/\!\!\!{\partial}+\gamma^0\hat{\mu})\psi\;,    
\label{lquark}
\end{eqnarray}
where $\hat{\mu}$ is the quark chemical potential matrix. It is diagonal in flavor and color space, whose components are denoted by $\mu_{ia}$. They are not all independent,
but can be expressed in terms of four chemical potentials $\mu_B$, $\mu_Q$, 
$\mu_3$, and $\mu_8$, which are associated with baryon number, electric charge and color charges $Q_3$ and $Q_8$. We will discuss this in more detail in Sect.~\ref{chempot1}.

We next consider the scalar-quark interactions terms.
The quark field is written as $\psi=\psi_L+\psi_R$, where the left-handed field
transforms as $\psi_L\rightarrow L\psi_L$ and the right-handed field transforms as
$\psi_R\rightarrow R\psi_R$. It is then clear that the term
\begin{eqnarray}
{\cal L}_{\rm scalar-quark}&=&
-{1\over2}g\bar{\psi}[\sigma+i\gamma^5\tau_a\pi_a]\psi
=-g\bar{\psi}_L\Sigma\psi_R-g\bar{\psi}_R\Sigma^{\dagger}\psi_L
\;,
\label{lsq}
\end{eqnarray}
where $g$ is the Yukawa coupling, is invariant under $SU(2)_L\times SU(2)_R$.
The sum of Eqs.~(\ref{lscalar}), ~(\ref{lquark}), and~(\ref{lsq}) constitute the 
standard two-flavor quark-meson model. It is equivalent to the $O(4)$-symmetric quark-meson model, where $\vec{\phi}=(\sigma,\pi_1,\pi_2,\pi_3)^T$ is an $O(4)$ vector~\footnote{For example, the mass term in the $O(4)$ model is written as ${1\over2}m^2\vec{\phi}^T\cdot\vec{\phi}$.}. The symmetry group $SU(2)_L\times SU(2)_R$ can be extended to
$U(2)_L\times U(2)_R=U(1)\times U(1)_A\times SU(2)_L\times SU(2)_R$ by extending the particle content in the scalar sector.
The new particles are the parity partners, namely $\eta$, $a_0$, and $a^{\pm}_0$.
The field $\Sigma$ is parameterized as
\begin{eqnarray}
\label{repr}
\Sigma&=&T_a(\sigma_a+i\pi_a)\;,
\end{eqnarray}
where $T_a$ are the four generators of the group $U(2)=SU(2)\times U(1)$,
$T_0={1\over2}\mathds{1}$, and $T_i={1\over2}\tau_i$, normalized to
$\langle T_aT_b\rangle={1\over2}\delta_{ab}$.
The new fields are $\sigma_i=a_i$ (for $i=1,2,3$) and $\pi_0=\eta$. 
The matrices in Eq.~(\ref{repr}) are
\begin{eqnarray}
T_a\sigma_a&=&{1\over2}\left(
\begin{array}{cc}
\sigma+a_0&\sqrt{2}a^+\\
\sqrt{2}a^-&\sigma-a_0\\
\end{array}\right)\;,    
\hspace{1cm}
T_a\pi_a={1\over2}\left(
\begin{array}{cc}
\eta+\pi^0&\sqrt{2}\pi^+\\
\sqrt{2}\pi^-&\eta-\pi^0\\
\end{array}\right)\;,
\end{eqnarray}
where we have defined 
$a^{\pm}={1\over\sqrt{2}}(a_1\mp ia_2)$.
It is known that the $U(1)_A$ symmetry is broken by instantons in the QCD
vacuum. In effective low-energy models, this effect is mimicked by a determinant term
proportional to $\langle\det\Sigma+\det\Sigma^{\dagger}\rangle$, which also 
explicitly breaks the $U(1)_A$.
The determinant term reads
\begin{eqnarray}
\langle\det\Sigma+\det\Sigma^{\dagger}\rangle={1\over2}(\sigma^2+\vec{\pi}^2-\eta^2-\vec{a}^2)\;,
\end{eqnarray}
and is independent of the term 
$\langle\Sigma^{\dagger}\Sigma\rangle$. Similarly, the term
\begin{eqnarray}
    \langle(\Sigma^{\dagger}\Sigma)^2\rangle &=& \frac{1}{8}(\sigma^2+\vec{\pi}^2 + \eta^2+\vec{a}^2)^2+ \frac{1}{2}\left[(\sigma^2+\vec{\pi}^2)(\eta^2+\vec{a}^2)- (\sigma\eta-\vec{\pi}\cdot\vec{a})^2\right]\;,
\end{eqnarray} 
is independent of $\langle\Sigma^{\dagger}\Sigma\rangle^2$ and must also be included in the 
Lagrangian. The additional terms in the Lagrangian are therefore
\begin{eqnarray}
\delta{\cal L}_{\rm scalar}^{}&=&c\left[\det\Sigma+\det\Sigma^{\dagger}\right]
-{\lambda_2}\langle(\Sigma^{\dagger}\Sigma)^2\rangle\;,
\label{extra0}
\end{eqnarray}
where $c$ and $\lambda_2$ are couplings.
Finally, the scalar-quark interaction term, Eq.~(\ref{lsq}), is then replaced by
\begin{eqnarray}
{\cal L}_{\rm scalar-quark}&=&
-g\bar{\psi}T_a[\sigma_a+i\gamma^5\pi_a]\psi\;.
\label{lsq1}
\end{eqnarray}
The sum of Eqs.~(\ref{lscalar}), ~(\ref{lquark}),~(\ref{extra0}), and~(\ref{lsq1})
constitute the extended two-flavor quark-meson model.

We finally consider the diquark sector and its coupling to the scalar sector.
As mentioned in the Introduction, there is an attractive interaction channel in the
one-gluon exchange approximation of the quark-quark scattering amplitude.
This is an antisymmetric antitriplet channel that is responsible for rendering
the Fermi surface unstable and the formation of diquark pairs. 
There is also a repulsive symmetric sextet channel, which gives rise to an induced pairing.
This induced pairing does not break any additional symmetries.
It turns out that this pairing is much weaker and we shall neglect it in the following.
However, it can in principle be included by introducing appropriate extra degrees of freedom.

It is most energetically
favorable that quarks form spin-zero Cooper pairs. Since a 
diquark pair is antisymmetric in Dirac indices, the Pauli principle dictates that it must also 
be antisymmetric in flavor. The form of the condensate is~\cite{alfordrev,shovkovy}
\begin{eqnarray}
\label{kondi}
\langle\Delta_a\rangle&\equiv&
\langle\bar{\psi}^b_i\gamma^5(\psi_j^c)^C\rangle\epsilon_{ij}\epsilon^{bca}\;,  
\end{eqnarray}
where 
$\psi^C=C\bar{\psi}^T$ and 
$C=i\gamma^2\gamma^0$ is the charge conjugation operator and 
the $\gamma^5$-matrix ensures that the condensate is a scalar and not a pseudoscalar. 
The condensate in Eq.~(\ref{kondi}) can point in an arbitrary direction in color space. However, using a global color rotation, one can always rotate the condensate so that it points in the third direction, $\langle\Delta_a\rangle\sim\delta_{a3}$ 
(conventionally associated with blue). 
The $(ur)$ quarks pair with the $(dg)$ quarks, the $(ug)$ quarks pair with $(dr)$ quarks, whereas the blue quarks do not participate in pairing.
 In the QMD model, diquark fields 
$\langle\Delta_a\rangle\equiv\bar{\psi}^b_i\gamma^5(\psi_j^c)^C\epsilon_{ij}\epsilon^{bca}$ 
are considered effective degrees
of freedom, described by propagating quantum fields, labeled by a color, and 
denoted by $\Delta=(\Delta_1,\Delta_2,\Delta_3)^T$. 

Next, we  consider the transformation properties of the diquark fields under 
$SU(2)_L\times SU(2)_R$ and $SU(3)$.
The term is written in terms of its left-handed and right-handed components,
\begin{eqnarray}
\label{splittie}
\bar{\psi}^b_i\gamma_5\epsilon_{ij}\epsilon_{abc}(\psi_j^c)^C&=&
i\bar{\psi}_{i,L}^b\gamma_5\gamma_2\epsilon_{ij}\epsilon_{abc}(\psi^c_{j,L})^*
+i\bar{\psi}^b_{i,R}\gamma_5\gamma_2\epsilon_{ij}\epsilon_{abc}(\psi^c_{j,R})^*\;,
\end{eqnarray}
where we used $\psi^C=i\gamma_2\psi^*$. Under $SU(2)_L\times SU(2)_R$, the first term in Eq.~(\ref{splittie}) transforms as 
\begin{eqnarray}
\label{leftterm}
i\bar{\psi}_{i,L}^b\gamma_5\epsilon_{ij}\epsilon_{abc}(\psi_{j,L}^c)^*
&\rightarrow&i\bar{\psi}_{k,L}^b\gamma_5\epsilon_{ij}(L^*)^i_k(L^*)^j_l
\epsilon_{abc}(\psi_{l,L}^c)^*=
i\bar{\psi}_{k,L}^b\gamma_5\epsilon_{kl}\epsilon_{abc}(\psi_{l,L})^*\;,
\end{eqnarray}
showing that it is invariant. The invariance of the second term 
in Eq.~(\ref{splittie}) follows similarly.
is then invariant under $SU(2)_R$,
This implies that the diquark field 
is a singlet under this group. In the same manner one can show that $\Delta$ transforms as a triplet under $SU(3)_c$,
\begin{eqnarray}
\Delta&\rightarrow&U_c\Delta\;,    
\end{eqnarray}
where $U_c$ is an $SU(3)_c$ rotation.~\footnote{It is also common to define the
diquark field as the Hermitian conjugate of the definition above, 
implying that it is antitriplet
that transforms as $\Delta\rightarrow\Delta U_c^{\dagger}$.} 
The quark-diquark interaction terms that respect the full symmetry group
$SU(3)_c\times SU(2)_L\times SU(2)_R$
are therefore
\begin{eqnarray}
{\cal L}_{\rm quark-diquark}&=&
{1\over2}g_{\Delta}\bar{\psi}^C_b\Delta_a\gamma^5\tau_2
\epsilon_{abc}\psi_c
+{1\over2}g_{\Delta}\bar{\psi}_b\Delta_a^{\dagger}\gamma^5\tau_2\epsilon_{abc}\psi_c^C\;,
\label{qdq}
\end{eqnarray}
where $g_{\Delta}$ is the quark-diquark coupling. In NJL models, the coupling $g_{\Delta}$ is denoted by $G_{D}$ and referred to as the diquark coupling. We reserve this name to the self-coupling(s) of the diquarks.

From the above discussion it follows
that any power of $\Delta^{\dagger}_a\Delta_a$ is invariant under $SU(3)_c\times SU(2)_L\times SU(2)_R$.
It is then straightforward to write down the Lagrangian in the diquark sector
including renormalizable interactions only,
\begin{eqnarray}
{\cal L}_{\rm diquark}&=&
D_{\mu}\Delta^{\dagger}_aD^{\mu}\Delta_a
-m_{\Delta}^2\Delta^{\dagger}_a\Delta_a
-{\lambda_{\Delta}}(\Delta_a^{\dagger}\Delta_a)^2\;,    
\label{diq}
\end{eqnarray}
where $D_{\mu}$ is the covariant derivative, including the appropriate chemical
potential of the diquark $a$, $m_{\Delta}$ is the diquark mass parameter, and 
$\lambda_{\Delta}$ is the diquark self-coupling.

Finally, we discuss the terms in the scalar-diquark sector, i.e. terms that involve
the (pseudo)scalars and the diquarks. The operators of dimension four are simply
given by products of dimension-two operators from each sector. This yields
the additional terms in the Lagrangian
\begin{eqnarray}
\label{sidq}
{\cal L}_{\rm scalar-diquark}&=&\lambda_3\langle\Sigma^{\dagger}\Sigma\rangle\Delta_a^{\dagger}\Delta_a  +\lambda_{\det}(\det\Sigma+\det\Sigma^{\dagger})\Delta_a^{\dagger}\Delta_a\;,
\end{eqnarray}
where $\lambda_3$ and $\lambda_{\det}$ are coupling constants.

The sum of 
Eqs.~(\ref{lscalar}), ~(\ref{lquark}),~(\ref{extra0}), (\ref{lsq1}) and~(\ref{qdq})--~(\ref{sidq}) constitute the extended quark-meson diquark model.
If one ignores the parity doublet degrees of freedom, the Lagrangian is given by the
sum of Eqs.~(\ref{lscalar}), ~(\ref{lquark}),~(\ref{lsq}),~(\ref{qdq})--(\ref{sidq}), where the second term in Eq.~(\ref{sidq}) can be omitted since it is
proportional to the first term.

\section{Three-flavor quark-meson-diquark model}
In this section, we discuss the Lagrangian and the symmetries of the three-flavor
quark-meson diquark model. The Lagrangian is
\begin{eqnarray}
{\cal L}&=&{\cal L}_{\rm scalar} +{\cal L}_{\rm quark}+
{\cal L}_{\rm scalar-quark}+ {\cal L}_{\rm diquark} 
+{\cal L}_{\rm scalar-diquark}
+{\cal L}_{\rm quark-diquark}\;.
\label{totallag}
\end{eqnarray}
As in the two-flavor case, the scalar and pseudo-scalars
are written in terms of $\Sigma$,
\begin{eqnarray}
\Sigma&=&T_a(\sigma_a+i\pi_a)\;.    
\end{eqnarray}
where the two $3\times3$ matrices are
\begin{eqnarray}
T_a\sigma_a&={1\over\sqrt{2}}\left(
\begin{array}{ccc}
{1\over\sqrt{2}}a_0^0+{1\over\sqrt{6}}\sigma^8+{1\over\sqrt{3}}\sigma_0&a_0^+&\kappa^+\\
a_0^-&-{1\over\sqrt{2}}a_0^0+{1\over\sqrt{6}}\sigma^8+{1\over\sqrt{3}}\sigma_0&\kappa^0\\
\kappa^-&\bar\kappa^0&-{2\over\sqrt{3}}\sigma^8+{1\over\sqrt{3}}\sigma_0\\
\end{array}\right)\;,    \\
T_a\pi_a&={1\over\sqrt{2}}\left(
\begin{array}{ccc}
{1\over\sqrt{2}}\pi^0+{1\over\sqrt{6}}\pi_8+{1\over\sqrt{3}}\pi_0&\pi^+&K^+\\
\pi^-&-{1\over\sqrt{2}}\pi^0+{1\over\sqrt{6}}\pi_8+{1\over\sqrt{3}}\pi_0&K^0\\
K^-&\bar K^0&-{2\over\sqrt{3}}\pi_8+{1\over\sqrt{3}}\pi_0\\
\end{array}\right)\;,    
\end{eqnarray}
Here, $T_a$ ($a=0,2,...8$) are the nine generators of the group $U(3)=U(1)\times SU(3)$. $T_0=\sqrt{1\over6}\mathds{1}$ and $T_a={1\over2}\lambda_a$ ($a=1,2,...8$), where $\lambda_a$ are the Gell-Mann matrices with normalization $\langle\lambda_a\lambda_b\rangle=2\delta_{ab}$. The charged pions are $\pi^{\pm}={1\over\sqrt{2}}(\pi_1\mp i\pi_2)$ and $\pi^0=\pi_3$. The $\eta$ and $\eta^{\prime}$
are admixtures of $\pi_0$ and $\pi_8$, and 
$K^{\pm}={1\over\sqrt{2}}(\pi_4\mp i\pi_5)$ and 
$K^0/\bar{K}^0={1\over\sqrt{2}}(\pi_6\mp i\pi_7)$. The parity partner of the pion is $a_0$,
$a_0^0=\sigma_3$, and $a_0^{\pm}={1\over\sqrt{2}}(\sigma_1\mp i\sigma_2)$.
The parity partner of the kaon is identified with $\kappa$.
The $\sigma$ and the $f_0$ are admixtures
of $\sigma_0$ and $\sigma_8$.

As in the two-flavor case, the matrix $\Sigma$ is the building block for invariants.
In complete analogy, the terms are
\begin{eqnarray}
\nonumber
{\cal L}_{\rm scalar}&=&\langle(D_{\mu}\Sigma)^{\dagger}D^{\mu}\Sigma\rangle-m^2\langle\Sigma^{\dagger}\Sigma\rangle-\lambda_1\langle(\Sigma^{\dagger}\Sigma)\rangle^2-\lambda_2\langle(\Sigma^{\dagger}\Sigma)^2\rangle 
+c\langle\det\Sigma^{\dagger}+\det\Sigma\rangle
\\ &&
+\langle H(\Sigma^{\dagger}+\Sigma)\rangle\;,
\label{scalar3f}
\end{eqnarray}
where $H$ is the matrix
\begin{eqnarray}
H&=&h_0T_0+h_3T_3+h_8T_8\;.    
\end{eqnarray}
The last term in Eq.~(\ref{scalar3f}) explicitly breaks the symmetries. A nonzero $h_0$
breaks it down to $SU(3)_V$, a nonzero $h_3$ breaks isospin, and a nonzero $h_8$ breaks 
it down to $SU(2)_V\times U(1)_A$.
The determinant term breaks the $U(1)_A$ symmetry.

In analogy to the two-flavor case, we introduce the quark field $\psi$
\begin{eqnarray}
\psi&=&(\psi_u^r,\psi_d^r,\psi_s^r\psi_u^g,\psi_d^g,\psi_s^g,\psi_u^b,\psi_d^b,\psi_s^b)^T\;.    
\end{eqnarray}
Thus, $\psi$ is a triplet in both color and flavor space.
The quark term in the Lagrangian is 
\begin{eqnarray}
{\cal L}_{\rm quark}&=&
\bar{\psi}(i/\!\!\!{\partial}+\gamma^0\hat{\mu})\psi\;,
\label{lquark2}
\end{eqnarray}
with $\hat{\mu}$ being the quark chemical potential matrix.
Again, it is diagonal in flavor and color space and can be expressed in terms 
of $\mu_B$, $\mu_Q$, $\mu_3$, and $\mu_8$.

The scalar-quark interaction term is a generalization of the two-flavor 
case, Eq.~(\ref{lsq})
\begin{eqnarray}
{\cal L}_{\rm scalar-quark}&=&-g\bar{\psi}T_a\left(\sigma_a+i\gamma^5\pi_a\right)\psi\;.    
\end{eqnarray}
We next turn to ${\cal L}_{\rm diquark}$.
The left-handed and right-handed diquark fields are
\begin{eqnarray}
\Delta_{L,i}^a=\epsilon_{ijk}\epsilon^{abc}(\psi_{j,L}^b)^\dagger\gamma^0\gamma^2\gamma^5(\psi_{k,L}^c)^*
\;,
\hspace{1cm}
\Delta_{R,i}^a=\epsilon_{ijk}\epsilon^{abc}(\psi_{j,R}^b)^\dagger\gamma^0\gamma^2\gamma^5(\psi_{k,R}^c)^*
\;,
\end{eqnarray}
where the superscript $a$ is a color index and the subscript $i$ is a flavor index.
The left-handed and right-handed diquarks are arranged in matrices as
\begin{eqnarray}
\Delta_L=
\left(
\begin{array}{ccc}
\Delta^1_{L,1}&\Delta^2_{L,1}&\Delta^3_{L,1}\\
\Delta^1_{L,2}&\Delta_{L,2}^2&\Delta^3_{L,2}\\
\Delta^1_{L,3}&\Delta^2_{L,3}&\Delta_{L,3}^3
\end{array}\right)\;,
\hspace{2cm}
\Delta_R=
\left(
\begin{array}{ccc}
\Delta^1_{R,1}&\Delta^2_{R,1}&\Delta^3_{R,1}\\
\Delta^1_{R,2}&\Delta^2_{R,2}&\Delta^3_{R,2}\\
\Delta^1_{R,3}&\Delta^2_{R,3}&\Delta^3_{R,3}
\end{array}\right)\;.
\label{matrisur}
\end{eqnarray}
The left-handed and right-handed diquarks transform under left-handed, right-handed
flavor rotations and color rotations as~\cite{sonstep1} 
\begin{eqnarray}
\Delta_L\rightarrow L\Delta_LU_c^T\;,
\hspace{1cm}
\Delta_R\rightarrow R\Delta_RU_c^T\;.
\end{eqnarray}
This implies that $\Delta_L\Delta_L^{\dagger}\rightarrow L\Delta_L\Delta^{\dagger}_LL^{-1}$ and $\Delta_R\Delta_R^{\dagger}\rightarrow R\Delta_R\Delta_R^{\dagger}R^{-1}$. Thus, 
$\Delta_L^{\dagger}\Delta_L$ transforms as $\Sigma^{\dagger}\Sigma$, implying that there
are two distinct dimension-four operators for the left-handed fields as well as the
right-handed fields. Including the cross term, we obtain
\begin{eqnarray}
\nonumber
{\cal L}_{\rm diquark}&=&\langle (D_{\mu}\Delta_L)^{\dagger}D^{\mu}\Delta_L\rangle
-m_{\Delta}^2\langle\Delta^{\dagger}_L\Delta_L\rangle
-\lambda_1^{\Delta}\langle\Delta^{\dagger}_L\Delta_L\rangle^2
-\lambda_2^{\Delta}\langle(\Delta^{\dagger}_L\Delta_L)^2\rangle
\\ &&
+L\rightarrow R
-\lambda_3^{\Delta}\langle\Delta^{\dagger}_L\Delta_L\rangle
\langle\Delta^{\dagger}_R\Delta_R\rangle\;.
\end{eqnarray}
In the scalar-diquark sector, there are several terms of dimension four.
The first is the product between the dimension-two operators in each sector.  
The transformation properties of $\Sigma\Sigma^{\dagger}$ and 
$\Delta_L^{\dagger}\Delta_L$ show that the trace of their product is also invariant.
Due to the transformation of the products of $\Delta_L$, it is seen that 
we must multiply $\Sigma\Sigma^{\dagger}$ with $\Delta_L^{\dagger}\Delta_L$
and $\Sigma^{\dagger}\Sigma$ with $\Delta_R^{\dagger}\Delta_R$. 
There is one more term to consider, namely
\begin{eqnarray}
\epsilon_{ijk}\epsilon_{lmn}
\Sigma_{il}\Sigma_{jm}(\Delta^a_{R,n})^{\dagger}\Delta_{L,k}^a + h.c\;.
\end{eqnarray}
This term is reminiscent of an invariant determinant term of a $3\times3$ matrix $M$, with
$6\det M=\epsilon_{ijk}\epsilon^{abc}M_a^iM_b^jM_c^k$~\cite{su3}.
Under $SU(3)_L\times SU(3)_R$, the 
term transforms as~\footnote{Summing over the color index $a$ implies that the
object transforms as a singlet under $SU(3)_c$.}
\begin{eqnarray}
\nonumber
\epsilon_{ijk}\epsilon_{lmn}
\Sigma_{il}\Sigma_{jm}(\Delta^a_{R,n})^{\dagger}\Delta_{L,k}^a&\rightarrow&
\epsilon_{ijk}\epsilon_{lmn}L_i^oL_j^pL_k^q(R^*)_l^r(R^*)_m^s(R^*)_n^t
\Sigma_{or}\Sigma_{ps}(\Delta^a_{R,t})^{\dagger}\Delta_{L,q}^a
\\
&=&
\epsilon_{opq}\epsilon_{rst}\det(L)\det(R^{\dagger})
\Sigma_{or}\Sigma_{ps}(\Delta^a_{R,t})^{\dagger}\Delta_{L,q}^a\;,
\end{eqnarray}
where we have used that $\epsilon_{ijk}$ is invariant under any $SU(3)_X$ with
$X=L,R,c$. Thus, this term is invariant and is necessary to cancel certain divergences arising from quark loops.
Adding the terms, we obtain
\begin{eqnarray}
\nonumber
{\cal L}_{\rm scalar-diquark}
&=&-\lambda_3\langle\Sigma^{\dagger}\Sigma\rangle\langle\Delta^{\dagger}_L\Delta_L\rangle
-\lambda_4\langle\Sigma\Sigma^{\dagger}\Delta^{\dagger}_L\Delta_L\rangle
+L\rightarrow R
\\ &&
-\lambda_5\epsilon_{ijk}\epsilon_{lmn}
\Sigma_{il}\Sigma_{jm}\Delta_{L,k}^a(\Delta^a_{R,n})^{\dagger} + h.c
\;.
\end{eqnarray}
Finally, we consider the quark-diquark terms. For three flavors, the quark condensates
are $\sim \epsilon_{ijk}\epsilon^{abc}$. The Lagrangian is then a straightforward generalization
of the two-flavor case, including separating left-handed and right-handed fields
\begin{eqnarray}
\nonumber
{\cal L}_{\rm quark-diquark}&=&    
-{1\over2\sqrt{2}}g_{\Delta}(\bar{\psi}_{L, j}^b)^C\Delta^a_{L,i})\gamma^5\epsilon^{abc}\epsilon_{ijk}\psi_{L,k}^c
\\ &&
-{1\over2\sqrt{2}}g_{\Delta}\bar{\psi}_{L,j}^b(\Delta^a_{L,i})^{\dagger}\gamma^5\epsilon^{abc}\epsilon_{ijk}(\psi_{L,k}^c)^C-{L\rightarrow R}\;.
\end{eqnarray}
The extra factor of ${1\over\sqrt{2}}$ is only for convenience.

\section{Chemical potentials and charge neutrality}
\label{chempot1}
Noether's theorem states that for every continuous symmetry of the action, there exists a conservation law. The conserved quantity is denoted by $Q$ and is 
the conserved charge associated with the symmetry. 
For a given continuous symmetry, we introduce a
chemical potential $\mu_Q$. The grand-canonical Hamiltonian is $H_{\mu}=H-\mu_QQ$.
In the Lagrangian, the chemical potential enters as the zeroth component of the
gauge field in the covariant derivative.
The expectation value of the charge density $n_Q$ is given by
\begin{eqnarray}
n_Q&=&-{\partial\Omega\over\partial\mu_Q}\;,  
\end{eqnarray}
where $\Omega$ is the thermodynamic potential.
It is possible to introduce a chemical potential $\mu_i$ for each conserved charge
$Q_i$. In the present case, we have a chemical potential for each quark flavor and color, i.e nine independent chemical potentials. However, two different chemical
potentials $\mu_{Q_1}$ and $\mu_{Q_2}$
can be introduced simultaneously, only if the corresponding charges commute,
i.e $[Q_i,Q_j]=0$~\cite{kapusta,weldon}. Thus, the number of independent chemical potentials that can be introduced is given by the dimension of the Cartan subalgebra, i. e. the rank of the
symmetry group. 

In the present case, we  introduce four chemical potentials
$\mu_B$, $\mu_e$, $\mu_3$, and $\mu_8$ that correspond to baryon number $n_B$, electric charge density, and the two color charge densities 
\begin{eqnarray}
n_3={1\over2}(n_r-n_g)\;,
\hspace{1cm}
n_8={1\over2\sqrt{3}}(n_r+n_g-2n_b)
\end{eqnarray}
The quark chemical potential matrix can be expressed in terms of these four chemical potentials as
\begin{eqnarray}
\mu_{ij,ab}&=&(\mbox{$1\over3$}\mu_B\delta_{ij}-\mu_eQ_{ij})\delta_{ab}
+\delta_{ij}\left[\mu_3(T_3)_{ab}+
{2\over\sqrt{3}}\mu_8
(T_8)_{ab}\right] \;,
\label{chempot}
\end{eqnarray}
where $i,j$ are flavor indices and $a,b$ are color indices, and
$Q_{ij}$ are the matrix elements of the matrix
$Q={\rm diag}({2\over3},-{1\over3})$ for two flavors and $Q={\rm diag}({2\over3},-{1\over3},-{1\over3})$ for three flavors.
Specifically, we find 
\begin{eqnarray}
\mu_{ur}&=&\mu-{2\over3}\mu_e+{1\over2}\mu_3+{1\over3}\mu_8\;,\\   
\mu_{ug}&=&\mu-{2\over3}\mu_e-{1\over2}\mu_3+{1\over3}\mu_8\;,\\   
\mu_{ub}&=&\mu-{2\over3}\mu_e-{2\over3}\mu_8\;,\\
\mu_{dr}&=&\mu+{1\over3}\mu_e+{1\over2}\mu_3+{1\over3}\mu_8\;,\\   
\mu_{dg}&=&\mu+{1\over3}\mu_e-{1\over2}\mu_3+{1\over3}\mu_8\;,\\   
\mu_{db}&=&\mu+{1\over3}\mu_e-{2\over3}\mu_8\;,\\
\mu_{sr}&=&\mu+{1\over3}\mu_e+{1\over2}\mu_3+{1\over3}\mu_8\;,\\   
\mu_{sg}&=&\mu+{1\over3}\mu_e-{1\over2}\mu_3+{1\over3}\mu_8\;,\\   
\mu_{sb}&=&\mu+{1\over3}\mu_e-{2\over3}\mu_8\;.
\end{eqnarray}
For convenience, we also introduce the following average chemical potentials
\begin{eqnarray}
  \bar{\mu}_{ud}&=&   {1\over2}(\mu_{ur}+\mu_{dg}) ={1\over2}\left(\mu_{dr}+\mu_{ug}\right)
=\mu-{1\over6}\mu_e+{1\over3}\mu_8  
  \;, \\
\bar{\mu}_{us}&=&
    {1\over2}\left(\mu_{ur}+\mu_{sb}\right) = {1\over2}\left(\mu_{ub}+\mu_{sr}\right)
=\mu-{1\over6}\mu_e+{1\over4}\mu_3-{1\over6}\mu_8    
    \;, \\
\bar{\mu}_{ds} &=&    {1\over2}\left(\mu_{dg}+\mu_{sb}\right) = {1\over2}\left(\mu_{db}+\mu_{sg}\right)
=\mu+{1\over3}\mu_e-{1\over4}\mu_3-{1\over6}\mu_8 
\;.    
\end{eqnarray}
Since we mainly have macroscopic objects such as hybrid stars in mind, it is
important to address the issue of charge neutrality. Stable bulk matter must be
neutral with respect to all gauged charges, spontaneously broken or not~\cite{shovkovy,alfordrev}. A net charge will generate electric fields so that
the energy of the system no longer is an extensive quantity.
An important difference between the QCD and the QMD model is that $SU(N_c)$ is a local symmetry in the former and a global symmetry in the latter.
In QCD, color electric neutrality is enforced dynamically. The zeroth component
of the gauge fields, $A_0^a$, acts as a chemical potential whose values are driven
to values such that $Q_3=Q_8=0$.
In the QMD model (and the NJL model), there are no dynamical gauge fields, so charge
neutrality must be imposed by hand.
The neutrality constraints are
\begin{eqnarray}
\label{ladningi}
{\partial\Omega\over\partial\mu_e}=0\;,
\hspace{2cm}
{\partial\Omega\over\partial\mu_3}=0\;,
\hspace{2cm}
{\partial\Omega\over\partial\mu_8}=0\;.    
\end{eqnarray}
Electric charge neutrality can be ensured if we add a background of (massless)
electrons. The electron contribution to the thermodynamic potential is
\begin{eqnarray}
{\Omega}_1^{e}&=&-{4\mu_e^4\over3(4\pi)^2}\;.    
\end{eqnarray}
These equations must be solved together with the gap equations
\begin{eqnarray}
\label{gappio}
{\partial\Omega\over\partial\phi_u}=0\;,
\hspace{2cm}
{\partial\Omega\over\partial\phi_d}=0\;,
\hspace{2cm}
{\partial\Omega\over\partial\phi_s}=0\;.    
\end{eqnarray}
Eqs.~(\ref{ladningi}) and~(\ref{gappio}) can be solved with respect to
the gaps and $\mu_e$, $\mu_3$, and $\mu_8$.
Substituting the solutions 
into $-\Omega$, yields the pressure $p$ as a function of the remaining free chemical
potential, that is $\mu_B$.

The neutrality constraints above are local, which means that they are imposed at
each point in space. This is appropriate for uniform 
phases~\cite{glen}. 
In the context of a hybrid star, it corresponds to the case where the core consists of  
quark matter, whereas the outer region consists of confined matter.
For a given baryon density, the phase with the higher pressure wins, and the phase transition
takes place when the pressure of the two phases is the same.
However, this is not the only possibility. A hybrid star may consist of a quark matter core
surrounded by a shell of mixed phase, and an outer layer of hadronic matter.~\footnote{For less massive stars, the pure quark phase may be absent, so that the core is in the mixed phase.}
The mixed phase consists of quark matter and hadronic matter in phase equilibrium with each other.
The mixed phase is in a charge-separated state consisting of domains with opposite charges.
Each phase is locally charged, but the local charge integrates 
to zero, i. e. the system is globally neutral.
Coulomb repulsion will prevent regions of equal charge from growing arbitrarily large.
On the other hand, the energy will act in the opposite direction. The 
resulting geometry of the mixed phase corresponds to a minimum of the sum of Coulomb and surface energies.

\section{Pion condensation}
\label{pionkond}
In this section, we discuss pion condensation in some detail.
For simplicity, we limit ourselves to two flavors and a vanishing baryon chemical potential. Since we consider $\mu_B=0$, diquarks play no role and the QMD model reduces to the QM model. 
Extensions to three flavors can be found in~\cite{iso3f}.
As mentioned in the Introduction,
there are recent high-precision lattice results that can be compared with the predictions of the QM model at $\mu_B=0$. 
In addition, there are interesting results from 
the NJL model~\cite{isofarias1} within the Medium Separation Scheme (MSS)~\cite{batti,mss} as well 
as perturbation theory at large $\mu_I$~\cite{minato}.

In the pion-condensed phase, there is a nonzero pion condensate 
in addition to the quark condensates.
The transition from the vacuum phase to a BEC phase takes place 
for an isospin chemical potential equal to the pion pole mass.
The ground state $\Sigma_0$ can then be written as
\begin{eqnarray}
\Sigma_0&=&
T_0\bar{\sigma}_0+T_3\bar{a}_0+iT_1\rho=
{1\over\sqrt{2}}\left(
\begin{array}{cc}
\phi_u&{i\over\sqrt{2}}\rho\\
{i\over\sqrt{2}}\rho&\phi_d\\
\end{array}\right)\;,
\end{eqnarray}
where $\phi_u={1\over\sqrt{2}}(\bar{\sigma}_0+\bar{a}_0)$, $\phi_d={1\over\sqrt{2}}(\bar{\sigma}_0-\bar{a}_0)$, and $\rho = \langle\pi_1\rangle$
is the pion condensate.
In the isospin limit, $\phi_u=\phi_d$ in the vacuum since $h_3=0$. However, we allow
$\phi_u\neq\phi_d$ since the isospin symmetry is broken in the BEC phase.
The tree-level thermodynamic potential is
\begin{eqnarray}
\nonumber
\Omega_0^{\rm BEC/BCS}&=&{1\over2}m^2(\phi_u^2+\phi_d^2+\rho^2)-{1\over2}\mu_I^2\rho^2
-h(\phi_u+\phi_d)-c\phi_u\phi_d - {1\over2}c\rho^2
\\ && 
\nonumber
+{1\over4}\lambda_1(\phi_u^2+\phi_d^2+\rho^2)^2
+{1\over4}\lambda_2(\phi_u^2+{1\over2}\rho^2)^2+{1\over4}\lambda_2(\phi_d^2+{1\over2}\rho^2)^2
\\ &&
+{1\over4}\lambda_2(\phi_u-\phi_d)^2\rho^2
\;,
\label{tripje}
\end{eqnarray}
where $h={1\over\sqrt{2}}h_0$ and \(\mu_I = \mu_u-\mu_d\).
The light quark masses are given by
\begin{eqnarray}
m_u&=&{1\over2}g(\bar{\sigma}_0+\bar{a}_0)={1\over\sqrt{2}}g\phi_u\;,
\\
m_d&=&
{1\over2}g(\bar{\sigma}_0-\bar{a}_0)={1\over\sqrt{2}}g\phi_d\;.
\end{eqnarray}
The inverse quark propagator $S^{-1}$ reads
\begin{eqnarray}
S^{-1}&=&    
\left(\begin{array}{cc}
p\!\!\!/-m_u+\gamma_0\mu_u&{1\over{2}}ig\gamma^5\rho\\
{1\over{2}}ig\gamma^{5}\rho&p\!\!\!/-m_d+\gamma_0{\mu}_d\\
\end{array}\right)\;.
\end{eqnarray}
After going to Euclidean space,
the one-loop contribution to the thermodynamic potential, from quark loops is
\begin{eqnarray}
\label{integral}
\Omega_1^{\rm BEC/BCS}&=&-N_cT\Lambda^{-2\epsilon}\sum_n\int_p\log\det S^{-1}(p_0=i\omega_n)\;.
\end{eqnarray}
At zero temperature the Matsubara frequency sum becomes an integral over \(p_0\). The one-loop contribution to the thermodynamical potential then takes the simpler form
\begin{eqnarray}
\label{integral}
\Omega_1^{\rm BEC/BCS}&=&-N_c\Lambda^{-2\epsilon}\int_{-\infty}^{\infty}{dp_0\over2\pi}\int_p\log\det[\mathbb{I}ip_0+M^{\pm}]\;,
\end{eqnarray}
where a sum of $\pm$ is implied, the integral $\int_p$ is defined in Appendix A and $M^{\pm}$ is the matrix
\begin{eqnarray}
M^{\pm}&=&\begin{pmatrix}
        -m_u+\mu_{u} & \pm p & 0 & {i\over2}g\rho\\
        \pm p & m_u+\mu_{u} & -{i\over2}g\rho & 0\\
        0 & {i\over2}g\rho & -m_d+\mu_{d} &\pm p\\
        -{i\over2}g\rho & 0 & \pm p & m_d+\mu_{d}
    \end{pmatrix}\;.    
\end{eqnarray}
In the remainder of this section, we consider the case 
$\mu_u=-{\mu}_d$, that is
nonzero isospin chemical potential $\mu_I$, but vanishing baryon chemical potential $\mu_B$. $\Omega_1^{\rm BEC/BCS}$ requires renormalization. This is
discussed in Appendix~\ref{pionren}.
The result is 
\begin{eqnarray}
    \Omega_{0+1}^{\rm BEC/BCS}&=&{1\over2}m_{0}^2(\phi_{u,0}^2+\phi_{d,0}^2+\rho^2)-{1\over2}\mu_I^2\rho^2
-h_{0}(\phi_{u,0}+\phi_{d,0})\nonumber
\\ && \nonumber
-c_{0}\phi_{u,0}\phi_{d,0}
-{1\over2}c_{0}\rho^2
+{1\over4}\lambda_{1,0}(\phi_{u,0}^2+\phi_{d,0}^2+\rho^2)^2
\\ && \nonumber
+{1\over4}\lambda_{2,0}(\phi_{u,0}^2+{1\over2}\rho^2)^2
+{1\over4}\lambda_{2,0}(\phi_{d,0}^2+{1\over2}\rho^2)^2
\\ &&
\nonumber
+{1\over4}\lambda_{2,0}(\phi_{u,0}-\phi_{d,0})^2\rho^2
+{N_cm_+^4\over(4\pi)^2}\left[\log{\Lambda_0^2\over m_+^2}+{3\over2}\right]
+{N_cm_-^4\over(4\pi)^2}\left[\log{\Lambda_0^2\over m_-^2}+{3\over2}\right]
\\ &&
-{N_c\mu_I^2\rho^2g_0^2\over2(4\pi)^2}\log{\Lambda_0^2\over g_0^2\rho^2/4}
+\Omega_{0+1}^{\rm BEC/BCS, fin}\;,
\label{compli}
\end{eqnarray}
where \(m_\pm^2\) is defined in Eq. \eqref{eqn:mass_splitting_pion}, \(\Lambda_0\) is defined in Eq. \eqref{refscale2} and \(m_0^2,\;\phi_0,\;\rho,\; h_0,\; c_0,\; g_0,\; \lambda_{1,0}\), and \( \lambda_{2,0}\) are given by inserting for \(\Lambda = \Lambda_0\) in Eqs. \eqref{ronny1}--\eqref{ronny8}.
The pressure is given by minus the thermodynamic potential evaluated at the
solution of the gap equation for $\rho_0$, 
\begin{eqnarray}
p&=&-\Omega_{0+1}^{\rm BEC/BCS}(\mu_I,\rho_0)\;.    
\end{eqnarray}
The energy density is given by the Legendre transform
\begin{eqnarray}
\epsilon&=&-p-\mu_I{\partial\Omega_{0+1}\over\partial\mu_I}\;,
\end{eqnarray}
where the last term is also evaluated at the solution of the gap equation.
From these expressions, we can calculate numerically the speed of sound
squared $c_s=\sqrt{dp\over d\epsilon}$.

Before we present the results for the pion-condensed phase using the QM model, we will briefly discuss chiral perturbation theory ($\chi$pt) and Bose condensation.
Chiral perturbation theory is a low-energy effective theory for QCD, which is 
based on global symmetries and relevant degrees of freedom~\cite{gasser1,gasser2}.
For two-flavor QCD, the degrees of freedom are the (pseudo)Goldstone bosons, namely
the pions. For three-flavor QCD, we have in additional charges and neutral kaons
as well the eta particle. 
Pion condensation in chiral perturbation theory was first discussed by Son and Stephanov in~\cite{sonstep}.
They showed that the pressure to leading order in $\chi$pt is~\footnote{The leading-order results are the same for two and three flavors.} 
\begin{eqnarray}
p&=&{1\over2}f_{\pi}^2\mu_I^2\left[1-{m_{\pi}^2\over\mu_I^2}\right]^2\;,
\hspace{1cm}\mu_I\geq m_{\pi}\;,
\end{eqnarray}
where the pressure in the vacuum is zero.
The energy density can be expressed in terms of the pressure as
\begin{eqnarray}
\epsilon&=&-p+2\sqrt{p(p+2m_{\pi}^2f_{\pi}^2)}\;.    
\end{eqnarray}
The speed of sound is
\begin{eqnarray}
c_s^2&=&
{dp\over d\epsilon}=
{\mu_I^4-m_{\pi}^4\over\mu_I^4+3m_{\pi}^4}\;.    
\end{eqnarray}
The speed of sound increases with the isospin chemical potential and approaches
the speed of light for large $\mu_I$. 
This result is at odds with expectation, namely that the 
speed of sound approaches the conformal limit of QCD.
As explained in the introduction, this is due to the fact that the degrees of freedom of $\chi$pt are bosonic and the pressure is proportional to $f_{\pi}^2\mu_I^2$.
For sufficiently small values of isospin $\mu_I\simeq m_{\pi}$, it is expected that the system behaves non-relativistically. Writing $\mu_I=m_{\pi}+\mu_{\rm NR}$
and expanding the NLO result for the energy density
in powers of $\mu_{\rm NR}/m_{\pi}$, we obtain
\begin{eqnarray}
\epsilon&=&m_{\pi}n_I+{2\pi n_I^2a\over m_{\pi}}\left[1+{128\over15\sqrt{\pi}}\sqrt{n_Ia^3}\right]\;,  
\label{nre}
\end{eqnarray}
where $n_I=-{\partial\Omega\over\partial\mu_I}$ is the isospin density and $a\simeq{m_{\pi}\over16\pi f_{\pi}^2}$ is the $s$-wave scattering length.
This is the classic result for the dilute Bose gas by Lee, Huang, and Yang~\cite{lee,huang}, where the second term in Eq.~(\ref{nre}) is the small correction to Bogoliubov's mean-field result. Note that this correction is proportional to the so-called dimensionless gas parameter $\sqrt{n_Ia^3}$.

\begin{figure}[htb!]
    \centering
    \includegraphics[width=\linewidth]{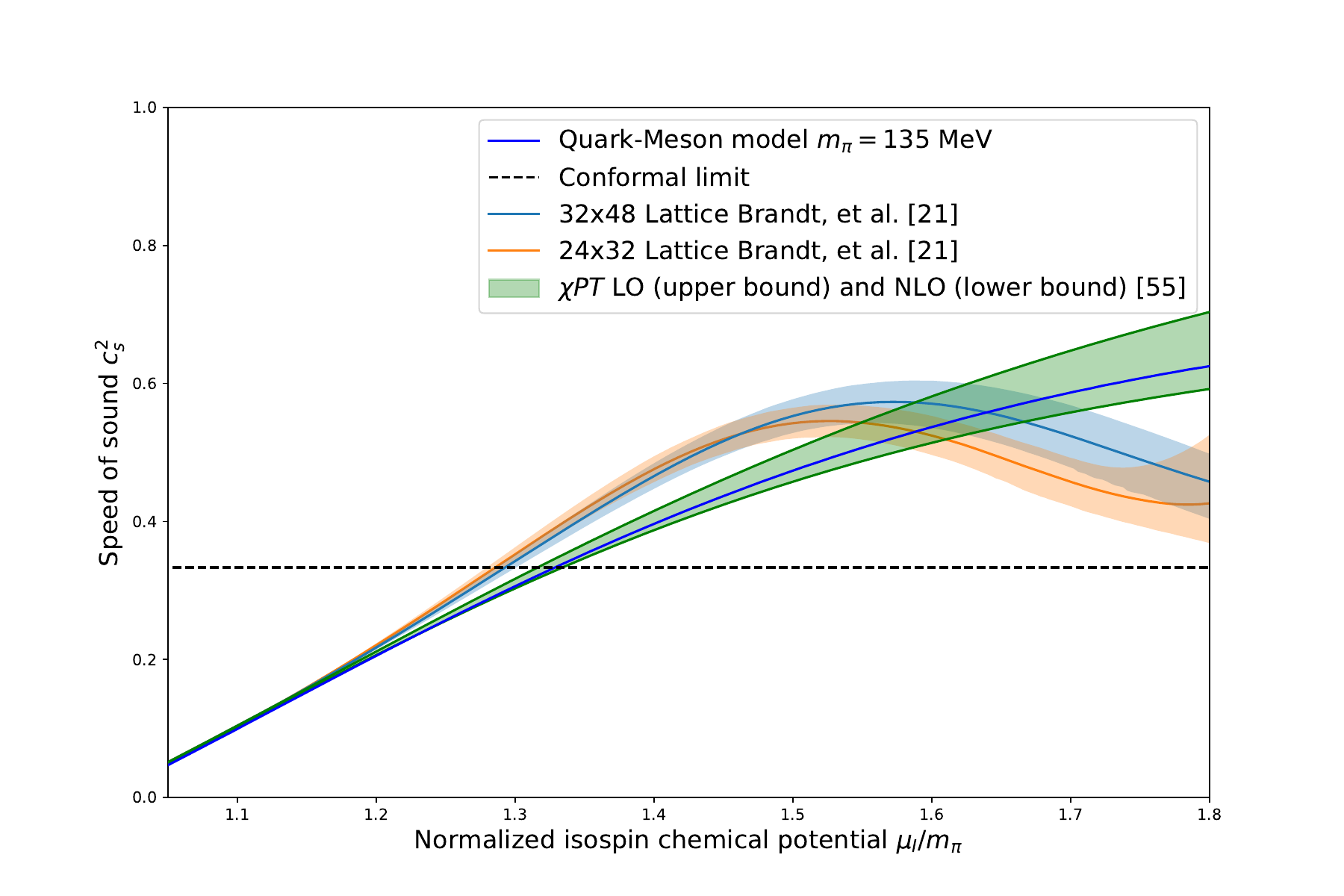}
    \caption{Speed of sound squared in the pion-condensed phase versus $\mu_I/m_{\pi}$. See main text for details.}
    \label{speedoiso1}
\end{figure}

In Refs.~\cite{pionus0,pionus}, the authors performed NLO calculations
of the pressure, energy density, and speed of sound
in two- and three-flavor $\chi$pt at zero temperature. 
The results are expressed in terms of a number of low-energy constants, which are
running couplings evaluated at a certain scale $\Lambda$, for example
$\Lambda=m_{\pi}$ for two flavors and $\Lambda=m_{\rho}$ for three flavors. There are large experimental uncertainties associated with these constants, which
propagates into uncertainty bands for $p$, $\epsilon$, and $c_s^2$.
The results for $c_s^2$ as a function of $\mu_I/m_{\pi}$ are shown in Fig.~\ref{speedoiso1}. 
The upper curve is the leading-order $\chi$pt result,
whereas the lower curve is the next-to-leading order result. The green band is the region between the two curves. The blue curve is the result of a QM-model calculation obtained in~\cite{kojo,brandtqm} as well. Note that the blue curve lies entirely within the
band obtained in $\chi$pt. Finally, we show the results of lattice
simulations using two different lattices~\cite{isobrandt}, given by the
light blue and yellow bands. In the lattice simulations.
In our calculations, we used the values $m_{\pi}=135$ MeV and
$f_{\pi}=90$ MeV, which are the values used in Ref.~\cite{isobrandt}.

\begin{figure}[htb!]
    \centering
    \includegraphics[width=\linewidth]{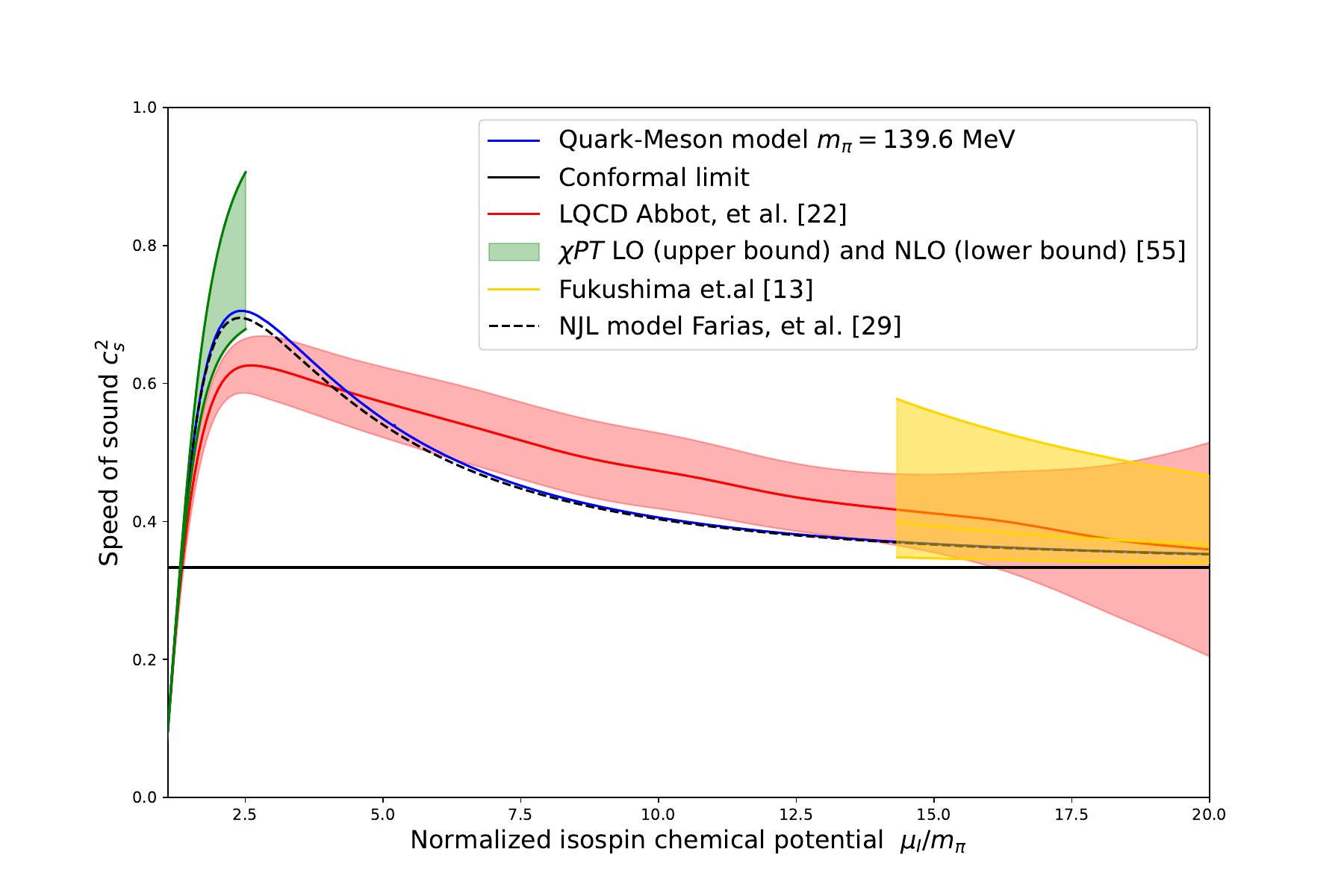}
    \caption{Speed of sound squared in the pion-condensed phase versus $\mu_I/m_{\pi}$. See main text for details.}
    \label{speedoiso2}
\end{figure}

In Fig.~\ref{speedoiso2}, we show the result for $c_s^2$ as a
function of $\mu_I/m_{\pi}$ for a much larger range of $\mu_I$. The QM result
is the blue line and the green band is again the $\chi$pt result. In this case, we
used the values $m_{\pi}=139.6$ MeV and $f_{\pi}=92$ MeV, also used in the
lattice simulations~\cite{abbott}.
The band was constructed from 2000 bootstrap samples with 
$\pm$ two standard deviations from the mean.

We also show the result for the speed of sound of a recent NJL-model calculation
in~\cite{isofarias1}. We first notice that the result is very close to the
result obtained using the quark-meson model. The calculation is based on the so-called Medium Separation Scheme. In this scheme, the medium effects
are disentangled from the divergent vacuum terms. In this way, one avoids the
regularization artifacts that plague the NJL model when it is regularized in a conventional way using e.g. a sharp UV cutoff. For example, using a three-dimensional
cutoff $\Lambda$, the pion condensate vanishes for $\mu_I\sim \Lambda$, which is clearly 
unphysical~\cite{isofarias1}.~\footnote{There is another recent improvement of the NJL model  based on renormalization-group ideas~\cite{gholami1}.}
Finally, we show the perturbative calculations of Ref.~\cite{minato}.
In their paper, they calculate the pressure, trace anomaly, and speed of sound at finite isospin using the Cornwall-Jackiw-Tomboulis formalism, including
a pion condensate. The results are plagued with large scale variations. Varying the renormalization scale $\Lambda=x\mu_I$ from $x=1$ to $x=3$, the yellow band results. The scale variation decreases somewhat with $\mu_I$, and the entire band is above the conformal limit.
The speed of sound shows an interesting structure, namely a peak around $\mu_I=335$ MeV
and a monotonous decrease towards the conformal limit, $c_s^2={1\over3}$.
It is remarkable that the QM model and the NJL model reproduce the peak and qualitatively agree with lattice simulations over a wide range of isospin chemical potential,
and also agree with the perturbative calculations at large $\mu_I$.

We can understand the relaxation of the speed of sound to the conformal limit from above by considering the expression for the thermodynamic potential at large $\mu_I$.
This yields the renormalized thermodynamic potential for large $\mu_I$
\begin{eqnarray}
\Omega_{0+1}^{\rm BEC/BCS}&=&-{1\over2}\mu_I^2\rho_0^2 - {N_c\mu_I^2g_0^2\rho_0^2\over2(4\pi)^2}
\log{\Lambda_0^2\over g_0^2\rho_0^2/4}
-{N_c\mu_I^4\over6(4\pi)^2}\;.    
\end{eqnarray}
This is sufficient to calculate the pion condensate $\rho_0$ for large $\mu_I$.
The solution to the equation ${\partial\Omega_{0+1}\over\partial\rho_0}=0$ is
\begin{eqnarray}
g_0^2\rho_0^2&=&{4\Lambda_0^2}e^{{(4\pi)^2\over N_cg_0^2}-1}\;.    
\end{eqnarray}
Thus, for sufficiently large $\mu_I$, the gap approaches a constant. This is in contrast to the tree-level behavior. As noted in~\cite{kojo}, the gap grows linearly with
$\mu_I$. The change in behavior is due to the quark loops. It is generic for the model in our approximation since the superconducting gaps behave in the same way.
For large values of $\mu_I$, the pressure, energy density, and speed of sound are
\begin{eqnarray}
p&=&{N_c\mu_I^4\over6(4\pi)^2}+{N_c\mu_I^2g_0^2\rho_0^2\over2(4\pi)^2}
\;,\\
\epsilon&=&{N_c\mu_I^4\over2(4\pi)^2}+{N_c\mu_I^2g_0^2\rho_0^2\over2(4\pi)^2}
\;,\\
c_s^2&=&{\mu_I^2+{3\over2}g_0^2\rho_0^2\over3\mu_I^2+{3\over2}g^2\rho_0^2}\;.
\label{lastly}
\end{eqnarray}
Eq.~(\ref{lastly}) shows that the speed of sound approaches the conformal limit from above.

\section{Color superconductivity}
In this section, we discuss color superconductivity in the QMD model.
We first consider the 2SC phase, where two of three colors and flavors are pairing.
Since the three-flavor case is only marginally more complicated than the two-flavor
case, we focus on the former.

The ground state $\Sigma_0$ is
\begin{eqnarray}
\Sigma_0&=&T_0\bar{\sigma}_0+T_3\bar{\sigma}_3+T_8\bar{\sigma}_8\;,
\end{eqnarray}
where the barred quantities are the expectations of the fields $\sigma_0$, $\sigma_3$, and $\sigma_8$.
It is customary to use the linear combinations
\begin{eqnarray}
\phi_u&=&\sqrt{{2\over3}}\bar{\sigma}_0+\bar{\sigma}_3+\sqrt{{1\over3}}\bar{\sigma}_8\;,\\
\phi_d&=&\sqrt{{2\over3}}\bar{\sigma}_0-\bar{\sigma}_3+\sqrt{{1\over3}}\bar{\sigma}_8\;,\\
\phi_s&=&\sqrt{{1\over3}}\bar{\sigma}_0-\sqrt{{2\over3}}\bar{\sigma}_8\;.
\end{eqnarray}
In this basis, the ground state and the quark masses are
\begin{eqnarray}
\label{grscalar}
\Sigma_0&=&{1\over2}{\rm diag}(\phi_u,\phi_d,\sqrt{2}\phi_s)\;,
\\
m_{u}&=&{1\over2}g\phi_u\;,\\
m_d&=&{1\over2}g\phi_d\;,\\
m_s&=&{1\over\sqrt{2}}g\phi_s\;.
\end{eqnarray}
The quark contribution to the thermodynamic potential is
\begin{eqnarray}
\Omega_1^{\rm}&=&-{1\over2}\int_{-i\infty}^{i\infty}{dp_0\over2\pi}\log\det S^{-1}\;,
\end{eqnarray}
where $S$ is the quark propagator in the Gorkov basis. The factor of ${1\over2}$
is to compensate for the doubling of fermionic degrees of freedom.
The inverse propagator reads
\begin{eqnarray}
S^{-1}&=& \begin{pmatrix}
        p\!\!\!/-m_i+\gamma^0\mu_{ai} &i\gamma^5g_{\Delta}\Delta\\
       i\gamma^5g_{\Delta}\Delta & p\!\!\!/-m_i-\gamma^0\mu_{ai}\\
    \end{pmatrix}\;,    
\end{eqnarray}
where 
\begin{eqnarray}
\Delta&=&\Delta_{ud}\epsilon_{ij3}\epsilon^{ab3}+\Delta_{us}\epsilon_{ij2}\epsilon^{ab2}+  
\Delta_{ds}\epsilon_{ij1}\epsilon^{ab1}\;.  
\end{eqnarray}
Since the determinant is invariant under an even number of row/column exchanges we can put the total inverse propagator in block diagonal form. After some algebra, the thermodynamic potential can be written as
\begin{eqnarray}
\Omega_1&=&
-\frac{1}{2}\sum_{i=1,\pm}^7
\int_{-\infty}^{\infty}{dp_0\over2\pi}\int_p\log\det[ip_0\mathds{1}-M_i^{\pm}]\;,
\label{omegafun}
\end{eqnarray}
where the $4\times4$ matrices $M_1^{\pm}-M_6^{\pm}$ are defined as
\begin{eqnarray}
    M_1^{\pm} &=& \begin{pmatrix}
        -m_u-\mu_{ug} & \pm p & 0 & \Delta_{ud}\\
       \pm p & m_u-\mu_{ug} & -\Delta_{ud} & 0\\
        0 & -\Delta_{ud} & -m_d+\mu_{dr} &\pm p\\
        \Delta_{ud} & 0 &\pm p & m_d+\mu_{dr}
    \end{pmatrix}\;,\\
    M_2^{\pm} &=& \begin{pmatrix}
        -m_d - \mu_{dr} &\pm p & 0 & \Delta_{ud}\\
        \pm p & m_d - \mu_{dr} & -\Delta_{ud} & 0\\
        0 & -\Delta_{ud} & -m_u+\mu_{ug} &\pm  p\\
        \Delta_{ud} & 0 &\pm p & m_u + \mu_{ug}
    \end{pmatrix}\;,\\
    M_3^{\pm} &=& \begin{pmatrix}
        -m_u+\mu_{ub} &\pm p & 0 & \Delta_{us}\\
        \pm p & m_u+\mu_{ub} & -\Delta_{us} & 0\\
        0 & -\Delta_{us} & -m_s - \mu_{sr} &\pm p\\
        \Delta_{us} & 0 &\pm p & m_s - \mu_{sr}
    \end{pmatrix}\;,\\
    M_4^{\pm} &=& \begin{pmatrix}
        -m_s+\mu_{sr} &\pm p & 0 & \Delta_{us}\\
        \pm p & m_s + \mu_{sr} & -\Delta_{us} & 0\\
        0 & -\Delta_{us} & -m_u-\mu_{ub} &\pm p\\
        \Delta_{us} & 0 &\pm p & m_u - \mu_{ub}
    \end{pmatrix}\;,\\
        M_5^{\pm} &=& \begin{pmatrix}
        -m_d+\mu_{db} &\pm p & 0 & -\Delta_{ds}\\
       \pm p & m_d + \mu_{db} & \Delta_{ds} & 0\\
        0 & \Delta_{ds} & -m_s-\mu_{sg} &\pm p\\
        -\Delta_{ds} & 0 &\pm p & m_s-\mu_{sg}
    \end{pmatrix}\;,\\
    M_{6}^{\pm} &=& \begin{pmatrix}
    -m_s + \mu_{sg} &\pm p & 0 & -\Delta_{ds}\\
    \pm p & m_s+\mu_{sg} & \Delta_{ds} & 0\\
    0 & \Delta_{ds} & -m_d-\mu_{db} &\pm p\\
    -\Delta_{ds} & 0 &\pm p & m_d-\mu_{db}
    \end{pmatrix}\;.
\end{eqnarray}    
The seventh matrix, $M_7^{\pm}$, is a $12\times12$ matrix that takes the form
\begin{eqnarray}
    M_7^{\pm} &=&\begin{pmatrix}
    M_7^{ur} & M_7^{\Delta_{ud}} & M_7^{\Delta_{us}}\\
    M_7^{\Delta_{ud}} & M_7^{dg} & M_7^{\Delta_{ds}}\\
    M_7^{\Delta_{us}} & M_7^{\Delta_{ds}} & M_7^{sb}
    \end{pmatrix}\;,
\end{eqnarray}
where the submatrices are defined as
\begin{eqnarray}
    M_7^{fc\pm} &=& \begin{pmatrix}
        -\mu_{fc}-m_f & \pm p & 0 & 0\\
       \pm p & -\mu_{fc}+m_f  & 0 & 0\\
        0 & 0 & \mu_{fc}-m_f &\pm p\\
        0 & 0 &\pm p & \mu_{fc}+m_f
    \end{pmatrix}\;,
    \hspace{1mm}
\end{eqnarray}
\begin{eqnarray}
    M_7^{\Delta_{f_1f_2}} = \begin{pmatrix}
        0 & 0 & 0 & -\Delta_{f_1f_2}\\
        0 & 0 & \Delta_{f_1f_2} & 0\\
        0 & \Delta_{f_1f_2} & 0 & 0\\
        -\Delta_{f_1f_2} & 0 & 0 & 0
    \end{pmatrix}\;,
\end{eqnarray}
and where $c$ is a color index and
$f$, $f_1$, and $f_2$ are flavor indices. 
Note that the matrices $M_i^{+}$ and $M_i^{-}$ give the same contribution to the thermodynamic potential. In the matrices above, we have allowed for three different gaps $\Delta_{ud}$,
$\Delta_{ds}$, $\Delta_{us}$. For asymptotically large $\mu$, we know that we are in the ideal
CFL phase with the gaps being equal. For lower values of $\mu$, we expect that the finite mass of the $s$ quark gives rise to a splitting of the condensate so that $\Delta_{us}=\Delta_{ds}\neq\Delta_{ud}$. NJL calculations suggest
that $\Delta_{ud}$ is the larger. For even smaller values of $\mu$, we enter the 2SC phase where only $\Delta_{ud}$ is nonzero, or we go directly to a phase of normal quark matter.
In Ref.~\cite{us3}, we showed that it is possible to renormalize the QMD model in the 
mean-field approximation for arbitrary gaps and quark masses.

\subsection{2SC phase}
In the standard 2SC phase, the $u$ and $d$ quarks form pairs and the diquark antitriplet develops a nonzero expectation value denoted by
$\Delta_{ud}$. We can always perform a color rotation so that the order parameter
takes the form $(0,0,\Delta_{ud})^T$. In this state, light quarks with colors
red and green quarks form pairs, while blue light quarks and $s$ quarks of any
color do not participate. Assuming massless light quarks, the symmetry breaking pattern is
\begin{eqnarray}
\nonumber
SU(3)_c\times SU(2)_L\times SU(2)_R\times U(1)_B\times U(1)_s
&\rightarrow&SU(2)_c\times SU(2)_L\times SU(2)_R
\\ &&
\times U(1)_{\tilde{B}}\times U(1)_s\;.
\end{eqnarray}
The $U(1)_s$ symmetry is due to the unpaired $s$ quarks. Except for this factor, this is
the same pattern as in two-flavor QCD. The residual $SU(2)$ symmetry is a red-green symmetry.
Baryon symmetry survives as a linear combination of the original baryon symmetry and 
the broken generator $T_8$ as $\tilde{B}=B-{2\over\sqrt{3}}T_8=(0,0,1)$.
Similarly, electromagnetism survives as a linear combination $\tilde{Q}=Q-{1\over\sqrt{3}}T_8={1\over2}(1,-1,0)$.
Also note that the global chiral symmetry is intact. Thus, there are no massless Goldstone bosons
in this phase. Instead, the breaking of the gauge symmetry gives rise to $5$ massive gluons
via the Higgs mechanism, which is the number of broken generators.
Assuming the diquarks take on the expectation value \(\Delta_L =-\Delta_R =  -{\rm diag}(0,0,\Delta_{ud})/\sqrt{2}\). 
The tree-level thermodynamic potential is
\begin{eqnarray}
\nonumber
    \Omega_0^{\rm2SC} &=& {1\over4}m^2(\phi_u^2+\phi_d^2+2\phi_s^2)-\frac{1}{2}h_u\phi_u-\frac{1}{2}h_d\phi_d-h_s\phi_s 
    + \frac{1}{16}\lambda_1(\phi_u^2+\phi_d^2+2\phi_s^2)^2 
    \\ &&
    + \frac{1}{16}\lambda_2(\phi_u^4+\phi_d^4+4\phi_s^4)\nonumber
    -\frac{1}{4}\sqrt{2}c\phi_u\phi_d\phi_s
    +\frac{1}{4}\lambda_3(\phi_u^2+\phi_d^2+2\phi_s^2)\Delta_{ud}^2
    \\ &&
    +\frac{1}{2}\lambda_4\phi_s^2\Delta_{ud}^2\nonumber -\frac{1}{2}\lambda_{5}\phi_u\phi_d\Delta_{ud}^2\nonumber 
    + \frac{1}{4}\left(2\lambda^{\Delta}_{1}+2\lambda_2^{\Delta}+\lambda_3^\Delta\right)\Delta_{ud}^4
    \\ &&
    +(m_\Delta^2 - 4\bar{\mu}_{ud}^2)\Delta_{ud}^2 
    \;,
\end{eqnarray}
where we have defined the symmetry-breaking parameters
\begin{eqnarray}
\label{hu}
h_u&=&{\sqrt{2\over3}}h_0 + h_3 + {1\over\sqrt{3}}h_8\;,\\    
h_d&=&{\sqrt{2\over3}}h_0 - h_3 + {1\over\sqrt{3}}h_8\;,\\    
h_s&=&{1\over\sqrt{3}}h_0-{\sqrt{2\over 3}}h_8\;.
\label{hs}
\end{eqnarray}
In the 2SC phase with two degenerate light quarks, one can obtain the 
spectrum explicitly,
\begin{eqnarray}
E_{\Delta^{\pm}}^{\pm}&=&\sqrt{(E\pm\bar{\mu}_{ud})^2+g_{\Delta}^2\Delta_{ud}^2}\pm\delta\mu\;,\\
E_{ub}^{\pm}&=&\sqrt{p^2+m_l^2}\pm\mu_{ub}\;;\\
E_{db}^{\pm}&=&\sqrt{p^2+m_l^2}\pm\mu_{db}\;;\\
E_{sa}^{\pm}&=&\sqrt{p^2+m_s^2}\pm\mu_{sa}\;,    
\end{eqnarray}
where $E=\sqrt{p^2+m_l^2}$ and $a={r,g,b}$.
Each energy $E_{\Delta^{\pm}}^{\pm}$ is doubly degenerate and the 
states belong to an $SU(2)_c$ doublet. The remaining states are singlets.
We have defined $\delta\mu={1\over2}(\mu_{dg}-\mu_{ur})={1\over2}(\mu_{dr}-\mu_{ug})$, which is called the mismatch parameters because it
measures the mismatch between the Fermi surfaces of the $u$ and $d$ quarks
(in terms of the electron chemical potential $\mu_e$).
The quasiparticle with dispersion relation $E_{\Delta^{\pm}}^+$ has a gap $g_{\Delta}\Delta_{ud}+\delta\mu$, while the
quasiparticle with dispersion relation $E_{\Delta^{\pm}}^-$
has a gap $g_{\Delta}\Delta_{ud}-\delta\mu$ as long as
$g_{\Delta}\Delta_{ud}>\delta\mu$. For $g_{\Delta}\Delta_{ud}<\delta\mu$, it is gapless. 

One can proceed as in the previous section by immediately isolating the divergences before integrating over $p_0$. Alternatively, we first integrate over $p_0$ obtaining the well known result for the thermodynamic potential for a single fermion with energy
$E\pm\mu$,
\begin{eqnarray}
\label{basic}
\Omega_1&=&-\Lambda^{-2\epsilon}\int_p\left[E+(\mu-E)\theta(\mu-E)
+(-\mu-E)\theta(-\mu-E)
\right]\;.    
\end{eqnarray}
The momentum integral of the first term is divergent, whereas the second is convergent because of the step function. In the case of ungapped quarks, the integral Eq.~(\ref{basic}) can be evaluated directly in dimensional regularization using
Eq.~(\ref{i0}). The integral involving gapped quarks requires a subtraction
term that is used to isolate the divergences. We write
\begin{eqnarray}
\Omega_1^{\rm 2SC}&=&\Omega_1^{\rm 2SC, div}+\Omega_1^{\rm 2SC, finite}+\Omega_1^{{\rm 2SC},\mu}\;,   
\end{eqnarray}
where
\begin{eqnarray}
\nonumber
\Omega_1^{\rm 2SC,div}&=&
-\Lambda^{-2\epsilon}\int_p\left[8\sqrt{p^2+m_l^2+g_{\Delta}^2\Delta_{ud}^2}
+{4\bar{\mu}_{ud}^2g_{\Delta}^2\Delta_{ud}^2\over(p^2+m_l^2+g_{\Delta}^2\Delta_{ud}^2)^{3\over2}}
+4\sqrt{p^2+m_l^2}
\right. \\ &&
\left.
+6\sqrt{p^2+m_s^2}
\right]\;,
\\
\label{2scf}
\Omega_1^{\rm 2SC,finite}&=&
-\int_p\left[2E_{\Delta^{\pm}}^{\pm}-8\sqrt{p^2+m_l^2+g_{\Delta}^2\Delta_{ud}^2}
-{4\bar{\mu}_{ud}^2g_{\Delta}^2\Delta_{ud}^2\over(p^2+m_l^2+g_{\Delta}^2\Delta_{ud}^2)^{3\over2}}
\right]\;,
\\
    \Omega_1^{2SC,\mu}&=& -2\int_p\left[2(\delta\mu - E_\Delta^\pm)\theta(\delta\mu-E_\Delta^\pm) + (\mu_{ub}-E)\theta(\mu_{ub}-E) + (\mu_{db}-E)\theta(\mu_{db}-E) \right.\nn\\
    &&\left.+(\mu_{sr}-E_s)\theta(\mu_{sr}-E_s) + (\mu_{sg}-E_s)\theta(\mu_{sg}-E_s)+ (\mu_{sb}-E_s)\theta(\mu_{sb}-E_s)\right]\;.
    \label{2scmy}
\end{eqnarray}
Note that we have set $\epsilon=0$ in 
Eqs.~(\ref{2scf})--(\ref{2scmy}) since these expressions are
finite and can be evaluated (numerically) directly in $d=3$ dimensions.

It is possible to obtain a number of analytical results deep into the 2SC phase,
where we can ignore the masses of the light quarks. 
The dispersion relations are then given by 
\begin{eqnarray}
\label{dis1}
E_{\Delta^{\pm}}^{\pm}&=&\sqrt{[p\pm\bar{\mu}_{ud}]^2+g_{\Delta}^2\Delta_{ud}^2}\pm\delta\mu\;,\\
\label{disp3}
E_{ub}^{\pm}&=&p\pm\mu_{ub}\;,\\
E_{db}^{\pm}&=&p\pm\mu_{db}\;,\\
E_{sa}^{\pm}&=&\sqrt{p^2+m_s^2}\pm\mu_{sa}\;.
\label{displ}
\end{eqnarray}
In this case, the finite
term $\Omega_1^{\rm 2SC, fin}$ can be calculated analytically,
\begin{eqnarray}
\nonumber
\Omega_1^{\rm 2SC, finite}&=&-
\int_p\left[4\sqrt{(p\pm\bar{\mu}_{ud})^2
+g_{\Delta}^2\Delta_{ud}^2}
-8\sqrt{p^2+g_{\Delta}^2\Delta_{ud}^2}-{4\bar{\mu}^2_{ud}g_{\Delta}^2\Delta_{ud}^2\over(p^2+\Delta_{ud}^2)^{3\over2}}\right]
\\
&=&
-{16\bar{\mu}^4_{ud}\over3(4\pi)^2}\;.
\label{mycont}
\end{eqnarray}
Thus, the subtraction term reduces to the thermodynamic potential of free massless quarks with average chemical potential $\bar{\mu}_{ud}$.
The finite-density contribution to the thermodynamic potential
of unpaired fermions of mass $m$ and chemical potential $\mu$ is
\begin{eqnarray}
\nonumber
\Omega_1^{\mu}&=&-{2\over3(4\pi)^2}\left[\mu\sqrt{\mu^2-m^2}(2\mu^2-5m^2)
-m^4\log{m\over\mu+\sqrt{\mu^2-m^2}}
\right]
\\
&=&-{4\over3(4\pi)^2}\left[\mu^4-3\mu^2m^2+m^4\left({9\over8}-{3\over2}\log{m\over2\mu}\right)\right]+{\cal O}(m^6/\mu^2)\;.    
\end{eqnarray}
The contribution $\Omega_1^{\rm 2SC,div}$ is evaluated 
using Eqs.~(\ref{i0})--(\ref{i2}) and the divergences are removed as explained
in Appendix~\ref{3fren}.  
This yields the renormalized thermodynamic potential  
\begin{eqnarray}
\nonumber
\Omega_{0+1}^{\rm 2SC}&=&
{1\over2}m_{\ms}^2\phi_{s,\ms}^2
-h_{s,\ms}\phi_{s,\ms}
+{1\over4}(\lambda_{1,\ms}+\lambda_{2,\ms})\phi_{s,\ms}^4+{1\over2}(\lambda_{3,\ms}+\lambda_{4,\ms})\phi_{s,\ms}^2\Delta_{ud,\ms}^2
\\ &&
\nonumber
+{1\over4}(2\lambda_{1,\ms}^{\Delta}+2\lambda_{2,\ms}^{\Delta}+\lambda_{3,\ms}^{\Delta})\Delta_{ud,\ms}^4+
(m_{\Delta,\ms}^2-4\bar{\mu}^2_{ud})\Delta_{ud,\ms}^2
\\ &&
-{16\bar{\mu}_{ud}^2g_{\Delta,\ms}^2\Delta_{ud,\ms}^2\over(4\pi)^2}\log{\Lambda^2\over g_{\Delta,\ms}^2\Delta_{ud,\ms}^2}
+{4g_{\Delta,\ms}^4\Delta_{ud,\ms}^4\over(4\pi)^2}\left[\log{\Lambda^2\over g_{\Delta,\ms}^2\Delta_{ud,\ms}^2}
+{3\over2}\right]
\nonumber
\\ && \nonumber
+{3m_s^4\over(4\pi)^2}\left[\log{\Lambda^2\over m_s^2}
+{3\over2}\right]
-{16\bar{\mu}_{ud}^4\over3(4\pi)^2}
-{4\mu_{ub}^4\over3(4\pi)^2}-{4\mu_{db}^4\over3(4\pi)^2}-{4\mu_e^4\over3(4\pi)^2}
\\ &&\nonumber
-{2\over3(4\pi)^2}\sum_{a=r,g,b}\left[\mu_{sa}\sqrt{\mu_{sa}^2-m_s^2}(2\mu_{sa}^2-5m_s^2)
-m_s^4\log{m_s\over\mu_{sa}+\sqrt{\mu_{sa}^2-m_s^2}}
\right]\;,\\
&&
\end{eqnarray}
where we have added the contribution from free massless electrons.
In order to compare with the NJL model analysis of Refs.~\cite{markraja,reddy},
we consider the leading-order neutrality effects of the mass of the $s$ quark
deep in the 2SC phase. The thermodynamic potential can then be approximated by
\begin{eqnarray}
\nonumber
\Omega_{0+1}^{\rm 2SC}&=&
-4\bar{\mu}^2_{ud}\Delta_{ud,\ms}^2
-{16\bar{\mu}_{ud}^2g_{\Delta,\ms}^2\Delta_{ud,\ms}^2\over(4\pi)^2}
\log{\Lambda^2\over g_{\Delta,\ms}^2\Delta_{ud,\ms}^2}
-{16\bar{\mu}_{ud}^4\over3(4\pi)^2}
\nonumber
-{4\mu_{ub}^4\over3(4\pi)^2}-{4\mu_{db}^4\over3(4\pi)^2}
\\ 
&&-{4\over3(4\pi)^2}\sum_{a=r,g,b}\left[\mu_{sa}^4-3\mu_{sa}^2m_s^2+m_s^4\left({9\over8}-{3\over2}\log{m_s\over2{\mu}_{sa}}\right)\right]\;.
\end{eqnarray}
Deep in 2SC phase, the chemical potentials $\mu_e$, $\mu_3$, $\mu_8$
and the gap are all small quantities compared to $\mu$.
To the order at which we are working, the gap and neutrality conditions are
\begin{eqnarray}
{\partial\Omega_{0+1}^{\rm 2SC}\over\partial\Delta_{ud}}&=&-8{\mu}_{}^2
-{32{\mu}^2_{}g_{\Delta,\ms}^2\over(4\pi)^2}\left[\log{\Lambda^2\over g_{\Delta,\ms}^2\Delta_{ud,\ms}^2}-1\right]
=0\;,
\\ 
{\partial\Omega_{0+1}^{\rm 2SC}\over\partial\mu_e}&=&
{8\mu\over3(4\pi)^2}
\left[2g^2_{\Delta,\ms}\Delta_{ud,\ms}^2+3m_s^2-6\mu_e\mu
\right]=0\;,\\
{\partial\Omega_{0+1}^{\rm 2SC}\over\partial\mu_3}&=&
-{8\mu_3\mu^2\over(4\pi)^2}=0\;,
\\
{\partial\Omega_{0+1}^{\rm 2SC}\over\partial\mu_8}&=&
-{32{\mu}\over3(4\pi)^2}
\left[3{\mu}_{8}\mu+g^2_{\Delta,\ms}\Delta_{ud,\ms}^2\right]
=0\;,
\end{eqnarray}
whose solutions are
\begin{eqnarray}
g_{\Delta}^2\Delta_{ud,\ms}^2=\Lambda^2e^{{(4\pi)^2\over4g_{\Delta,\ms}^2}-1}\;,
\hspace{0.2cm}
\mu_e={1\over3}{g_{\Delta,\ms}^2\Delta_{ud,\ms}^2\over{\mu}}+{1\over2}{m_s^2\over{\mu}}\;,\hspace{0.2cm}
\mu_3=0\;,
\hspace{0.2cm}
\mu_8=-{1\over3}{g_{\Delta}^2\Delta_{ud,\ms}^2\over{\mu}}\;.
\label{ymsee}
\end{eqnarray}
The results for the chemical potentials above are the same as those obtained in the NJL analysis of Refs.~\cite{markraja,reddy}.
We note that $\mu_e$ and $\mu_8$ are nonzero also for massless $s$ quarks.
The reason is that only $u$ and $d$ quarks form pairs, and these pairs have
nonzero $Q_e$ and $Q_8$ charges.
For the same reason, $\mu_3$ vanishes in the 2SC phase, since the ground state 
has zero $Q_3$ charge. Therefore, most authors implement $\mu_3=0$ immediately in the calculation,
here we find it reassuring that we obtain this result, having allowed for nonzero $\mu_3$
from the start.
The thermodynamic potential then becomes
\begin{eqnarray}
\Omega_{0+1}^{\rm 2SC}&=&-{12{\mu}^4\over(4\pi)^2}-{12{\mu}^2m_s^2\over(4\pi)^2} -{(5-12\log{m_s\over2{\mu}})m_s^4\over2(4\pi)^2}-{16{\mu}^2g_{\Delta,0}^2\Delta_{ud}^2\over(4\pi)^2}\;. 
\end{eqnarray}
also in agreement with the analysis of the NJL model in~\cite{reddy}.

\subsection{CFL phase}
For very large baryon chemical potentials, we can ignore the mass of the $s$ quark, and the system is in the ideal CFL phase
In the ideal CFL phase, the ground state is given by
$\langle\Delta_{L,i}^a\rangle=-\langle\Delta_{R,j}^b\rangle=-\delta_i^{a}\delta_j^{b}\Delta$, i.e. 
all flavors and colors form pairs on an equal footing.
It is also neutral with respect to all charges, broken or not.
The matrices in Eq.~(\ref{matrisur}) are diagonal and it is clear that the ground state is invariant under $SU(3)_{L+R+c}$. The symmetry-breaking pattern is~\cite{alfordrev}
\begin{eqnarray}
SU(3)_c\times SU(3)_L\times SU(3)_R\times U(1)_B&\rightarrow&SU(3)_{L+R+c}\times Z_2\;.    
\end{eqnarray}
Note that in contrast to the 2SC phase, baryon number is broken in the CFL phase.
Again, there is an unbroken $U(1)_{\tilde{Q}}$ symmetry, which is a linear combination of the original electric charge generator and color generators. The rest of the color group is broken, as is the orthogonal gluon-photon generator, which leads to eight massive gauge bosons in the usual way.

For general quark masses and gaps, it is prohibitively difficult to find explicit expressions for the quasiparticle dispersion relations. 
The strategy we will use to isolate the ultraviolet divergences is as in section~\ref{pionkond}:
we expand the thermodynamic potential around zero chemical potentials and then integrate
over $p_0$. This gives rise to a number of terms whose ultraviolet behavior can be easily analyzed and used to construct suitable subtraction terms. These subtraction terms must again be sufficiently simple for them to be evaluated in dimensional regularization.
The thermodynamic potential is then renormalized in the usual way. The finite renormalized thermodynamic potential depends on three quark masses, three gaps, and three chemical potentials. This gives rise to nine coupled gap equations that must be solved simultaneously for each value of $\mu$. In Appendix~\ref{3fren}, we discuss the renormalization of $\Omega$ in some detail.

In certain cases, it is possible to obtain analytical results. For example, in the case of equal quark masses, a common
chemical potential, and equal gaps, the spectrum reduces to
\begin{eqnarray}
E_o&=&\sqrt{(E\pm\mu)^2+g_{\Delta}^2\Delta^2}\;,\\    
E_s&=&\sqrt{(E\pm\mu)^2+4g^2_{\Delta}\Delta^2}\;,
\end{eqnarray}
where $E=\sqrt{p^2+m_c^2}$ and $m_c={1\over2}g\phi$ is the common quark mass
(with $\phi_u=\phi_d=\sqrt{2}\phi_s=\phi$).
The subscripts indicate that the nine quasiparticles split into 
an $SU(3)$ octet and an $SU(3)$ singlet.
The gapped spectrum is of the same form as in the 2SC phase with $\delta\mu=0$.
For this reason, the finite-density term $\Omega_1^{\mu}$ vanishes.
Omitting the terms involving $h_u$, $h_d$, $h_s$, and $c$, 
the tree-level potential is
\begin{eqnarray}
\Omega_0^{\rm CFL}&=&
\nonumber
{3\over4}m^2\phi^2+{3\over16}(3\lambda_1+\lambda_2)\phi^4
+{3\over4}\left(3\lambda_3+\lambda_4-2\lambda_5\right)\phi^2\Delta^2
+3(m_{\Delta}^2-4\mu^2)\Delta^2
\\ &&
+{3\over4}\left(6\lambda_1^{\Delta}+2\lambda_2^{\Delta}+3\lambda_3^{\Delta}\right)\Delta^4\;.
\end{eqnarray}
As in the 2SC case, we split the quark contribution to the thermodynamic potential as
\begin{eqnarray}
\nonumber
\Omega_1^{\rm CFL,div}&=&-\Lambda^{-2\epsilon}\int_p\left[16\sqrt{p^2+{1\over4}g^2\phi^2+g_{\Delta}^2\Delta^2}+2\sqrt{p^2+{1\over4}g^2\phi^2+4g_{\Delta}^2\Delta^2}
\right. \\ && \left.
+{8\mu^2g_{\Delta}^2\Delta^2\over(p^2+{1\over4}g^2\phi^2+g_{\Delta}^2\Delta^2)^{3\over2}}
+{4\mu^2g_{\Delta}^2\Delta^2\over(p^2+{1\over4}g^2\phi^2+4g_{\Delta}^2\Delta^2)^{3\over2}}
\right]\;,\\
\Omega_1^{\rm CFL,finite}&=&-\int_p\left[16E_o+2E_s\right]-\Omega_1^{\rm CFL, div}\;,\\
\Omega_1^{{\rm CFL}, \mu}&=&0\;.
\end{eqnarray}
After renormalization, we obtain
\begin{eqnarray}
\nonumber
\Omega_{0+1}^{\rm CFL}&=&{3\over4}m_{\ms}^2\phi^2_{\ms}+{3\over16}(3\lambda_{1,\ms}+\lambda_{2,\ms})\phi_{\ms}^4+{3\over4}(3\lambda_{3,\ms}+\lambda_{4,\ms}-2\lambda_{5,\ms})\phi_{\ms}^2\Delta_{\ms}^2\\
\nonumber
&&+3({m_{\Delta,\ms}^2}-4\mu^2)\Delta_{\ms}^2+{3\over4}(6\lambda_{1,\ms}^{\Delta}+2\lambda_{2,\ms}^{\Delta}+3\lambda_{3,\ms}^{\Delta})\Delta_{\ms}^4
\\ 
&&
\nonumber
-{16{\mu}_{}^2g_{\Delta,\ms}^2\Delta_{\ms}^2\over(4\pi)^2}
\left[
2\log{\Lambda^2\over{1\over4}g_{\ms}^2\phi_{\ms}^2+ g_{\Delta,\ms}^2\Delta_{\ms}^2}+\log{\Lambda^2\over{1\over4}g_{\ms}^2\phi_{\ms}^2+4g_{\Delta,\ms}^2\Delta^2_{\ms}}
\right]\\
&&
\nonumber
+{8({1\over4}g^2_{\ms}\phi^2_{\ms}+g_{\Delta,\ms}^2\Delta_{\ms}^2)^2\over(4\pi)^2}
\left[
\log{\Lambda^2\over{1\over4}g_{\ms}^2\phi_{\ms}^2+ g_{\Delta,\ms}^2\Delta_{\ms}^2}
+{3\over2}
\right]\\
&&
+{({1\over4}g^2_{\ms}\phi^2_{\ms}+4g_{\Delta,\ms}^2\Delta_{\ms}^2)^2\over(4\pi)^2}
\left[
\log{\Lambda^2\over{1\over4}g_{\ms}^2\phi_{\ms}^2+ 4g_{\Delta,\ms}^2\Delta_{\ms}^2}
+{3\over2}
\right]\;.
\end{eqnarray}
Deep into the CFL phase, we can ignore the quark mass. In this case, the subtraction term
can be calculated analytically, cf Eq.~(\ref{mycont}). 
We can also ignore the $\Delta_{\ms}^4$-terms since they are suppressed.
The thermodynamic potential then reduces to
\begin{eqnarray}
\Omega_{0+1}^{\rm CFL}&=&
-12{\mu}^2_{}\Delta_{\ms}^2-
{12{\mu}_{}^4\over(4\pi)^2}
-{16{\mu}_{}^2g_{\Delta,\ms}^2\Delta_{\ms}^2\over(4\pi)^2}
\left[
2\log{\Lambda^2\over g_{\Delta,\ms}^2\Delta_{\ms}^2}+\log{\Lambda^2\over4g_{\Delta,\ms}^2\Delta_{\ms}^2}
\right]\;.
\end{eqnarray}
Now it is straightforward to calculate the gap, pressure, energy density, and speed of sound.
One finds
\begin{eqnarray}
\label{guppy}
g_{\Delta,\ms}^2\Delta_{\ms}^2&=&\Lambda^22^{-{2\over3}}e^{{(4\pi)^2\over4g_{\Delta,\ms}^2}-1}
\;,\\
p&=&{12\mu^4\over(4\pi)^2}+{48\mu^2g_{\Delta,\ms}^2\Delta_{\ms}^2\over(4\pi)^2}\;,\\
\epsilon&=&{36\mu^4\over(4\pi)^2}+{48\mu^2g_{\Delta,\ms}^2\Delta_{\ms}^2\over(4\pi)^2}\;,\\
c_s^2&=&{\mu^2+2g_{\Delta,\ms}^2\Delta_{\ms}^2\over3\mu^2+2g_{\Delta,\ms}^2\Delta_{\ms}^2}\approx{1\over3}\left[1+{4\over3}
{g_{\Delta,\ms}^2\Delta_{\ms}^2\over\mu^2}
\right]\;.
\end{eqnarray}
Note that the gap~(\ref{guppy}) is independent of the scale $\Lambda$.
The CFL gap is also $2^{-{1\over3}}$ times the 2SC gap, cf. Eq.~(\ref{ymsee}).

\subsection{Sample calculations}
The complete renormalized thermodynamic potential $\Omega$
is given by Eq.~(\ref{komplett}).
The last term $\Omega_1^{\rm CFL,fin}$, which is the difference between the unrenormalized
one-loop contribution $\Omega_1$ and the subtraction term, 
is by construction finite. Generally, it must be handled numerically.
In some limits, $\Omega_1^{\rm CFL,fin}$ simplifies. Above, 
it was shown that in the limit of vanishing quark masses, i.e. 
deep into the CFL phase, it reduces to ${4N_cN_f\mu^4\over3(4\pi)^2}$.
Perhaps a better approximation is to evaluate 
$\Omega_1^{\rm CFL,fin}$ at vanishing gaps. 
For a common chemical potential $\mu$, the term reduces to
\begin{eqnarray}
    \Omega_1^{\rm CFL,finite} &=& -2N_c\int_p\left[2(\mu-E)\theta(\mu-E) + (\mu-E_s)\theta(\mu-E_s)\right]\;.
\end{eqnarray}
This approximation is briefly discussed at the end of 
Appendix~\ref{omegarenorm}.
We have checked it in the 2SC phase against numerically exact results, and the agreement is excellent.

\begin{figure}[htb!]
    \centering
    \includegraphics[width=\linewidth]{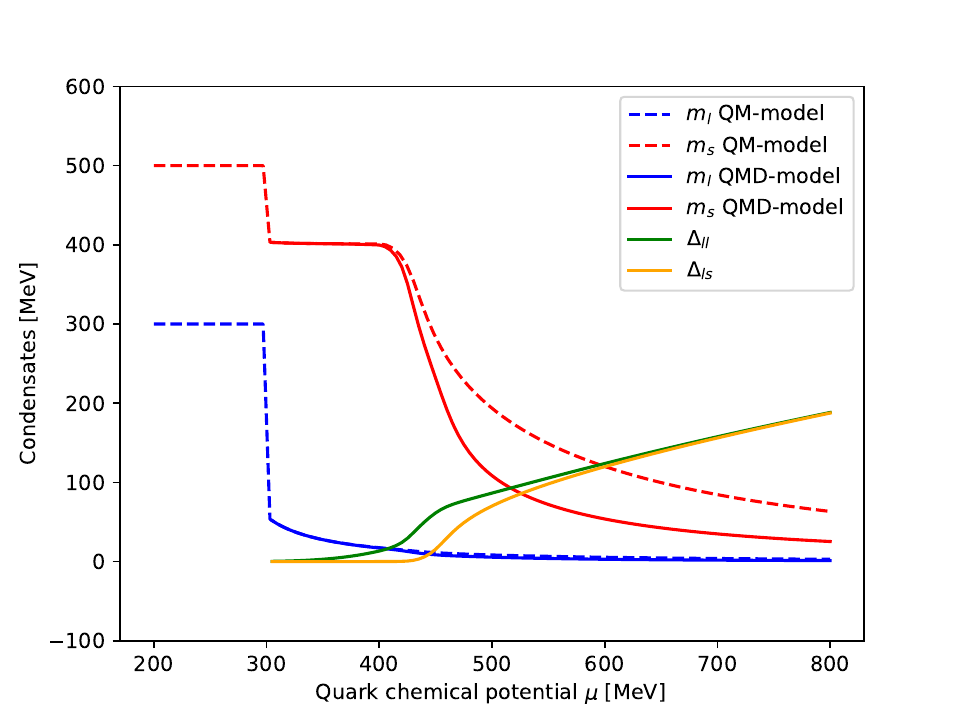}
    \caption{Light and $s$-quark masses (solid red and blue lines) and diquark condensates (solid green and yellow lines) as a function of quark chemical potential $\mu$.
    For comparison, light and $s$-quark masses in the absence of superconducting gaps.
    See main text for details.}
    \label{fig:CFL_condensates}
\end{figure}

For the experimentally determined parameters, we use \(m_\pi = 140\) MeV, \(f_\pi = 93\) MeV, $m_{\sigma}=500$ MeV
and $m_{\eta'}=548$ MeV~\footnote{Experimentally, $m_{\eta^{\prime}}=957.8$ MeV.
We choose to match to the mass of \(m_\eta\) since we have set $c=0$ in our calculations.}. The light quark masses is $m_q=300$ MeV
and the $m_s=500$ MeV.
For the parameters in the diquark sector we choose $m_\Delta = 1000$ MeV, $\lambda_{3} = 300$, $\lambda_{4} = -200$, $\lambda_{5} = 0$, $\lambda_{1,\Delta} = 100$, 
$\lambda_{2,\Delta} = 0$, and $g_\Delta = 3$. 
In Fig.~\ref{fig:CFL_condensates}, we show the
masses $m_l$ and $m_s$ of the light (solid blue line) and $s$ quarks
(solid red line) as functions of the quark chemical potential $\mu$.
We also show the gaps $\Delta_{ud}$ (solid green line) and
$\Delta_{ds}=\Delta_{us}$. In order to gauge the effects of the
gaps on the quark masses, we plot $m_l$ and $m_s$ assuming
$\Delta_{ud}=\Delta_{ds}=\Delta_{us}=0$ (dashed red and blue lines).
The transition from the vacuum phase to the 2SC quark matter takes
place at $\mu=300$ MeV. In other words, there is no phase of
normal quark matter for the particular values of the parameters
chosen here. At the same value of $\mu$, the quark masses make a large jump. The transition from the 2SC phase to the CFL phase
takes place at $\mu=450$ MeV. The light quark masses are almost
vanishing, whereas the mass of the $s$ quark drops quickly.
Moreover, we see that the gaps $\Delta_{ud}$ and 
$\Delta_{ds}=\Delta_{us}$ are essentially equal from $\mu=600$ MeV onward, showing that we approach the ideal CFL phase for large $\mu$.
Interestingly, the solid and dashed blue lines are on top of each
other, implying that the light quarks hardly are affected by the presence of superconducting gaps. Their mass has dropped to very low values before the gaps start increasing.
This is in contrast to the solid and dashed red lines, showing a significant difference 
for $\mu\geq450$ MeV.

\subsection{Quark-meson diquark model as a cutoff field theory}
We have used dimensional regularization and the $\overline{\rm MS}$ scheme to handle the
divergences that appear in the calculations. However, there is nothing that prevents us from using
a simple three- or four-dimensional cutoff $\Lambda$ to regulate the model.
We then keep $\Lambda$ finite and treat the QMD model as a cutoff field theory according to
the modern view of renormalization~\cite{lepage}.
Let us illustrate this by reconsidering the simplest case: the 2SC phase in two-flavor QCD
in the chiral limit, 
where we ignore the parity partners and use a common quark chemical potential $\mu$.
The thermodynamic potential is
\begin{eqnarray}
\nonumber
\Omega_{0+1}&=&m^2\phi^2+(m_{\Delta}^2-4\mu^2)\Delta^2+\lambda_1\phi^4
+\lambda_3\phi^2\Delta^2+\lambda_{\Delta}\Delta^4
\\ &&
-4\int{d^3p\over(2\pi)^2}\left[\sqrt{{(E_p\pm\mu)^2}+g_{\Delta}^2\Delta^2}+\sqrt{p^2+m_l^2}\right]\;,
\end{eqnarray}
where $E_p=\sqrt{p^2+m_l^2}$. We first integrate over angles and then over $p$, imposing
an ultraviolet cutoff $\Lambda$. In the limit of large $\Lambda$, the thermodynamic potential can be written as
\begin{eqnarray}
\nonumber
\Omega_{0+1}^{\rm 2SC}&=&
\left[m^2(\Lambda)-{6g^2\over(4\pi)^2}\Lambda^2\right]\phi^2(\Lambda)
+\left[m_{\Delta}^2(\Lambda)
-{16g_{\Delta}^2\over(4\pi)^2}\Lambda^2
\right]\Delta^2(\Lambda)
\\ &&
\nonumber
-4\left[1+{4g_{\Delta}^2\over(4\pi)^2}\left(\log{4\Lambda^2\over m_l^2+g_{\Delta}^2\Delta^2}-2\right)\right]\mu^2\Delta^2(\Lambda)
\\ &&
\nonumber
+\left[\lambda_1(\Lambda)
+{g^4\over8(4\pi)^2}\left(\log{4\Lambda^2\over m_l^2}-{1\over2}\right)
+{g^4\over4(4\pi)^2}\left(\log{4\Lambda^2\over m_l^2+g_{\Delta}^2\Delta^2}-{1\over2}\right)
\right]\phi^4(\Lambda)
\\ &&
\nonumber
+\left[\lambda_3(\Lambda)
+{2g^2g_{\Delta}^2\over(4\pi)^2}\left(\log{4\Lambda^2\over m_l^2+g_{\Delta}^2\Delta^2}-{1\over2}\right)
\right]\phi^2(\Lambda)\Delta^2(\Lambda)
\\ &&
+\left[\lambda_{\Delta}(\Lambda)
+{4g_{\Delta}^4\over(4\pi)^2}\left(\log{4\Lambda^2\over m_l^2+g_{\Delta}^2\Delta^2}-{1\over2}\right)
\right]\Delta^4(\Lambda)\;,
\label{rgpot}
\end{eqnarray}
where we have promoted all masses, couplings, and fields to parameters that are running with the cutoff $\Lambda$. Requiring the effective potential to be independent of $\Lambda$ gives
rise to a number of differential equations that govern how the parameters change with the cutoff. For example, the diquark field satisfies
\begin{eqnarray}
\label{cut1}
\Lambda{d\Delta^2(\Lambda)\over d\Lambda}&=&-{8g_{\Delta}^2(\Lambda)\Delta^2(\Lambda)\over(4\pi)^2}\;,
\end{eqnarray}
which follows from taking the derivative of the third term in Eq.~(\ref{rgpot}) with respect to
$\Lambda$. The equation for the quark-diquark follows by noting that neither the corresponding vertex nor the quark field depend on $\Lambda$, and therefore
\begin{eqnarray}
\Lambda{d\over d\Lambda}\left[g_{\Delta}^2(\Lambda)\Delta^2(\Lambda)\right]&=&0\;. 
\end{eqnarray}
This yields
\begin{eqnarray}
\label{cut2}
\Lambda{dg_{\Delta}^2(\Lambda)\over d\Lambda}&=&{8g_{\Delta}^4(\Lambda)\over(4\pi)^2}\;.    
\end{eqnarray}
The solutions to Eqs.~(\ref{cut1}) and~(\ref{cut2}) are
\begin{eqnarray}
\Delta^2(\Lambda)&=&\left[1-{4g_{\Delta,0}^2\over(4\pi)^2}\log{\Lambda^2\over\Lambda_0^2}\right]\Delta_0^2\;.  
\\ 
g_{\Delta}^2(\Lambda)&=&{g_{\Delta,0}^2\over1-{4g_{\Delta,0}^2\over(4\pi)^2}\log{\Lambda^2\over\Lambda_0^2}}\;. 
\end{eqnarray}
Similar results are obtained in dimensional regularization.
The remaining $\Lambda$-dependent parameters can be found in the same way and substituted into the thermodynamic potential $\Omega_{0+1}^{\rm 2SC}$ as before.~\footnote{We note the term 
proportional to $\Lambda^2$ in the first term in Eq.~(\ref{rgpot}) which complicates the 
equation for $m^2(\Lambda)$. One of the virtues of dimensional regularization is that such terms are set to zero and only logarithmic divergences appear as poles in $\epsilon$.}
A related analysis was performed in~\cite{bjs}. The authors calculate the contributions to $\Omega_1$ for large
$\Lambda$, which can be easily obtained from
Eq.~(\ref{rgpot}). The terms that diverge in the limit 
$\Lambda\rightarrow\infty$ are eliminated by adding appropriately chosen counterterms $\delta m^2$, etc. After renormalization of the parameters, the limit $\Lambda\rightarrow\infty$ was taken.

\section{Goldstone bosons in the color superconducting phases}
In this section, we discuss the massless Nambu Goldstone bosons that appear in the color superconducting phases of QCD. We will also see that their number is different
in QCD and in the QMD model. The reason is that the local $SU(3)_c$ gauge symmetry
is replaced by a global $SU(3)_c$ symmetry. 
A local gauge symmetry is never broken spontaneously~\cite{elitzur}, meaning that the expectation value of any gauge non-invariant operator vanishes. Instead, the gauge bosons become massive via the Higgs mechanism~\cite{sonstep1,casal,sanni}.

The classification and properties
of massless excitations have a long history which dates back to the paper by
Nielsen and Chadha in the 1970s~\cite{chadha}. In Lorentz-invariant theories, there
are as many massless Goldstone modes as there are broken generators.
Introducing a finite chemical potential breaks Lorentz invariance~\footnote{Conventionally, the view is that a finite $\mu$ breaks Lorentz invariance explicitly. In recent years, a different view has been promoted
in which the symmetry is considered spontaneously broken by a time-dependent ground 
state ~\cite{nico1,nico2}.} 
and it is no longer guaranteed that the number of massless modes is equal to the number of broken generators. Work on classifying Goldstone bosons and their properties can be
found in~\cite{watanabe,hidaka,nicolis,ata} and a review in~\cite{brauner}.

Let us consider the 2SC phase in two-flavor QCD, where 
the symmetry-breaking pattern is
$SU(3)_c\rightarrow SU(2)_c$. 
The symmetry breaking is due to a nonzero expectation value of $\Delta_3=\Delta_0$, so the
GB modes are found among the diquark fields.
The quadratic part of the Lagrangian involving the diquarks is
\begin{eqnarray}
\nonumber
{\cal L}_2^{\rm diquarks}&=&
(\partial_{\mu}+2i\mu\delta_{0\mu})\Delta_a^{\dagger}
(\partial_{\mu}-2i\mu\delta_{0\mu})\Delta_a-
m_{\Delta}^2\Delta^{\dagger}_a\Delta_a
-{\lambda}_{\Delta}\Delta_0^2\left[\Delta_1^{\dagger}\Delta_1+\Delta_2^{\dagger}\Delta_2+\Delta_3^{\dagger}\Delta_3
\right. \\ &&\left.
+2\Delta_3^{\dagger}\Delta_3^{\dagger}+2\Delta_3\Delta_3
\right]\;.
\end{eqnarray}
It is convenient to write $\Delta_3$ in terms of the two real fields
$\phi_1$ and $\phi_2$, $\Delta_3=\Delta_0+{1\over\sqrt{2}}(\phi_1+i\phi_2)$. The quadratic Lagrangian can be written as
\begin{eqnarray}
\nonumber
{\cal L}_2^{\rm diquarks}&=&
(\partial_{\mu}+2i\delta_{0\mu}\mu)\Delta_1^{\dagger}
(\partial_{\mu}-2i\delta_{0\mu}\mu)\Delta_1
-(m_{\Delta}^2+2\lambda_{\Delta}\Delta_0^2)\Delta_1^{\dagger}\Delta_1\;.
\\
&&
\nonumber
+(\partial_{\mu}+2i\delta_{0\mu}\mu)\Delta_2^{\dagger}
(\partial_{\mu}-2i\delta_{0\mu}\mu)\Delta_2
-(m_{\Delta}^2+2\lambda_{\Delta}\Delta_0^2)\Delta_2^{\dagger}\Delta_2
+{1\over2}(\partial_{\mu}\phi_1)(\partial^{\mu}\phi_1)
\\&&
\nonumber
-{1\over2}(m_{\Delta}^2+6\lambda_{\Delta}\Delta_0^2)\phi_1^2
+{1\over2}(\partial_{\mu}\phi_2)(\partial^{\mu}\phi_2)
-{1\over2}(m_{\Delta}^2+2\lambda_{\Delta}\Delta_0^2)\phi_2^2
\\ &&
+2\mu(\phi_2\partial_0\phi_1-\phi_1\partial_0\phi_2)
\;.
\end{eqnarray}
The spectrum is
\begin{eqnarray}
E_{\Delta_1}^{\pm}(p)&=&\sqrt{p^2+4\mu^2}\pm2\mu\;,\\
E_{\Delta_2}^{\pm}(p)&=&\sqrt{p^2+4\mu^2}\pm2\mu\;,\\
E_{\Delta_3}^{\pm}(p)&=&\sqrt{p^2+12\mu^2-m_{\Delta}^2\pm\sqrt{16p^2\mu^2+(12\mu^2-m_{\Delta}^2)^2}}\;,
\end{eqnarray}
where we have used that $m_{\Delta}^2-4\mu^2+2\lambda_\Delta\Delta_0^2=0$ at the minimum
of the tree-level thermodynamic potential $\Omega_0$.
The three dispersion relations $E_{\Delta_i}^-$ with are all massless
Goldstone bosons. The relations $E_{\Delta_{1,2}}^-$ are both quadratic for small momenta $p$, while $E_{\Delta_3}^-$ is linear, 
\begin{eqnarray}
E_{\Delta_{1,2}}^-(p)&=&{p^2\over4\mu}+{\cal O}(p^4)
\;,\\
   E_{\Delta_3}^-(p)&=&\sqrt{{4\mu^2-m_{\Delta}^2\over12\mu^2-m_{\Delta}^2}}\,p+{\cal O}(p^3)\;.
\label{treecs}
\end{eqnarray}
The coefficient of the linear term is the speed of sound $c_s$
In the limit $\mu\rightarrow\infty$, the speed of sound approaches $c_s={1\over\sqrt{3}}$, which is the conformal limit of the speed of sound in QCD.
Note that $c_s$ approaches the limit from below. Once we include quark loops,
the approach is from above, as we shall see later.

We have seen that in the 2SC phase, there are three Goldstone bosons, two with a quadratic dispersion relation and one with a linear dispersion relation. 
The linear modes are referred to as type-A GBs, and the quadratic modes are called
type-B GBs. Counting rules for the numbers of type-A and type-B
NG modes were derived in~\cite{watanabe,hidaka}. Denoting
these numbers by $n_A$ and $n_B$, respectively, the rules are
\begin{eqnarray}
n_A&=&{\rm dim}\,G-{\rm dim\,}H-{\rm rank}\,\rho\;,\\
n_B&=&{1\over2}{\rm rank}\,\rho\;.
\end{eqnarray}
where $G$ is the full symmetry group, $H$ is the unbroken subgroup, and 
$\rho$ is the commutator matrix with elements
\begin{eqnarray}
\rho_{ab}&=&\lim_{V\rightarrow\infty}{i\over V}\langle0|[T_a,T_b]|0\rangle\;,    
\end{eqnarray}
where $\langle0|A|0\rangle$ is the ground-state expectation value of $A$ and
$V$ is the spatial volume of the system. 
In the 2SC phase with two flavors, $G=SU(3)_c$ and
$H=SU(2)_c$. It is clear that $\rho_{ab}=0$ if $T_a$ and $T_b$ are unbroken generators. The only nonzero matrix elements are
$\rho_{45}=-\rho_{54}$ 
and $\rho_{67}=-\rho_{76}$. 
Thus, the rank of the matrix $\rho$ is four. This implies that $n_A=1$ and $n_B=2$
in accordance with our calculations. In QCD, the five broken generators implies
that five of the eight gluons acquire a mass via the Higgs mechanism.
The commutators with nonzero expectation values 
can be written in terms of the generators $T_3$ and $T_8$,
\begin{eqnarray}
    [T_4,T_5]=\frac{i}{2}(T_3+\sqrt{3}T_8)\;,
    \hspace{1cm}
    [T_6,T_7]=\frac{i}{2}(T_3-\sqrt{3}T_8)\;.
\end{eqnarray}
Since $T_3$ is unbroken in the 2SC phase, the nonzero matrix elements is 
a direct consequence of the $Q_8$ charge in the ground state.
If we enforce charge neutrality, $Q_8=0$, it immediately follows that 
${\rm rank}\,\rho=0$ and $n_B=0$, as discussed in the context
of kaon condensation~\cite{kaonboys}.
Naively. one may think that this would lead to five GB, each with a linear dispersion relation.
However, introducing a nonzero chemical potential $\mu_8$ breaks the 
global $SU(3)_c$ symmetry down to $SU(2)_c\times U(1)_{Q_8}$.
A nonzero expectation value of $\Delta_3$ breaks the 
$U(1)_{Q_8}$-symmetry leading to a single linear Goldstone boson. This was first 
observed in the context of the NJL model~\cite{GBblas,GBebert}.

Using the same methods as in the previous section, it can be shown
that the speed of sound in the two-flavor case in the 
large-$\mu$ limit is given by~\cite{us}
\begin{eqnarray}
\label{sos}
c_s^2&=&{\mu^2 + g_{\Delta}^2\Delta^2_0\over3\mu^2 + g_{\Delta}^2\Delta_0^2}\;.  
\end{eqnarray}
There is another way to obtain this result, namely via Son's effective theory for
the $U(1)$ Goldstone boson~\cite{sonlow}. The idea is that the low-energy effective theory~\footnote{This theory is valid for momenta $p$ well below the gap.} for this mode is given in terms of the pressure of the system.
If the pressure is a function of a chemical potential $\mu$, $p(\mu)$, the effective Lagrangian reads
\begin{eqnarray}
{\cal L}&=&p\left(\mu\rightarrow\sqrt{(\partial_0\phi-\mu)^2+(\partial_i\phi)(\partial^i\phi)}\right)\;,    
\end{eqnarray}
where $\phi$ denotes the massless Goldstone boson.
Making the above substitution in the expression for the pressure 
$p={8\mu^4\over(4\pi)^2}+{16\mu^2g_{\Delta}^2\Delta_0^2\over(4\pi)^2}$, and 
expanding in powers of the derivatives of the fields, generates a constant term as well as linear, quadratic, and higher-order terms. The constant is just the pressure.
The linear term is a total derivative and can be ignored if one is not interested, for example, in vortices. The quadratic term gives rise to the free dispersion relation, while higher-order terms are self-interaction terms of $\phi$.
Note that these interactions are derivatively coupled and the GB is therefore weakly
interacting. 
After rescaling the field 
\begin{eqnarray}
\varphi&=&{\sqrt{6}\mu\over\pi}\sqrt{1+{g_{\Delta}^2\Delta_0^2\over3\mu^2}}\phi\;,  
\end{eqnarray}
the quadratic Lagrangian is
\begin{eqnarray}
{\cal L}_2&=&{1\over2}(\partial_0\varphi)^2+
{\mu^2+g_{\Delta}^2\Delta_0^2\over3\mu^2+g_{\Delta}^2\Delta_0^2}(\partial_i\varphi)(\partial^i\varphi)\;.
\end{eqnarray}
In momentum space, the dispersion relation is
\begin{eqnarray}
E^2(p)&=&{\mu^2+g_{\Delta}^2\Delta_0^2\over3\mu^2+g_{\Delta}^2\Delta_0^2}p^2\;.
\end{eqnarray}
The GB propagates with the speed $c_s$ given by the square root of Eq.~(\ref{sos}).
We note in passing that the tree-level result for the speed of sound Eq.~(\ref{treecs}) can be obtained in the same manner. The behavior and the
approach to the conformal limit are very different from the one-loop result
Eq.~(\ref{sos}). We can also obtain the interaction terms of $\varphi$. Expanding
to fourth order in the field, we find
\begin{eqnarray}
{\cal L}_{3+4}&=&
-{\pi^2\over3\sqrt{6}\mu^3}\left(1-{g^2_{\Delta}\Delta_0^2\over2\mu^2}\right)
\partial_0\varphi(\partial_{\mu}\varphi)(\partial^{\mu}\varphi)
+{\pi^2\over72\mu^4}\left(1-{2g^2_{\Delta}\Delta_0^2\over3\mu^2}\right)
(\partial_{\mu}\varphi)(\partial^{\mu}\varphi)^2\;.
\label{eftqcd}
\end{eqnarray}
Similar results were obtained for the CFL phase in~\cite{manuel1,manuel2}, where the authors studied the shear and bulk viscosity of dense QCD matter.




\section{Summary and Outlook}
In the present paper, we have discussed various aspects of the two- and three-flavor quark-meson diquark model at zero temperature and finite density. 
Let us summarize the most important results:
\begin{enumerate}
    \item We have discussed the properties of the two- and three flavor quark-meson diquark model as an effective low-energy renormalizable model for QCD. The effective degrees of freedom, symmetries, and renormalizabilty dictate the terms in the QMD Lagrangian.
    \vspace{3mm}
    \item We have renormalized the model in the mean-field approximation for general quark masses, chemical potentials, and superconducting gaps using the
    $\overline{\rm MS}$ renormalization scheme. 
     By solving the renormalization group equations for the running masses and couplings, we obtained a scale-independent thermodynamic potential.
    It is then possible to
    describe normal quark matter, the 2SC and CFL phases.
    \vspace{3mm}
    \item
    We have combined the  $\overline{\rm MS}$ and  $\overline{\rm OS}$ renormalization schemes in order to express the running parameters in the scalar sector in terms of meson pole masses and decay constants. The remaining couplings in the diquark sector
    are free. Although the parameter space is very large, preliminary
    calculations suggest that the results are most sensitive to the diquark mass and the diquark coupling.
    \vspace{3mm}
    \item We have analytically studied the properties of the 2SC and CFL phases at large $\mu$ and found expressions for pressure, energy density, and speed of sound. The speed of sound approaches a constant that depends only on the diquark coupling.
    The dominant contributions to the pressure and energy density
    are the Stefan-Boltzmann results with the leading corrections that depend on the gap. Similar results are obtained in the 
    pion-condensed phase at large $\mu_I$.
    \vspace{3mm}
    \item We have classified the GBs that appear in the 2SC phase and shown that their type and numbers agree with general counting rules.
    We have derived the effective low-energy theory for the GB 
    associated with the breaking of the $U(1)_{Q_8}$ symmetry in the 2SC phase using Son's prescription.
    \vspace{3mm} 
    \item 
    We have presented numerical results for the sound speed in the pion-condensed phase. The agreement with lattice and NJL model calculations at finite $\mu_I$ is remarkable. 
    We have performed a sample calculation of the quark phases at finite $\mu$.
    Choosing reasonable
    values for the diquark mass and couplings in the diquark sector, 
    it is possible to obtain physically sound results for the phase
    structure, quark masses, and gaps as functions of $\mu$.
\end{enumerate}
In this paper, we have laid the foundation for extensive numerical work. Let us briefly mention three directions for future work:
\begin{enumerate}
\item Phases and EoS. For a given set of masses and couplings, and fixed 
$\mu$, we must solve the gap six equations and the three neutrality conditions simultaneously. For a given value of $\mu$,
the system consists of normal quark matter or color superconducting matter, either in the 2SC phase or the CFL phase.
For each value of $\mu$, we can calculate $p$ and $\epsilon$, and 
thus obtain the EoS for deconfined matter.

\vspace{3mm}
\item Hybrid stars. 
The QMD model includes a number of free parameters 
in the Lagrangian. It would be interesting to see if it is possible to construct massive neutron stars with quark cores that fit current observational constraints by varying the free parameters in the diquark sector. This can be done using a Maxwell or Gibbs construction involving a nuclear equation state as briefly discussed. One can also use interpolation~\cite{baymrev}.
In this approach, one trusts the nuclear EoS up to a certain density 
$\rho_{\rm low}$
and the QMD EoS down to a certain density $\rho_{\rm high}$,
and smoothly interpolates the EOS between these two densities. 

\item Finite temperature. So far, we have worked exclusively at $T=0$. Temperature effects
are, however, in principle straightforward to include in the mean-field approximation. The phase diagram in the $\mu$--$T$
plane is of intrinsic interest, with the possibility of different melting patterns of the condensates giving rise to several new phases.  
    
\end{enumerate}


Finally, there is one aspect of dense QCD and color superconductivity that we have not yet discussed, namely the effects of magnetic fields on the pairing dynamics~\cite{ferrer,ferrer2,noronja,harmen,faya,faya0,yu,mss2}.
Strongly interacting matter in extreme magnetic fields The interest in dense QCD in the presence of a magnetic field is mainly driven by applications in astrophysics, where their presence plays an important role. For example, in neutron stars, magnetic fields on the surface can be up to
$B=10^{12}$ G. For magnetars, the magnetic fields can be 2--3 orders of magnitude larger than that~\cite{duncan}. In stellar interiors, the magnitude of the magnetic fields might be even higher than these values, perhaps up to $B=10^{18}$ G.

In the 2SC phase, $U(1)_{\rm em}$ is broken; however, there is a 
$\tilde{U}(1)_{\rm em}$, which is unbroken. The generator is
\begin{eqnarray}
\tilde{Q}&=&Q-{1\over\sqrt{3}}T_8\;.    
\end{eqnarray}
The corresponding rotated gauge field $\tilde{A}_{\mu}$ is a linear combination of the
photon in the vacuum and the eighth gluon.
A constant magnetic background $\tilde{B}$ is given by the vector potential
$\tilde{A}_{\rm ext}=(0,0,Bx,0)$.
While the vacuum respects this symmetry, the quark quasiparticles carry
$\tilde{Q}$ charge: $\tilde{Q}(u_r)=\tilde{Q}(u_g)={1\over2}$, 
$\tilde{Q}(u_b)=1$,  $\tilde{Q}(d_r)=\tilde{Q}(d_g)=-{1\over2}$, and 
$\tilde{Q}(d_b)=0$. In order to calculate the thermodynamic potential in the mean-field
approximation, it is convenient to know the form of the quasiparticle spectrum. This problem is reminiscent of the standard problem of
a particle in a constant magnetic background, although the superconducting gap
complicates the calculations. In the chiral limit and with a quark 
chemical potential $\mu$, the spectrum is~\cite{yu}
\begin{eqnarray}
E_0&=&\sqrt{p^2+g^2\phi_0^2}\;,\hspace{1cm}\text{for}\,\,\tilde{Q}=0\;,\\
E_{\tilde{Q}}&=&\sqrt{(E_q\pm\mu)^2+g_{\Delta}^2\Delta^2}
\;,\hspace{1cm}\text{for}\,\,\tilde{Q}=1,\,\pm{1\over2}\;,
\end{eqnarray}
where 
\begin{eqnarray}
E_q&=&\sqrt{p_z^2+g^2\phi^2+2|\tilde{Q}B|n}\;.    
\end{eqnarray}
In the original calculations using the NJL model with conventional
regularization, the gaps show oscillatory behavior induced by the
magnetic field. This has been interpreted as de Haas–van Alphen oscillations, due to the discrete nature of the dispersion relation.
However, in~\cite{mss2}, this interpretation has been questioned in view of the regularization artefacts of the NJL model. Using
the combined Magnetic Field Independent Regularization-- Medium Separation Scheme (MFIR-MSS scheme), these oscillations are suppressed. In view of this, it would be of interest to revisit
the problem of color superconductivity in a strong magnetic background using the QMD model.

We end this section with a remark regarding large fields. For sufficiently large magnetic fields, the ground state is not given by 
constant superconducting gaps. For example, in the 2SC phase, the
$U(1)_A$ symmetry is spontaneously broken.~\footnote{It is also explicitly broken by instantons, but this effect decreases with chemical potential $\mu$.} The effective Lagrangian for the GB, denoted by $\eta$ in~\cite{sonmag}, has domain walls as 
a static solution of the equation of motion.
For sufficiently large magnetic fields, $eB\geq 10^{17}$--$10^{18}$ G,
the domain walls are stable: the free-energy gain arising from the
interaction with the magnetic field outweighs the surface-energy cost
of creating the walls. This interaction terms is due to the axial anomaly.

{


\authorcontributions{First draft was written by JOA. Numerical work has been carried out by MN. Plots have been generated by MN.}

\funding{This research received no external funding}

\dataavailability{Code and data can be found at https://github.com/mathiaspno/QM-model.}

\acknowledgments{We thank Ricardo Farias for sharing the data of Ref.~\cite{isofarias1}, Bastian Brandt for sharing the data of Ref.~\cite{isobrandt}, and Kenji Fukushima for sharing the data of Ref.~\cite{minato}.}

\conflictsofinterest{The authors declare no conflicts of interest. } 



\abbreviations{Abbreviations}{
The following abbreviations are used in this manuscript:
\\

\noindent 
\begin{tabular}{@{}ll}
BCS& Bardeen-Cooper-Schrieffer\\
BEC& Bose-Einstein condensation/condensed\\
CFL&color-flavor-locked \\
$\chi$pt& chiral perturbation theory\\
EoS &equation of state \\
$\overline{\rm MS}$ scheme&
modified minimal subtraction scheme
\\
GB & Goldstone boson \\
MFIR &Magnetic Field Independent Regularization \\
MSS scheme&Medium Separatation Scheme \\
NJL & Nambu-Jona-Lasinio\\
NQM &normal quark matter\\
${\rm OS}$ scheme &on-shell scheme\\
QM&quark-meson \\
QMD & quark-meson diquark\\
QCD & quantum chromodynamics \\
2SC phase&2 flavor color superconducting phase\\
UV& ultraviolet 
\end{tabular}
}

\setcounter{equation}{0}
\appendix

\renewcommand{\thesection}{A}
\section{Integrals}
In the calculations, we encounter a couple of ultraviolet divergent integrals
in three dimensions that must be regularized. 
We use dimensional regularization in $d$ dimensions, where the integral is defined as
\begin{eqnarray}
\int_p&=&\left({e^{\gamma}\Lambda^{2}\over4\pi}\right)^{\epsilon}\int{d^dp\over(2\pi)^d}\;,
\end{eqnarray}
where $d=3-2\epsilon$ and  $\Lambda$ is the renormalization scale associated with the $\overline{\rm MS}$-scheme.
The specific integrals needed are
\begin{eqnarray}
\label{i0}
\int_p\sqrt{p^2+m^2}&=&-{m^4\over2(4\pi)^2}\left[{1\over\epsilon}+\log{\Lambda^2\over m^2}+{3\over2}+{\cal O}(\epsilon)\right]\;,\\
\int_p{1\over(p^2+m^2)^{3\over2}}&=&{4\over(4\pi)^2}\left[{1\over\epsilon}+\log{\Lambda^2\over m^2}+{\cal O}(\epsilon)\right]\;.
\label{i2}
\end{eqnarray}
There are also a few divergent vacuum integrals in four dimensions that need to be evaluated in dimensional regularization with $d=4-2\epsilon$,
\begin{eqnarray}
\nonumber
A(m^2)&=&
\int_k{1\over k^2-m^2}
\\ &=&
\label{defa}\frac{im^2}{(4\pi)^2}\left[\frac{1}{\epsilon} + \log\frac{\Lambda^2}{m^2}+1+{\cal O}(\epsilon)\right]\;,\\
\nonumber
B_{q_1q_2}(p^2)&=&
\label{bdef}
\int_k\frac{1}{(k+p)^2-m_{q_1}^2}\frac{1}{k^2-m_{q_2}^2}\\
    &=&\frac{i}{(4\pi)^2}\left[\frac{1}{\epsilon}+\frac{1}{2}\log\frac{\Lambda^2}{m_{q_1}^2}+\frac{1}{2}\log\frac{\Lambda^2}{m_{q_2}^2} + C_{q_1q_2}(p^2,m_{q_1}^2,m_{q_2}^2)+{\cal O}(\epsilon)\right]\;,\\
B_{q_1q_2}^{\prime}(p^2)&=&{i\over(4\pi)^2}\left[C^{\prime}(p^2,m_{q_1}^2,m_{q_2}^2)+{\cal O}(\epsilon)\right]\;,    
\end{eqnarray}
where 
\begin{eqnarray}
\nonumber
    C_{q_1q_2}(p^2,m_{q_1}^2,m_{q_2}^2) &=& 2+\frac{1}{2}\frac{m_{q_1}^2-m_{q_2}^2}{p^2}\log\frac{m_{q_2}^2}{m_{q_1}^2}
\\ &&    
\nonumber
    -\frac{\mathcal{G}(p^2)}{p^2}\left[\arctan\left(\frac{p^2 + m_{q_1}^2 -m_{q_2}^2}{\mathcal{G}(p^2)}\right) + \arctan\left(\frac{p^2+m_{q_2}^2-m_{q_1}^2}{\mathcal{G}(p^2)}\right)\right]
\;,\\
&& 
\label{cdef}
\\
\label{cdef1}
   \mathcal{G}(p^2) &=& \sqrt{[(m_{q_1}+m_{q_2})^2-p^2][p^2-(m_{q_2}-m_{q_1})^2]}\;,\
\end{eqnarray}
We have defined the combination of the functions \(B_{q_1q_2}\) and \(C_{q_1q_2}\) in a slightly different manner than done in ~\cite{threeflavors}. This notation is used so that the symmetry 
$q_1\leftrightarrow q_2$ is apparent.
\appendix
\renewcommand{\thesection}{B}
\setcounter{equation}{0}
\setcounter{figure}{0}
\renewcommand{\theequation}{B\arabic{equation}}
\renewcommand{\thefigure}{B\arabic{figure}}

\section[s]{Parameter fixing of the quark-meson model}
\label{matsj}
In this appendix, we explain how one can determine the parameters of
the three-flavor quark-meson model in terms of physical quantities such as quark and meson masses and meson decay constants. This has been discussed in the literature before; however, to be self-contained, we provide the interested reader with some details. The strategy is based on the combination of the on-shell scheme and the $\overline{\rm MS}$ scheme. 
In the Lagrangian Eq.~(\ref{totallag}), there are eight parameters that are 
determined by matching to physical observables and quark masses: $m^2$, $\lambda_1$,
$\lambda_2$, $h_u$, $h_d$ $h_s$, $g$, and $c$. We will need as many observables
as there are parameters in the model.
We set $c=0$ for convenience, thereby ignoring instanton effects and the mass splitting between
$\eta$ and $\eta^{\prime}$. We also work in the isospin limit $h_u=h_d=h_l$. 
This leaves us with six parameters, and we therefore need six observables.
We will choose the pion decay constant $f_{\pi}$, 
the quark masses $m_l$ and $m_s$, and the meson masses $m_{\pi}$, $m_{\sigma}$, and $m_{\eta^{\prime}}$.
The kaon decay constant $f_K$ and the remaining meson masses are then determined by various relations and are predictions of the model. 

The thermodynamic potential in the vacuum depends on the three expectation values 
$\phi_u$, $\phi_d$, and $\phi_s$.
The gap equations for $\phi_u$, $\phi_d$, and $\phi_s$ are
\begin{eqnarray}
{\partial\Omega_0\over\partial\phi_u}=
{\partial\Omega_0\over\partial\phi_d}=
{\partial\Omega_0\over\partial\phi_s}=0\;.
\end{eqnarray}
The solutions to these equations for $T=0$ and $\mu_B=\mu_e=\mu_3=\mu_8=0$
are denoted by $\phi_u^0$, $\phi_d^0$, and $\phi_s^0$, and 
give the vacuum values of the condensates. According to the partially conserved axial-vector current relation (PCAC), these condensates can be expressed in terms of the
pion and kaon decay constants as
\begin{eqnarray}
    \phi_u^0 &=& f_{\pi^\pm}+f_{K^\pm}-f_{K^0}\;,\\
    \phi_d^0 &=& f_{\pi^\pm}+f_{K^0}-f_{K^\pm}\;,\\
    \phi_s^0 &=&{1\over\sqrt{2}}
    \left( f_{K^0}+f_{K^\pm} -f_{\pi^\pm}\right)\;,
\end{eqnarray}
where the subscripts of the decay constants refer to the different mesons.
The quark masses are
\begin{eqnarray}
m_u={1\over2}g\phi_u^0\;,
\hspace{1cm}
m_d={1\over2}g\phi_d^0\;,
\hspace{1cm}
m_s={1\over\sqrt{2}}g\phi_s^0\;.
\end{eqnarray}
The decay constants can be expressed in terms of the quark masses
\begin{eqnarray}
    f_{K^0} &=& f_{\pi^\pm}\frac{m_d+m_s}{m_u+m_d} \quad {\rm and }\quad f_{K^\pm} = f_{\pi^\pm}\frac{m_u+m_s}{m_u+m_d}\;.
\end{eqnarray}
Since we are working in the isospin limit, the light quark condensates are equal, $\phi_u^0=\phi_d^0=f_{\pi}$.
The light quark masses in the vacuum are then denoted by $m_q$.
Moreover, there is no distinction between the decay constants $f_{K^0}$ and $f_{K^\pm}$, which implies that  and $\sqrt{2}\phi_s^0=(2f_k-f_{\pi})$.
The matching of the quark masses and the pion decay constant determines the Yukawa coupling $g$
and the condensates $\phi_l^0$ and $\phi_s^0$. This in turn determines $f_K$. Using the expressions for the masses, the tree-level gap equations can be written as
\begin{eqnarray}
\label{rel1}
h_l&=&f_{\pi}m_{\pi}^2\;,\\
h_s&=&{1\over\sqrt{2}}(2f_Km_K^2-f_{\pi}m_{\pi}^2)\;.
\label{hsdef}
\end{eqnarray}
Thus, the symmetry-breaking parameters are expressed in terms of the
pion and kaon masses and their decay constants. The kaon mass $m_K$
appearing in Eq.~(\ref{hsdef}) is eliminated by first expressing it in terms of
$m^2$, $\lambda_1$, $\lambda_2$ and the expectation values $\phi_l$ and $\phi_s$,
and then using the expressions 
Eqs.~(\ref{comp1})--(\ref{eqn:lambda2_tree_level})
to express $h_s$ in terms of the three observable meson masses and two decay constants that we have chosen.

In the on-shell scheme, the counterterms
are exactly canceling the quantum correction to the $n$-point functions, while
in the $\overline{\rm MS}$-scheme, the counterterms cancel the divergent terms.
Thus, in the latter scheme, the parameters are scale dependent and
relations such as  Eqs.~(\ref{rel1})--(\ref{hsdef}) are valid only at tree level and receive radiative corrections. By combining the two renormalization schemes, we can
express the running parameters in terms of physical quantities and the
renormalization scale. We would like to emphasize that tree-level matching of the
parameters is inconsistent with calculations of the thermodynamic potential including loops. A consistent calculation requires that the matching of the parameters is done in the same approximation as the determination of the thermodynamic potential. In the present case, this implies that we need to determine the $n$-point functions including a single quark loop. 

Let us first consider the one-point function $\Gamma^{(1)}_l$
for the light fields $\phi_u^0=\phi_d^0=\phi_l^0$, which is their
gap equation. At tree level, the equation of motion is $\Gamma^{(1)}=m_{\pi}^2\phi_l^0-h_l=0$.
Including the quark loop, we find
\begin{eqnarray}
\Gamma_l^{(1)}&=&m^2_{\pi}\phi_l^0-h_l-2N_cig^2\Lambda^{-2\epsilon}\phi_l^0A(m_q^2)-\delta\Gamma_l^{(1)}=0\;,    
\label{vertex}
\end{eqnarray}
where the third term is the loop correction,
$A(m^2)$ is the loop integral defined in Eq.~(\ref{defa}), and 
$\delta\Gamma^{(1)}_l$ is the counterterm. 
In the on-shell scheme, the counterterm
exactly cancels against the loop correction, and we are left with the tree-level relation.
Note that Eq.~(\ref{vertex}) is written in $d$ dimensions, so we use
$\phi_u^0$ instead of $f_{\pi}$, where the relation between them is $\phi_l^0=\Lambda^{-\epsilon}f_{\pi}$. The counterterm can therefore be written as
\begin{eqnarray}
\delta\Gamma_l^{(1)}&=&-2ig^2N_c\Lambda^{-3\epsilon}f_{\pi}A(m_q^2)\;,  \end{eqnarray}
Since $\Gamma_l^{(1)}=h_l-m_{\pi}f_{\pi}$, we can relate the counterterms of $h_l$,
$\delta m_{\pi}$, and $\delta f_{\pi}$, as 
\begin{eqnarray}
\delta h_l&=&[\delta m_{\pi}^2f_{\pi}+m_{\pi}^2\delta f_{\pi}]\Lambda^{-\epsilon}    
+\delta\Gamma_l^{(1)}\;.
\end{eqnarray}
The determination of $\delta h_l$ requires the expressions of $\delta m_{\pi}^2$
and $\delta f_{\pi}$, which we shall discuss next.
The light quark mass is given by $m_l=m_q={1\over2}gf_{\pi}$ and since there is no self-energy correction to the quark mass in our approximation, 
$2\delta m_l=\delta gf_{\pi}+g\delta f_{\pi}=0$. Similarly, since there is no loop
correction to the quark-pion vertex, we obtain 
$\delta g\phi_l+{1\over2}g\phi_l\delta Z_{\pi}=0$.
Combining these expressions, we find $\delta f_{\pi}={1\over2}f_{\pi}Z_{\pi}$, which yields
\begin{eqnarray}
\delta h_l&=&\left[\delta m_{\pi}^2+\mbox{$1\over2$}m_{\pi}^2\delta Z_{\pi}\right]f_{\pi}\Lambda^{-\epsilon}
-2ig^2N_c\Lambda^{-3\epsilon}f_{\pi}A(m_q^2)\;.
\end{eqnarray}
The two counterterms $\delta m_{\pi}^2$ and $\delta Z_{\pi}$ require the 
calculation of the pion self-energy correction. 
The two-point function can be written as 
\begin{eqnarray}
\Gamma_{\pi}^{(2)}(p^2)&=&p^2-m_{\pi}^2-\Sigma_{\pi}(p^2)-\Sigma_{\pi}^{\rm ct}(p^2)\;, 
\end{eqnarray}
where $\Sigma_{\pi}(p^2)$ is the self-energy and $\Sigma_{\pi}^{\rm ct}(p^2)$
is the counterterm. The self-energy and counterterm can be written in the form
\begin{eqnarray}
\Sigma_{\pi}(p^2)&=&\Sigma_{\pi}^{\rm 1PI}(p^2)+\Sigma_{\pi}^{\rm tadpole}
\;,\\
\Sigma_{\pi}^{\rm ct}(p^2)&=&(p^2-m_{\pi}^2)\delta Z_{\pi}-\delta m_{\pi}^2+
\Sigma_{\pi}^{\rm ct, tadpole}\;,
\end{eqnarray}
where the superscripts indicate the different contributions.
In the on-shell scheme, the counterterms exactly cancel the loop corrections, implying that
\begin{eqnarray}
\delta m_{\pi}^2&=&-\Sigma_{\pi}^{\rm 1PI}(m_{\pi}^2)\;,\\
\Sigma_{\pi}^{\rm ct,tadpole}&=&\Sigma_{\pi}^{\rm tadpole}\;.
\end{eqnarray}
We do not need to know the explicit expression for the tadpole counterterms. Finally, since we require that the pole of the propagator have unit
residue, we find
\begin{eqnarray}
\delta Z_{\pi}&=&\Sigma_{\pi}^{\rm1PI\prime}(m_{\pi}^2)\;.    
\end{eqnarray}
As explained above, a consistent calculation requires that we include the quark loops in the calculation of the self-energies.
The contribution to the self-energy from a quark loop depends on the
external momentum $p$, the masses of the 
two quarks $f_1$ and $f_2$, and whether the meson is
a scalar or a pseudo-scalar. The self-energy is then
denoted by $\Sigma_{\pm}^{f_1f_2}(p^2)$ with a plus for scalar and minus for
pseudo-scalar. After some calculations, one finds
\begin{eqnarray}
 \Sigma_{\pi}^{\rm1PI}(p^2)&=&
 2ig^2N_c\Lambda^{-2\epsilon}\left[2A(m_{q}^2)-{1\over2}p^2B_{ll}(p^2)\right]\;,\\
    -\frac{d\Sigma_{\pi}^{\rm1PI}}{dp^2}(p^2) &=& iN_cg^2\Lambda^{-2\epsilon}\left[B_{q_1q_2}(p^2)+p^2B_{ll}'(p^2)\right]\;,
\end{eqnarray}
where $B_{q_1q_2}(p^2)$ is defined
in Eq.~(\ref{bdef}). 
Using the expressions for $A(m^2)$ and $B_{q_1q_2}(p^2)$, the final results for the counterterms $\delta m_{\pi}^2$ and $\delta Z_{\pi}$ are
\begin{eqnarray}
\delta m_{\pi}^2&=&-{2g^2N_c\Lambda^{-2\epsilon}\over(4\pi)^2}\left\{
\left[{1\over\epsilon}+\log{\Lambda^2\over m_q^2}+1\right]m_q^2
-{1\over2}m_{\pi}^2\left[{1\over\epsilon}+\log\frac{\Lambda^2}{m_q^2} + C_{ll}(m_{\pi}^2)\right]
\right\}\;,\\    
\delta Z_{\pi}&=&-{g^2N_c\Lambda^{-2\epsilon}\over(4\pi)^2}\left[{1\over\epsilon}+\log{\Lambda^2\over m_q^2}+C_{ll}(m_{\pi}^2)+m_{\pi}^2C^{\prime}_{ll}(m_{\pi}^2)\right]\;.
\end{eqnarray}
We now have all the terms needed to determine $\delta h_l$ in the on-shell scheme,
\begin{eqnarray}
\delta h_l&=&{g^2\Lambda^{-2\epsilon}N_ch_l\over2(4\pi)^2}\left[{1\over\epsilon}+\log{\Lambda^2\over m_q^2}+C_{ll}(m_{\pi}^2)-m_{\pi}C_{ll}^{\prime}(m_{\pi}^2)
+{\cal O}(\epsilon)
\right]\;.
\end{eqnarray}
In the $\overline{{\rm MS}}$-scheme,the counterterm $\delta h_{l,\ms}$
is given by the divergent term only
\begin{eqnarray}
\delta h_{l,\ms}&=&{g^2\Lambda^{-2\epsilon}N_ch_l\over2(4\pi)^2\epsilon}\;. 
\end{eqnarray}
In the $\overline{{\rm MS}}$-scheme, the relation between the bare and 
renormalized parameter is
\begin{eqnarray}
h_l&=&\Lambda^{-\epsilon}(h_{l,\ms}+\delta h_{l,\ms})\;.    
\end{eqnarray}
Moreover, the bare parameter $h_l$ is the same in the two renormalization schemes, which yields
\begin{eqnarray}
\nonumber
h_{l,\ms}(\Lambda)&=&h_l+\delta h_{l}-\delta h_{l,\ms}\\
&=&h\left\{1+{g^2N_c\over2(4\pi)^2}\left[\log{\Lambda^2\over m_q^2}
+C_{ll}(m_{\pi}^2)-m_{\pi}C_{ll}^{\prime}(m_{\pi}^2)
\right]\right\}\;.
\end{eqnarray}
Thus, we have expressed the running parameter $h_{l,\ms}(\Lambda)$ entirely in terms of 
physical quantities in the vacuum, $m_{\pi}$, $f_{\pi}$, and
$m_q$,
and the renormalization scale.

We next consider the Yukawa coupling $g$.
The counterterm $\delta g^2$ can now be easily obtained by noting that there 
is no wavefunction renormalization of the quark fields and 
no loop corrections to the quark-meson vertices in our mean-field approximation.
This implies $\delta g^2=-\delta Z_{\pi}$. 
Using $g_{\ms}^2+\delta g_{\ms}^2=g^2+\delta g_{\os}^2=g^2-\delta Z_{\pi}$, we obtain
\begin{eqnarray}
g_{\ms}^2&=&{m_q^2\over f_{\pi}^2}\left[1+{m_q^2N_c\over f_{\pi}^2(4\pi)^2}\left(\log{\Lambda^2\over m_q^2}+C_{ll}(m_{\pi}^2)+m_{\pi}^2C^{\prime}(m_{\pi}^2)\right)\right]\;.    
\end{eqnarray}
We now turn to the renormalization of the parameters $m^2$, $\lambda_1$, 
and $\lambda_2$. To determine the corresponding counterterms 
$\delta m^2$, $\delta\lambda_1$, and $\delta\lambda_2$ requires that we calculate
the mass counterterms $\delta m_{\sigma}^2$ and $\delta m^2_{\eta^{\prime}}$, in addition 
to the counterterms that we have already determined.
In the isospin limit, the mass parameter $m^2$, the couplings $\lambda_1$ and $\lambda_2$ simplify significantly and read
\begin{eqnarray}
\label{comp1}
    m^2 &=& \frac{2(m_\sigma^2-m_{\eta'}^2)(m_\sigma^2 - 3m_\pi^2)\phi_l^4 +(3m_{\eta'}^2-m_\sigma^2)(m_\sigma^2-m_\pi^2)\phi_s^4}{4(m_{\eta'}^2-m_\sigma^2)\phi_l^4 + 2(m_\sigma^2+7m_\pi^2-8m_{\eta'}^2)\phi_l^2\phi_s^2 + 2(m_\sigma^2-m_\pi^2)\phi_s^4}\nn\\ 
    &&- \frac{(6m_{\eta'}^4 + 9m_{\eta'}^2m_\pi^2 - 12m_\pi^4 - 5m_{\eta'}^2m_\sigma^2 + m_\pi^2m_\sigma^2+m_\sigma^4)\phi_l^2\phi_s^2}{4(m_{\eta'}^2-m_\sigma^2)\phi_l^4 + 2(m_\sigma^2+7m_\pi^2-8m_{\eta'}^2)\phi_l^2\phi_s^2 + 2(m_\sigma^2-m_\pi^2)\phi_s^4}\;,
    \label{eqn:m2_tree_level}
    \\
    \nonumber
    \lambda_1 &=& \frac{[(2m_{\eta'}^2-3m_\pi^2+m_\sigma^2)\phi_l^2 + (m_\pi^2-m_\sigma^2)\phi_s^2][(m_{\eta'}^2-m_\sigma^2)\phi_l^2 + (m_\sigma^2+2m_\pi^2-3m_{\eta'}^2)\phi_s^2]}{2(m_{\eta'}^2-m_\sigma^2)\phi_l^6 + (3m_\sigma^2+7m_\pi^2-10m_{\eta'}^2)\phi_l^4\phi_s^2 + 8(m_{\eta'}^2-m_\pi^2)\phi_l^2\phi_s^4 + (m_\pi^2-m_\sigma^2)\phi_s^6}\;,\\
    \label{eqn:lambda1_tree_level}\\
    \lambda_2 &=& \frac{2(m_{\eta^{\prime}}^2-m_\pi^2)}{\phi_s^2-\phi_l^2}\;.
    \label{eqn:lambda2_tree_level}
\end{eqnarray}
The tree-level scalar and pseudo-scalar mass matrices are determined by the
second partial derivatives of the thermodynamic potential as
\begin{eqnarray}
\label{massmatrix}
    (m_{\sigma\sigma}^2)_{ab} = 
    \frac{\partial^2{\Omega_0}}{\partial\sigma_a\sigma_b}
    \bigg|_{\phi_{l,s}=\phi_{l,s}^0}
\hspace{1cm}    
    (m_{\pi\pi}^2)_{ab} = \frac{\partial^2 {\Omega_0}}{\partial\pi_a\pi_b}
      \bigg|_{\phi_{l,s}=\phi_{l,s}^0}\;.
\end{eqnarray}
We list them in Appendix~\ref{massapp}. Some of the submatrices in Eq.~(\ref{massmatrix}) are not diagonal. The physical masses
are then obtained by diagonalization. The tree-level masses for the non-diagonal \(\sigma\) and \(\eta'\) are for example given by
\begin{eqnarray}
    m_\sigma^2 &=& \frac{1}{2}((m_{\sigma\sigma}^2)_{00} + (m_{\sigma\sigma}^2)_{88}) - \frac{1}{2}\sqrt{((m_{\sigma\sigma}^2)_{00}-(m_{\sigma\sigma}^2)_{88})^2+4((m_{\sigma\sigma}^2)_{08})^2}\;,\\
    m_{\eta'}^2 &=& \frac{1}{2}((m_{\pi\pi}^2)_{00} + (m_{\pi\pi}^2)_{88}) + \frac{1}{2}\sqrt{((m_{\pi\pi}^2)_{00}-(m_{\pi\pi}^2)_{88})^2+4((m_{\pi\pi}^2)_{08})^2}\;.
\end{eqnarray}
The self energies for the physical \(\sigma\) and \(\eta'\), $\Sigma_{\sigma}(p^2)$ and $\Sigma_{\eta^{\prime}}(p^2)$
are written in terms of the fundamental self-energies. One finds
\begin{eqnarray}
\nonumber
    \Sigma_\sigma(p^2) &=& \frac{1}{2}\left[\Sigma_{\sigma}^{00}(p^2) + \Sigma_{\sigma}^{88}(p^2)- \frac{1}{\sqrt{(m_{\sigma,00}^2-m_{\sigma,88}^2)^2 + 4(m_{\sigma,08}^2)^2}}\right.\\
    &&\left.\times\left((m_{\sigma,00}^2-m_{\sigma,88}^2)(\Sigma_{\sigma}^{00}(p^2) - \Sigma_{\sigma}^{88}(p^2))+4m_{\sigma, 08}^2\Sigma_{\sigma}^{08}(p^2)\right)\right]\;,\\ \nonumber
    \Sigma_{\eta'}(p^2) &=& \frac{1}{2}\left[\Sigma_{\pi}^{00}(p^2) + \Sigma_{\pi}^{88}(p^2) + \frac{1}{\sqrt{(m_{\pi,00}^2-m_{\pi,88}^2)^2 + 4(m_{\pi,08}^2)^2}}\right.\\
    &&\left.\times\left((m_{\pi,00}^2-m_{\pi,88}^2)(\Sigma_{\pi}^{00}(p^2) - \Sigma_{\pi}^{88}(p^2))+4m_{\pi, 08}^2\Sigma_{\pi}^{08}(p^2)\right)\right]\;.
\end{eqnarray}
The relevant self-energies are given by
\begin{eqnarray}
\Sigma_{\pi}^{11}(p^2) &=& 2iN_cg^2\left[A(m_l^2) - p^2B_{ll}(p^2)\right]\;\\
    \Sigma_{\sigma}^{00}(p^2) &=& \frac{2iN_cg^2}{3}\left[2A(m_l^2) - (p^2-4m_l^2)B_{ll}(p^2) + A(m_s)-\frac{1}{2}(p^2 - 4m_s^2)B_{ss}(p^2)\right]\;,\\
    \Sigma_{\sigma}^{88}(p^2) &=& \frac{iN_cg^2}{3}\left[2A(m_l^2) - (p^2-4m_l^2)B_{ll}(p^2) + 4A(m_s^2)-2(p^2-4m_s^2)B_{ss}(p^2) \right]\;,\\
    \Sigma_{\sigma}^{08}(p^2) &=& \frac{\sqrt{2}iN_cg^2}{3}\left[2A(m_l^2) - (p^2-4m_l^2)B_{ll}(p^2)-2A(m_s^2) + (p^2-4m_s^2)B_{ss}(p^2)\right]\;,\\
\Sigma_{\pi}^{00}(p^2) &=&{2iN_cg^2\over3}\left[
2A(m_l^2)-p^2B_{ll}(p^2) + A(m_s^2) - {1\over2}p^2B_{ss}(p^2)
\right]
\;,\\
\Sigma_{\pi}^{88}(p^2) &=&{iN_cg^2\over3}\left[
2A(m_l^2)-p^2B_{ll}(p^2)-4A(m_s^2)+2p^2B_{ss}(p^2)\right]
\;, \\
\Sigma_{\pi}^{08}(p^2) &=& \frac{\sqrt{2}iN_cg^2}{3}\left[2A(m_l^2) - p^2B_{ll}(p^2) - 2A(m_s^2) + p^2B(p^2)\right]
\;,
\end{eqnarray}
where the quark masses are evaluated in the vacuum.

The remainder of the calculations is, in principle, straightforward, albeit rather tedious. The final expressions for the counterterms are very long and will
therefore not be listed here.


\renewcommand{\thesection}{C}
\setcounter{equation}{0}
 \renewcommand{\theequation}{C\arabic{equation}}

\section{Scalar and pseudo-scalar masses}
\label{massapp}
For completeness, we list the elements of the mass matrices.
The elements of the scalar mass matrix are
\begin{eqnarray}
  \nonumber
    (m_{\sigma\sigma}^2)_{00} &=& m^2+\frac{\lambda_1}{6}(5\phi_u^2+5\phi_d^2+10\phi_s^2+4\phi_u\phi_d + 4\sqrt{2}\phi_u\phi_s+4\sqrt{2}\phi_d\phi_s)
    \\ &&
    +\frac{\lambda_2}{2}(\phi_u^2+\phi_d^2+2\phi_s^2)\;,\\
    (m_{\sigma\sigma}^2)_{11} &=& (m_{\sigma\sigma}^2)_{22}= m^2+\frac{\lambda_1}{2}(\phi_u^2+\phi_d^2+2\phi_s^2) + \frac{\lambda_2}{2}(\phi_u^2+\phi_u\phi_d+\phi_d^2)
    \;,\\
    (m_{\sigma\sigma}^2)_{33} &=& m^2+\lambda_1\left(\phi_u^2-\phi_u\phi_d+\phi_d^2+\phi_s^2\right) + \frac{3\lambda_2}{4}(\phi_u^2+\phi_d^2)\\
    (m_{\sigma\sigma}^2)_{44} &=& (m_{\sigma\sigma}^2)_{55} =m^2+\frac{\lambda_1}{2}(\phi_u^2+\phi_d^2+2\phi_s^2)+\frac{\lambda_2}{2}(\phi_u^2+\sqrt{2}\phi_u\phi_s+2\phi_s^2)
    \;,
    \\
    (m_{\sigma\sigma}^2)_{66} &=& (m_{\sigma\sigma}^2)_{77} =m^2+\frac{\lambda_1}{2}(\phi_u^2+\phi_d^2+2\phi_s^2)+\frac{\lambda_2}{2}(\phi_d^2+\sqrt{2}\phi_d\phi_s+2\phi_s^2)\\
   \nonumber
    (m_{\sigma\sigma}^2)_{88} &=& m^2 + \frac{\lambda_1}{6}(4\phi_u^2 + 4\phi_d^2+14\phi_s^2+2\phi_u\phi_d-4\sqrt{2}\phi_u\phi_s-4\sqrt{2}\phi_d\phi_s) 
    \\ &&
    + \frac{\lambda_2}{4}(\phi_u^2+\phi_d^2+8\phi_s^2)\;,\\
    (m_{\sigma\sigma}^2)_{03} &=& (m_{\sigma\sigma}^2)_{30}=\frac{\lambda_1}{\sqrt{6}}(\phi_u-\phi_d)(\phi_u+\phi_d+\sqrt{2}\phi_s)+\frac{3\lambda_2}{2\sqrt{6}}(\phi_u^2-\phi_d^2)\\
   \nonumber
    (m_{\sigma\sigma}^2)_{08} &=& (m_{\sigma\sigma}^2)_{80}= \frac{\lambda_1}{3\sqrt{2}}(\phi_u+\phi_d+\sqrt{2}\phi_s)(\phi_u+\phi_d-2\sqrt{2}\phi_s)
    \\ &&
    + \frac{\lambda_2}{2\sqrt{2}}(\phi_u^2+\phi_d^2-4\phi_s^2)\;,\\
    (m_{\sigma\sigma}^2)_{38} &=& (m_{\sigma\sigma}^2)_{83}= \frac{\lambda_1}{2\sqrt{3}}(\phi_u-\phi_d)(\phi_u+\phi_d-2\sqrt{2}\phi_s) + \frac{\sqrt{3}\lambda_2}{4}(\phi_u^2-\phi_d^2)\;.
\end{eqnarray}
Note that $(m^2_{\sigma\sigma})_{03}=(m^2_{\sigma\sigma})_{30}=0$ for 
$\phi_u^0=\phi_d^0$. 
The elements of the pseudo-scalar mass matrix are
\begin{eqnarray}
    (m_{\pi\pi}^2)_{00} &=& m^2+\frac{\lambda_1}{2}(\phi_u^2+\phi_d^2+2\phi_s^2) + \frac{\lambda_2}{6}(\phi_u^2+\phi_d^2+2\phi_s^2)\;,\\
    (m_{\pi\pi}^2)_{11} &=& (m_{\pi\pi}^2)_{22} = m^2+\frac{\lambda_1}{2}(\phi_u^2+\phi_d^2+2\phi_s^2) + \frac{\lambda_2}{2}(\phi_u^2-\phi_u\phi_d+\phi_d^2)\;,\\
     (m_{\pi\pi}^2)_{33} &=& m^2 + \frac{\lambda_1}{2}(\phi_u^2+\phi_d^2+2\phi_s^2) + \frac{\lambda_2}{4}(\phi_u^2+\phi_d^2)\;,\\
     (m_{\pi\pi}^2)_{44} &=& (m_{\pi\pi}^2)_{55} =m^2+\frac{\lambda_1}{2}(\phi_u^2+\phi_d^2+2\phi_s^2) + \frac{\lambda_2}{2}(\phi_u^2-\sqrt{2}\phi_u\phi_s+2\phi_s^2)\;,\\
      (m_{\pi\pi}^2)_{66} &=& (m_{\pi\pi}^2)_{77} = m^2+\frac{\lambda_1}{2}(\phi_u^2+\phi_d^2+2\phi_s^2) + \frac{\lambda_2}{2}(\phi_d^2-\sqrt{2}\phi_d\phi_s + 2\phi_s^2)\;,\\   
      (m_{\pi\pi}^2)_{88} &=& m^2 + \frac{\lambda_1}{2}(\phi_u^2+\phi_d^2+2\phi_s^2) + \frac{\lambda_2}{12}(\phi_u^2+\phi_d^2+8\phi_s^2)\;,\\
      (m_{\pi\pi}^2)_{03} &=& (m_{\pi\pi}^2)_{30}=\frac{\lambda_2}{2\sqrt{6}}(\phi_u^2-\phi_d^2)\\
      (m_{\pi\pi}^2)_{08} &=& (m_{\pi\pi}^2)_{80} = \frac{\lambda_2}{6\sqrt{2}}(\phi_u^2+\phi_d^2-4\phi_s^2)\;,\\
      (m_{\pi\pi}^2)_{38} &=& (m_{\pi\pi}^2)_{83} =\frac{\lambda_2}{4\sqrt{3}}(\phi_u^2-\phi_d^2)\;.
\end{eqnarray}
Note that $(m^2_{\pi\pi})_{03}=(m^2_{\pi\pi})_{30}=0$ for 
$\phi_u^0=\phi_d^0$. The mass matrices are not diagonal, so the fields $(\sigma_a,\pi_a)$ do not correspond
to the mass eigenstates: $\eta$ and $\eta^{\prime}$ are admixtures of
$\pi_0$ and $\pi_8$, and $\sigma$ and $f$ are admixtures of $\sigma_0$
and $\sigma_8$.

\renewcommand{\thetable}{C1}

\setcounter{equation}{0}
 \renewcommand{\theequation}{D\arabic{equation}}

\renewcommand{\thesection}{D}

\section{Renormalization of the thermodynamic potential in the pion-condensed
phase}
\label{pionren}
In this Appendix, we renormalize the thermodynamic potential in the
pion-condensed phase. We work in the isospin limit $h_u=h_d=h$.
The four-dimensional momentum integral in Eq.~(\ref{integral}) is ultraviolet divergent and requires regularization. However, it is too complicated to be calculated analytically
using e.g. dimensional regularization. We therefore adopt another strategy.
From the tree-level thermodynamic potential, it is clear that the terms that require renormalization are at most quadratic in $\mu_I$. This implies that by expanding the quark-loop contribution $\Omega_1^{\rm BEC/BCS}$ in powers of $\mu_I$, we can isolate the divergent terms. 
This yields
\begin{eqnarray}
\label{dividi}
\Omega_1^{\rm BEC/BCS,div}&=&-2N_c
\Lambda^{-2\epsilon}\int_{-\infty}^{\infty}{dp_0\over2\pi}\int_p
\left[\log[p_0^2+p^2+m^2_{\pm}]
-{\mu_I^2(p^2-p_0^2-g^2\rho^2/4)\over  2(p_0^2+p^2+g^2\rho^2/4)^2}
\right]\;,\\
\Omega_{1}^{\rm BEC/BCS,fin}&=&\Omega_{1}^{\rm BEC/BCS}-\Omega_{1}^{\rm BEC/BCS,div}\;.
\end{eqnarray}
where the masses $m_{\pm}^2$ are
\begin{eqnarray}
m_{\pm}^2={1\over2}(m_u^2+m_d^2+{1\over2}g^2\rho^2)\pm{1\over2}|m_u-m_d|
\sqrt{(m_u+m_d)^2+g^2\rho^2}\;.
\label{eqn:mass_splitting_pion}
\end{eqnarray}
The integral in Eq.~(\ref{becfin}) is finite and  must be evaluated numerically.
Integrating Eq.~(\ref{dividi}) over $p_0$ yields

\begin{eqnarray}
\nonumber
\Omega_1^{\rm BEC/BCS,div}&=&-2N_c\Lambda^{-2\epsilon}\int_p\left[\sqrt{p^2+m_+^2} + \sqrt{p^2+m_-^2}
+{\mu_I^2g^2\rho^2\over 16(p^2+g^2\rho^2/4)^{3\over2}}
\right]\;.
\label{becfin}
\end{eqnarray}

Using the integrals in Appendix A, we obtain
\begin{eqnarray}
\nonumber
\Omega_1^{\rm BEC/BCS,div}&=&{N_c\Lambda^{-2\epsilon}\over(4\pi)^2\epsilon}    
\left[
(m_u^2+{1\over4}g^2\rho^2)^2+(m_d^2+{1\over4}g^2\rho^2)^2+{1\over2}(m_u-m_d)^2g^2\rho^2
\right]
\\ &&
\nonumber
+{N_cm_+^4\over(4\pi)^2}\left[\log{\Lambda^2\over m_+^2}+{3\over2}\right]
+{N_cm_-^4\over(4\pi)^2}\left[\log{\Lambda^2\over m_-^2}+{3\over2}\right]
\\ &&
-{N_c\mu_I^2g^2\rho^2\Lambda^{-2\epsilon}\over2(4\pi)^2}\left[{1\over\epsilon}+\log{\Lambda^2\over g^2\rho^2/4}\right]\;.
\end{eqnarray}
Renormalization is carried out by making the substitutions 
\begin{eqnarray}
\label{reli1}
m^{2}&\rightarrow&m_{\ms}^2+\delta m^2_{\ms}\;,\\
c&\rightarrow&c_{\ms}+\delta c_{\ms}\;,\\
h_{}&\rightarrow&\Lambda^{-\epsilon}(h_{\ms}+\delta h_{\ms})\;,\\
\lambda_1&\rightarrow&\Lambda^{2\epsilon}(\lambda_{1,\ms}+\delta\lambda_{1,\ms})\;,\\
\lambda_2&\rightarrow&\Lambda^{2\epsilon}(\lambda_{2,\ms}+\delta\lambda_{2,\ms})\;,\\
\phi_f^2&\rightarrow&(1+\delta Z_{\pi})\phi_{f,\ms}^2\;,\\
\rho^2&\rightarrow&(1+\delta Z_{\pi})\rho_{\ms}^2\;,
\label{reli2}
\end{eqnarray}
in the tree-level
thermodynamic potential $\Omega_0^{\rm BEC/BCS}$, Eq.~(\ref{tripje}).
The counterterms in the $\overline{\rm MS}$ scheme are~\cite{kumari} 
\begin{eqnarray}
\label{masscount}
\delta m_{\ms}^2&=&{N_cm^2g^2\Lambda^{-2\epsilon}\over(4\pi)^2\epsilon}\;,\\    
\delta c_{\ms}&=&{N_ccg^2\Lambda^{-2\epsilon}\over(4\pi)^2\epsilon}\;,\\    
\delta h_{\ms}&=&{N_cg^2h\Lambda^{-\epsilon}\over2(4\pi)^2\epsilon}\;\\
\delta\lambda_{1,\ms}&=&{2N_c\lambda_1g^2\Lambda^{-4\epsilon}\over(4\pi)^2\epsilon}
\;,\\    
\delta\lambda_{2,\ms}&=&{N_cg^2(2\lambda_2-g^2)\Lambda^{-4\epsilon}\over(4\pi)^2\epsilon}\;,\\    
\delta Z_{\pi}^{\ms}&=&-{N_cg^2\Lambda^{-2\epsilon}\over(4\pi)^2\epsilon}\;.  
\end{eqnarray}
Note the various factors of the scale $\Lambda$ in Eqs.~(\ref{reli1})--~(\ref{reli2}), which ensure that the running couplings also have their canonical dimensions for $d+1\neq4$. 
After renormalization, the one-loop thermodynamic potential reads
\begin{eqnarray}
\nonumber
\Omega_{0+1}^{\rm BEC/BCS}&=&{1\over2}m_{\ms}^2(\phi_{u,\ms}^2+\phi_{d,\ms}^2+\rho_{\ms}^2)-{1\over2}\mu_I^2\rho_{\ms}^2
-h_{\ms}(\phi_{u,\ms}+\phi_{d,\ms})
\\ && \nonumber
-c_{\ms}\phi_{u,\ms}\phi_{d,\ms}
-{1\over2}c_{\ms}\rho_{\ms}^2
+{1\over4}\lambda_{1,\ms}(\phi_{u,\ms}^2+\phi_{d,\ms}^2+\rho_{\ms}^2)^2
\\ && \nonumber
+{1\over4}\lambda_{2,\ms}(\phi_{u,\ms}^2+{1\over2}\rho_{\ms}^2)^2
+{1\over4}\lambda_{2,\ms}(\phi_{d,\ms}^2+{1\over2}\rho_{\ms}^2)^2
\\ &&
\nonumber
+{1\over4}\lambda_{2,\ms}(\phi_{u,\ms}-\phi_{d,\ms})^2\rho_{\ms}^2
+{N_cm_+^4\over(4\pi)^2}\left[\log{\Lambda^2\over m_+^2}+{3\over2}\right]
+{N_cm_-^4\over(4\pi)^2}\left[\log{\Lambda^2\over m_-^2}+{3\over2}\right]
\\ &&
-{N_c\mu_I^2g^2\rho^2\over2(4\pi)^2}\log{\Lambda^2\over g^2\rho^2/4}
+\Omega_{0+1}^{\rm BEC/BCS, fin}
\label{isopot2}
\;.
\end{eqnarray}
In the two-flavor case, the running couplings were calculated in~\cite{kumari}
using the techniques in Appendix \ref{matsj}. They read
\begin{eqnarray}
\nonumber
m_{\ms}^2(\Lambda)&=&m^2\left[1+{g^2N_c\over(4\pi)^2}\log{\Lambda^2\over m_q^2}\right]
+{g^2N_c\over2(4\pi)^2}\left[2m_{\pi}^2C_{ll}(m_{\pi}^2)
+m_{\eta}^2C_{ll}(m_{\eta}^2)
\right. \\ && \left.
\label{ronny1}
-(m_{\sigma}^2-4m_q^2)C_{ll}(m_{\sigma}^2)-4m_q^2\right]
\;,\\
c_{\ms}(\Lambda)&=&c\left[1+{g^2N_c\over(4\pi)^2}\log{\Lambda^2\over m_q^2}\right]
+{g^2N_c\over2(4\pi)^2}\left[m_{\eta}^2C_{ll}(m_{\eta}^2)-m_{\pi}^2C_{ll}(m_{\pi}^2)\right]\;,
\\
h_{\ms}(\Lambda)&=&h\left\{1+{g^2N_c\over2(4\pi)^2}\left[\log{\Lambda^2\over m_q^2}+C_{ll}(m_{\pi}^2)-m_{\pi}^2C_{ll}^{\prime}(m_{\pi}^2)\right]\right\}\;,
\\  
\nonumber
\lambda_1^{\ms}(\Lambda)&=&
\lambda_1\left[1+{2g^2N_c\over(4\pi)^2}\log{\Lambda^2\over m_q^2}\right]
+{g^2N_c\over2(4\pi)^2f_{\pi}^2}\left[
(m_{\sigma}^2-4m_q^2)C_{ll}(m_{\sigma}^2)-(m_{a_0}^2-4m_q^2)C_{ll}(m_{a_0}^2)
\right. \\ && \left.
+m_{\eta}^2C_{ll}(m_{\eta}^2)-m_{\pi}^2C_{ll}(m_{\pi}^2)
+2\lambda_1f_\pi^2(C_{ll}(m_\pi^2) + m_\pi^2C_{ll}'(m_\pi^2))\right]
\;,\\
\nonumber
\lambda_2^{\ms}(\Lambda)&=&\lambda_2\left[1+\left(2-{g^2\over\lambda_2}\right){g^2N_c\over(4\pi)^2}\log{\Lambda^2\over m_q^2}\right]
+{g^2N_c\over(4\pi)^2f_{\pi}^2}\left\{(m_{a_0}^2-4m_q^2)C_{ll}(m_{a_0}^2)
\right. \\&& \left.
-m_{\eta}^2C_{ll}(m_{\eta}^2)+\lambda_2f_{\pi}^2\left[C_{ll}(m_{\pi}^2)+m_{\pi}^2C^{\prime}_{ll}(m_{\pi}^2)\right]
\right\}
\;,\\
\phi_{f,\ms}^2(\Lambda)&=&\phi_{f,0}^2\left\{1-{g^2N_c\over(4\pi)^2}\left[\log{\Lambda^2\over m_q^2}+C_{ll}(m_{\pi}^2)+m_{\pi}^2C_{ll}^{\prime}(m_{\pi}^2)\right]\right\}\;,\\
\rho_{\ms}^2(\Lambda)&=&\rho_{0}^2\left\{1-{g^2N_c\over(4\pi)^2}\left[\log{\Lambda^2\over m_q^2}+C_{ll}(m_{\pi}^2)+m_{\pi}^2C_{ll}^{\prime}(m_{\pi}^2)\right]\right\}\;,\\
g^2_{\ms}(\Lambda)&=&
g^2\left\{1+{g^2N_c\over(4\pi)^2}\left[\log{\Lambda^2\over m_q^2}+C_{ll}(m_{\pi}^2)+m_{\pi}^2C_{ll}^{\prime}(m_{\pi}^2)\right]\right\}\;,\label{ronny8}
\end{eqnarray}
where the function $C_{q_1q_2}(p^2)$ is defined in Eq.~(\ref{cdef})
and $m_q$ is the light quark mass in the vacuum $m_q={1\over2}gf_{\pi}$.
The parameters on the right-hand side are expressed in terms of 
the meson masses as
\begin{eqnarray}
m^2&=&m_{\pi}^2+{1\over2}(m_{\eta}^2-m_{\sigma}^2)\;,\\
c&=&{1\over2}(m_{\eta}^2-m_{\pi}^2)\;,\\
\lambda_1&=&{m_{\sigma}^2+m_{\eta}^2-m_{a_0}^2-m_{\pi}^2\over2f_{\pi}^2}\;,\\
\lambda_2&=&{m_{a_0}^2-m_{\eta}^2\over f_{\pi}^2}\;,\\
g^2&=&{4m_q^2\over f_{\pi}^2}\;.
\end{eqnarray}
The renormalized parameters satisfy renormalization group equations that can be easily derived by noting that the bare parameters are independent of $\Lambda$. 
For example,
\begin{eqnarray}
\Lambda{d\over d\Lambda}(m_{\ms}^2+\delta m^2_{\ms})&=&0\;.    
\end{eqnarray}
Using the expression for the mass counterterm, Eq.~(\ref{masscount}), we obtain
in the limit $\epsilon\rightarrow0$
\begin{eqnarray}
\Lambda{dm_{\ms}^2(\Lambda)\over\ d\Lambda}&=&{2N_cm_{\ms}^2g_{\ms}^2(\Lambda)\over(4\pi)^2}\;.    
\end{eqnarray}
This equation can be solved for the running mass parameter $m_{\ms}^2(\Lambda)$ as can the other renormalization group equations.
The solutions to these equations are
\begin{eqnarray}
\label{sol1}
m^2_{\ms}(\Lambda)&=&{m_0^2\over{1-{g^2_{0}N_c\over(4\pi)^2}\log{\Lambda^2\over\Lambda_0^2}}}\;, \\
c_{\ms}(\Lambda)&=&{c_0\over\left[1-{N_cg_0^2\over(4\pi)^2}\log{\Lambda^2\over\Lambda_0^2}\right]}\;,
\\ 
\lambda_1^{\ms}(\Lambda)&=&{\lambda_{1,0}\over{\Big(1-{g_0^2N_c\over(4\pi)^2}\log{\Lambda^2\over\Lambda_0^2}}\Big)^2}\;, \\
\lambda_2^{\ms}(\Lambda)&=&{\lambda_{2,0}-{g_0^4N_c\over(4\pi)^2}\log{\Lambda^2\over\Lambda_0^2}\over{\Big(1-{g_0^2N_c\over(4\pi)^2}\log{\Lambda^2\over\Lambda_0^2}}\Big)^2}\;, \\
g^2_{\ms}(\Lambda)&=&{g_0^2\over{1-{g_0^2N_c\over(4\pi)^2}\log{\Lambda^2\over\Lambda_0^2}}}\;, \\
h_{\ms}(\Lambda)&=&{h_0\over{}
\sqrt{1-{g_0^2N_c\over(4\pi)^2}\log{\Lambda^2\over\Lambda_0^2}}}\;,
\label{sol4}
\\
\phi^2_{f,\ms}(\Lambda)&=&\left[1-{g_0^2N_c\over(4\pi)^2}\log{\Lambda^2\over\Lambda_0^2}
\right]\phi_{f,0}^2\;,\\
\rho^2_{\ms}(\Lambda)&=&\left[1-{g_0^2N_c\over(4\pi)^2}\log{\Lambda^2\over\Lambda_0^2}
\right]\rho_{0}^2\;,
\label{sollast}
\end{eqnarray}
where $\Lambda_0$ is a reference scale and the constants $m_0^2$, 
$c_0$, $\lambda_{1,0}$, 
$\lambda_{2,0}$, $g_0^2$, $h_0$, $\phi_{f,0}^2$, and $\rho_0^2$ are the values of the running parameters at the scale $\Lambda_0$. The reference scale can be chosen at will. A convenient choice is that it satisfies the equation
\begin{eqnarray}
\label{refscale}
\log{\Lambda_0^2\over m_q^2}+
C(m_{\pi}^2)+m_{\pi}^2C_{ll}^{\prime}(m_{\pi}^2)&=&0\;.
\label{refscale2}
\end{eqnarray}
The running parameters Eqs.~(\ref{sol1})--~(\ref{sollast})
are inserted into Eq.~(\ref{isopot2}). The $\Lambda$-dependence is canceled
and we are left with the reference scale $\Lambda_0$. This scale
is eliminated using Eq.~(\ref{refscale2}). The final result is Eq.~(\ref{compli}).

\setcounter{equation}{0}
 \renewcommand{\theequation}{E\arabic{equation}}

\renewcommand{\thesection}{E}
\section{Renormalization of the three-flavor thermodynamic potential}
\label{omegarenorm}
In this section, we renormalize the thermodynamic potential for three flavors
with arbitrary quark masses, chemical potentials, and superconducting gaps.
\label{3fren}
\begin{eqnarray}
\nonumber
    \Omega_0^{\rm CFL} &=& {1\over4}m^2(\phi_u^2+\phi_d^2+2\phi_s^2)-h_u\phi_u-h_d\phi_d-\sqrt{2}h_s\phi_s + \frac{1}{16}\lambda_1(\phi_u^2+\phi_d^2+2\phi_s^2)^2 
    \\ && \nonumber
    + \frac{1}{16}\lambda_2(\phi_u^4+\phi_d^4+4\phi_s^4)\nonumber
    -\frac{1}{4}\sqrt{2}c\phi_u\phi_d\phi_s
   \\ && \nonumber
    +\frac{1}{4}\lambda_3(\phi_u^2+\phi_d^2+2\phi_s^2)(\Delta_{ud}^2+\Delta_{us}^2+\Delta_{ds}^2)
    +\frac{1}{4}\lambda_4[\phi_u^2\Delta_{ds}^2+\phi_d^2\Delta_{us}^2 +2\phi_s^2\Delta_{ud}^2]\nonumber\\
    && -\frac{1}{2}\lambda_{5}(\phi_u\phi_d\Delta_{ud}^2 + \sqrt{2}\phi_u\phi_s\Delta_{us}^2+\sqrt{2}\phi_d\phi_s\Delta_{ds}^2)\nonumber +(m_\Delta^2 - 4\bar{\mu}_{ud}^2)\Delta_{ud}^2 
    \\ &&\nonumber
    +(m_\Delta^2 - 4\bar{\mu}_{us}^2)\Delta_{us}^2
    + (m_\Delta^2 - 4\bar{\mu}_{ds}^2)\Delta_{ds}^2 + \frac{1}{4}\left(2\lambda^{\Delta}_{1}+\lambda_3^\Delta\right)(\Delta_{ud}^2+\Delta_{us}^2+\Delta_{ds}^2)^2
    \\ &&
    + \frac{1}{2}\lambda^{\Delta}_{2}(\Delta_{ud}^4+\Delta_{us}^4+\Delta_{ds}^4)\;,
\end{eqnarray}
where we have defined the symmetry-breaking parameters
$h_u$, $h_d$, and $h_s$ i Eqs.~(\ref{hu})--(\ref{hs}).
The quark-loop contribution to the effective potential is given by
Eq.~(\ref{omegafun}). As explained before, we isolate the 
ultraviolet divergences by expanding around zero chemical potentials 
and constructing suitable subtraction terms. After integrating over $p_0$, the divergent part of the thermodynamic potential is
\begin{eqnarray}
    \Omega_1^{\rm CFL, div}&=&
\nonumber
-\Lambda^{-2\epsilon}\int_p\left\{\left[
4\sqrt{p^2+m_u^2+g_{\Delta}^2\Delta_{ud}^2}+4\sqrt{p^2+m_u^2+g_{\Delta}^2\Delta_{us}^2}
\right.\right.
\\ && \nonumber
-2\sqrt{p^2+m_u^2} + u\leftrightarrow d + u\leftrightarrow s
\\ &&
\nonumber
+\left[
-{(m_u-m_d)^2g_{\Delta}^2\Delta_{ud}^2\over(p^2+g_{\Delta}^2\Delta_{ud}^2)^{3\over2}}
+{4\bar{\mu}_{ud}^2g_{\Delta}^2\Delta_{ud}^2\over(p^2+g_{\Delta}^2\Delta_{ud}^2)^{3\over2}} + u\rightarrow d + u\rightarrow s
\right]\\ &&\left.
-{g_{\Delta}^4(\Delta_{ud}^2\Delta_{us}^2+\Delta_{ud}^2\Delta_{ds}^2+\Delta_{us}^2\Delta_{ds}^2)\over(p^2+M^2)^{3\over2}}
\right\}\;,
\label{subti}
\end{eqnarray}
where $M$ is an arbitrary mass, which is needed to avoid introducing infrared divergences in Eq.~(\ref{subti}). 
The $M$-dependence drops out in the final result.
The chemical potentials and gaps
are symmetric under flavor permutations, $\bar{\mu}_{du}=\bar{\mu}_{ud}$ etc.
Performing the integrals using equations \eqref{i0} and \eqref{i2}, we find
\begin{eqnarray}
\nonumber
    \Omega_1^{\rm CFL, div}&=& \Lambda^{-2\epsilon}\left\{\frac{2(m_u^2+g_\Delta^2\Delta_{ud}^2)^2}{(4\pi)^2}\left[\frac{1}{\epsilon}+\log\frac{\Lambda^2}{m_u^2+g_\Delta^2\Delta_{ud}^2}+\frac{3}{2}\right] \right.
    \\ &&\nonumber
    \left.
    + \frac{2(m_u^2+g_\Delta^2\Delta_{us}^2)^2}{(4\pi)^2}\left[\frac{1}{\epsilon}+\log\frac{\Lambda^2}{m_u^2+g_\Delta^2\Delta_{us}^2}+\frac{3}{2}\right] \right.\nonumber\\
    \nonumber
&& \left.-\frac{m_u^4}{(4\pi)^2}\left[\frac{1}{\epsilon}+\log\frac{\Lambda^2}{m_u^2}+\frac{3}{2}\right]+ u\leftrightarrow d + u \leftrightarrow s\right\} 
\\ && \nonumber
+ \left\{\frac{4(m_u-m_d)^2g_\Delta^2\Delta_{ud}^2}{(4\pi)^2}\left[\frac{1}{\epsilon}+\log\frac{\Lambda^2}{g_\Delta^2\Delta_{ud}^2}\right]\right.\nn\\
&&\left.- \frac{16\bar{\mu}_{ud}^2g_\Delta^2\Delta_{ud}^2}{(4\pi)^2}\left[\frac{1}{\epsilon}+\log\frac{\Lambda^2}{g_\Delta^2\Delta_{ud}^2}\right]+ u\rightarrow s + d \rightarrow s  \right\}\nonumber\\
&& +\frac{4g_\Delta^4(\Delta_{ud}^2\Delta_{us}^2 + \Delta_{ud}^2\Delta_{ds}^2+\Delta_{us}^2\Delta_{ds}^2)\Lambda^{-2\epsilon}}{(4\pi)^2}\left[\frac{1}{\epsilon}+\log\frac{\Lambda^2}{M^2}\right]\;,
\end{eqnarray}
The counterterms are generated by the tree-level thermodynamic potential as before. Except for $\delta c_{\ms}$, the other counterterms in the scalar sector
are the same as in the two-flavor case~\cite{kumari,threeflavors}. This counterterm
reads
\begin{eqnarray}
\delta c_{\ms}&=&{3N_ccg^2\Lambda^{-3\epsilon}\over2(4\pi)^2\epsilon}\;,
\end{eqnarray}
and the coupling $c_{\ms}$ runs as
\begin{eqnarray}
    c_{\ms}(\Lambda)&=&{c_0\over\left[1-{N_cg_0^2\over(4\pi)^2}\log{\Lambda^2\over\Lambda_0^2}\right]^{3\over2}}\;.
\end{eqnarray}
The new counterterms in the diquark sector are
\begin{eqnarray}
\label{mdelc}
\delta m_{\Delta,\ms}^2&=&{4g_{\Delta}^2m_{\Delta}^2\Lambda^{-2\epsilon}\over(4\pi)^2\epsilon}
\;,\\
    \delta\lambda_{3,\ms} &=& \frac{\left[3g^2\lambda_3+4g_\Delta^2\lambda_3-8g^2g_\Delta^2\right]\Lambda^{-4\epsilon}}{(4\pi)^2\epsilon}\;,
    \\
   \delta\lambda_{4,\ms} &=& \frac{\left[3g^2\lambda_4+4g_\Delta^2\lambda_4+8g^2g_\Delta^2\right]\Lambda^{-4\epsilon}}{(4\pi)^2\epsilon}\;,\\
    \delta\lambda_{5,\ms} &=& \frac{\left[3\lambda_5g^2+4\lambda_5g_\Delta^2 - 4g^2g_{\Delta}^{2}\right]\Lambda^{-4\epsilon}}{(4\pi)^2\epsilon}\;,
\\
    \delta\lambda^{\Delta}_{1,\ms} &=& \frac{4g_\Delta^2\left[2\lambda^{\Delta}_{1}-g_\Delta^2\right]\Lambda^{-4\epsilon}}{(4\pi)^2\epsilon}\;,\\
    \delta\lambda^{\Delta}_{2,\ms} &=& \frac{4g_\Delta^2\left[2\lambda^{\Delta}_{2}-g_\Delta^2\right]\Lambda^{-4\epsilon}}{(4\pi)^2\epsilon}\;,
\\
    \delta\lambda^{\Delta}_{3,\ms} &=& \frac{8g_\Delta^2\lambda_{3}^\Delta\Lambda^{-4\epsilon}}{(4\pi)^2\epsilon}\;,
    \\
    \label{dg}
 \delta g_{\Delta,\ms}^2&=&{4g_{\Delta}^4\Lambda^{-2\epsilon}\over(4\pi)^2\epsilon}\;,\\
  \delta Z^{\Delta}_{\ms}&=&-{4g_{\Delta}^2\Lambda^{-2\epsilon}\over(4\pi)^2\epsilon}\;.
  \label{bolge}
\end{eqnarray}
The counterterms in the diquark sector are determined by collecting
terms of the same type in the thermodynamic potential. For example, collecting
terms that are proportional to $\bar{\mu}_{ud}^2\Delta_{ud}^2$, one finds
\begin{eqnarray}
-4\bar{\mu}_{ud}^2\Delta_{ud}^2\left[1+\delta Z_{\Delta}+{4g_{\Delta}^2\Lambda^{-2\epsilon}\over(4\pi)^2\epsilon}\right]\;.    
\end{eqnarray}
Requiring that the pole in $\epsilon$ be eliminated
gives Eq.~(\ref{bolge}).
Since there is no one-loop correction to $m_{\Delta}^2\Delta_{ud}^2$,
we find $\delta m_{\Delta}^2\Delta_{ud}^2+m_{\Delta}^2\delta Z_{\Delta}\Delta_{ud}^2=0$, which gives Eq.~(\ref{mdelc}). The other counterterms
are determined in the same manner, except for $g_{\Delta}^2$.
$\delta g_{\Delta}^2$ is determined by noting that there is no loop correction to
the quark-diquark vertex and no wavefunction renormalization of the quark fields in the mean-field approximation. This yields 
$\delta g_{\Delta}^2=-g_{\Delta}^2\delta Z_{\Delta}$ and 
Eq.~(\ref{dg}).

The new running parameters satisfy renormalization group equations.
These are found using the fact that the bare parameters are independent of the
scale. The solutions to these equations are
\begin{eqnarray}
    m_{\Delta,\ms}^2(\Lambda)&=&{m_{\Delta,0}^2\over\left[1-{4g_{\Delta,0}^2\over(4\pi)^2}\log{\Lambda^2\over\Lambda_0^2}\right]}\;,\\
    \lambda_{3,\rm\ms}(\Lambda) &=& \frac{\lambda_{3,0}-\frac{8g_0^2g_{\Delta,0}^2}{(4\pi)^2}\log\frac{\Lambda^2}{\Lambda_0^2}}{\left[1-\frac{3g_{0}^2}{(4\pi)^2}\log\frac{\Lambda^2}{\Lambda_0^2}\right]\left[1-\frac{4g_{\Delta,0}^2}{(4\pi)^2}\log\frac{\Lambda^2}{\Lambda_0^2}\right]}\\
    \lambda_{4,\rm \ms}(\Lambda) &=& \frac{\lambda_{4,0}+\frac{8g_0^2g_{\Delta,0}^2}{(4\pi)^2}\log\frac{\Lambda^2}{\Lambda_0^2}}{\left[1-\frac{3g_{0}^2}{(4\pi)^2}\log\frac{\Lambda^2}{\Lambda_0^2}\right]\left[1-\frac{4g_{\Delta,0}^2}{(4\pi)^2}\log\frac{\Lambda^2}{\Lambda_0^2}\right]}\;,\\
    \lambda_{5,\rm \ms}(\Lambda) &=& \frac{\lambda_{5,0}-\frac{4g_0^2g_{\Delta,0}^2}{(4\pi)^2}\log\frac{\Lambda^2}{\Lambda_0^2}}{\left[1-\frac{3g_{0}^2}{(4\pi)^2}\log\frac{\Lambda^2}{\Lambda_0^2}\right]\left[1-\frac{4g_{\Delta,0}^2}{(4\pi)^2}\log\frac{\Lambda^2}{\Lambda_0^2}\right]}\;\\
        \lambda^{\Delta}_{1,\rm\ms}(\Lambda) &=& \frac{\lambda^{\Delta}_{1,0}-\frac{4g_{\Delta,0}^4}{(4\pi)^2}\log\frac{\Lambda^2}{\Lambda_0^2}}{\left[1-\frac{4g_{\Delta,0}^2}{(4\pi)^2}\log\frac{\Lambda^2}{\Lambda_0^2}\right]^2}\;,\\
    \lambda^{\Delta}_{2,\rm\ms}(\Lambda) &=& \frac{\lambda^{\Delta}_{2,0}-\frac{4g_{\Delta,0}^4}{(4\pi)^2}\log\frac{\Lambda^2}{\Lambda_0^2}}{\left[1-\frac{4g_{\Delta,0}^2}{(4\pi)^2}\log\frac{\Lambda^2}{\Lambda_0^2}\right]^2}\;,\\
    \lambda^{\Delta}_{3,\rm \ms}(\Lambda) &=& \frac{\lambda_{\Delta3,0}}{\left[1-\frac{4g_{\Delta,0}^2}{(4\pi)^2}\log\frac{\Lambda^2}{\Lambda_0^2}\right]^2}\;,\\
    \Delta_{\ms}^2(\Lambda)&=&\left[1-{4g_{\Delta,0}^2\over(4\pi)^2}\log{\Lambda^2\over\Lambda_0^2}\right]\Delta_{0}^2\;,
\end{eqnarray}
where the subscript indicates that a parameter is evaluated at the reference scale $\Lambda_0$.
The final renormalized thermodynamic potential in the CFL phase is
 \begin{eqnarray}
\nonumber
    \Omega^{\rm CFL}_{0+1} &=& \Omega_0^{\rm CFL} +  \left\{\frac{2(m_u^2+g_\Delta^2\Delta_{ud}^2)^2}{(4\pi)^2}\left[\log\frac{\Lambda^2}{m_u^2+g_\Delta^2\Delta_{ud}^2}+\frac{3}{2}\right]
    +\right.\\ &&\left.
    + \frac{2(m_u^2+g_\Delta^2\Delta_{us}^2)^2}{(4\pi)^2}\left[\log\frac{\Lambda^2}{m_u^2+g_\Delta^2\Delta_{us}^2}+\frac{3}{2}\right] \right.\nonumber\\
&& \left.-\frac{m_u^4}{(4\pi)^2}\left[\log\frac{\Lambda^2}{m_u^2}+\frac{3}{2}\right]+ u\leftrightarrow d + u \leftrightarrow s\right\} + \left\{\frac{4(m_u-m_d)^2g_\Delta^2\Delta_{ud}^2}{(4\pi)^2}\log\frac{\Lambda^2}{g_\Delta^2\Delta_{ud}^2}\right.\nn\\
&&\left.- \frac{16\bar{\mu}_{ud}^2g_\Delta^2\Delta_{ud}^2}{(4\pi)^2}\log\frac{\Lambda^2}{g_\Delta^2\Delta_{ud}^2}+ u\rightarrow s + d \rightarrow s  \right\}\nonumber\\
&& +\frac{4g_\Delta^4(\Delta_{ud}^2\Delta_{us}^2 + \Delta_{ud}^2\Delta_{ds}^2+\Delta_{us}^2\Delta_{ds}^2)}{(4\pi)^2}\log\frac{\Lambda^2}{M^2}\; + \Omega_{1}^{\rm CFL, fin}\;,
\label{komplett}
\end{eqnarray}
where the last term is the difference between the exact one-loop contribution to the thermodynamic potential and the subtraction term.
This term must be evaluated numerically.

Finally, we consider a few approximations that allow us to obtain
analytical and yet realistic results.
In the following, we assume that \(m_u = m_d = m_l, \Delta_{us}=\Delta_{ds} = \Delta_{ls}\), redefine \(\Delta_{ud}=\Delta_{ll}\) and a common chemical potential. As explained before, we isolate the 
ultraviolet divergences by expanding around zero chemical potentials 
and constructing suitable subtraction terms. After integrating over $p_0$, the divergent part of the thermodynamic potential is
\begin{eqnarray}
    \Omega_1^{\rm CFL, div} &=& -\int_p\left[4\sqrt{p^2+\hat m_+^2}+4\sqrt{p^2+\hat m_-^2} + 6\sqrt{p^2+m_l^2+\Delta_{ll}^2}\right.\nn\\
    &&+ 2\sqrt{p^2+\tilde m_+^2}+2\sqrt{p^2+\tilde m_-^2} + \frac{3\mu^2\Delta_{ll}^2}{(p^2+\Delta_{ll}^2)^{3/2}} + \frac{4\mu^2\Delta_{ls}^2}{(p^2+\Delta_{ls}^2)^{3/2}}\nn\\
    &&\left.+ \frac{\mu^2\bar\Delta_+^2}{(p^2+\bar\Delta_+^2)^{3/2}}+ \frac{\mu^2\bar\Delta_-^2}{(p^2+\bar\Delta_-^2)^{3/2}}\right]\;,
\end{eqnarray}
where 
\begin{eqnarray}
    \hat m_\pm^2 &=& \frac{1}{2}\left[m_l^2+m_s^2 +2\Delta_{ls}^2 \pm (m_l-m_s)\sqrt{(m_l+m_s)^2+4\Delta_{ls}^2}\right]\;,\\
    \tilde m_\pm^2 &=& \frac{1}{2}\left[m_l^2+m_s^2+\Delta_{ll}^2 + 4\Delta_{ls}^2\pm\sqrt{(m_l^2-m_s^2+\Delta_{ll}^2)^2 + 8[(m_l-m_s)^2+\Delta_{ll}^2]\Delta_{ls}^2}\right]\;,\\
    \bar\Delta_\pm^2 &=& \frac{1}{2}\left[\Delta_{ll}^2+4\Delta_{ls}^2 \pm\sqrt{\Delta_{ll}^4+8\Delta_{ll}^2\Delta_{ls}^2}\right]\;.
\end{eqnarray}
Performing the integrals using equations \eqref{i0} and \eqref{i2} we find
\begin{eqnarray}
    \Omega_1^{\rm CFL, div} &=& \frac{2(\hat m_\pm^2)^2}{(4\pi)^2}\left[\frac{1}{\epsilon}+\frac{3}{2}+\log\frac{\Lambda^2}{\hat m_\pm^2}\right]+\frac{3(m_l^2+\Delta_{ll}^2)^2}{(4\pi)^2}\left[\frac{1}{\epsilon}+\frac{3}{2}+\log\frac{\Lambda^2}{m_l^2+\Delta_{ll}^2}\right]\nn\\
    && + \frac{(\tilde m_\pm^2)^2}{(4\pi)^2}\left[\frac{1}{\epsilon} + \frac{3}{2}+\log\frac{\Lambda^2}{\tilde m_\pm^2}\right] - \frac{4\bar\Delta_\pm^2\mu^2}{(4\pi)^2}\left[\frac{1}{\epsilon}+\log\frac{\Lambda^2}{\bar\Delta_\pm^2}\right]\nn\\
    && - \frac{12\mu^2\Delta_{ll}^2}{(4\pi)^2}\left[\frac{1}{\epsilon}+\log\frac{\Lambda^2}{\Delta_{ll}^2}\right] - \frac{16\mu^2\Delta_{ls}^2}{(4\pi)^2}\left[\frac{1}{\epsilon}+\log\frac{\Lambda^2}{\Delta_{ls}^2}\right]\;.
\end{eqnarray}
The finite contribution $\Omega_1^{\rm CFL,finite}$
is numerically difficult, but simplifies significantly if we set
the gaps to zero in this term. It then reads
\begin{eqnarray}
    \Omega_1^{\rm CFL,finite} &=& -2N_c\int_p\left[2(\mu-E)\theta(\mu-E) + (\mu-E_s)\theta(\mu-E_s)\right]\;,
\end{eqnarray}
where
\begin{eqnarray}
    E = \sqrt{p^2+m_l^2}\quad {\rm and }\quad  E_s =\sqrt{p^2+m_s^2}\;.
\end{eqnarray}
This is the standard finite-density contribution for massive fermions 
of mass $m_l$ and $m_s$ with chemical potential $\mu$.

\begin{adjustwidth}{-\extralength}{0cm}

\reftitle{References}




\isAPAandChicago{}{%

}

\PublishersNote{}
\end{adjustwidth}
\end{document}